%% file: ms.tex
\documentclass[numberedappendix,twocolappendix,revtex4]{emulateapj}
\usepackage{apjfonts}
\usepackage{natbib}
\usepackage{graphics}
\usepackage{multirow}
\usepackage{amsbsy}
\usepackage{mathrsfs}
\usepackage{footmisc}
\usepackage{hyperref}
\usepackage{amsmath}
\usepackage{mathtools}
\usepackage{amssymb}

\hypersetup{
  colorlinks = true,
  urlcolor = black,
  linkcolor=black,
  citecolor=black
}
\input{defs}

\shorttitle{Matrix based $E/B$ Separation}

\begin{document}

\title{\biceptwo\ / \keckarray\ VII:
Matrix based $E/B$ Separation applied to \biceptwo\ and the \keckarray}

\author{\keckarray\ and \biceptwo\ Collaborations:
 P.~A.~R.~Ade\altaffilmark{1}}
\author{Z.~Ahmed\altaffilmark{2,3}}
\author{R.~W.~Aikin\altaffilmark{4}}
\author{K.~D.~Alexander\altaffilmark{5}}
\author{D.~Barkats\altaffilmark{5}}
\author{S.~J.~Benton\altaffilmark{6}}
\author{C.~A.~Bischoff\altaffilmark{5}}
\author{J.~J.~Bock\altaffilmark{7,8}}
\author{R.~Bowens-Rubin\altaffilmark{5}}
\author{J.~A.~Brevik\altaffilmark{7}}
\author{I.~Buder\altaffilmark{5}}
\author{E.~Bullock\altaffilmark{9}}
\author{V.~Buza\altaffilmark{5,10}}
\author{J.~Connors\altaffilmark{5}}
\author{B.~P.~Crill\altaffilmark{8}}
\author{L.~Duband\altaffilmark{11}}
\author{C.~Dvorkin\altaffilmark{10}}
\author{J.~P.~Filippini\altaffilmark{7,12}}
\author{S.~Fliescher\altaffilmark{9}}
\author{J.~Grayson\altaffilmark{2}}
\author{M.~Halpern\altaffilmark{13}}
\author{S.~Harrison\altaffilmark{5}}
\author{S.~R.~Hildebrandt\altaffilmark{7,8}}
\author{G.~C.~Hilton\altaffilmark{14}}
\author{H.~Hui\altaffilmark{7}}
\author{K.~D.~Irwin\altaffilmark{2,3}}
\author{J.~Kang\altaffilmark{2,3}}
\author{K.~S.~Karkare\altaffilmark{5}}
\author{E.~Karpel\altaffilmark{2}}
\author{J.~P.~Kaufman\altaffilmark{15}}
\author{B.~G.~Keating\altaffilmark{15}}
\author{S.~Kefeli\altaffilmark{7}}
\author{S.~A.~Kernasovskiy\altaffilmark{1}}
\author{J.~M.~Kovac\altaffilmark{5,10}}
\author{C.~L.~Kuo\altaffilmark{2,3}}
\author{E.~M.~Leitch\altaffilmark{16}}
\author{M.~Lueker\altaffilmark{7}}
\author{K.~G.~Megerian\altaffilmark{8}}
\author{T.~Namikawa\altaffilmark{2,3}}
\author{C.~B.~Netterfield\altaffilmark{6,17}}
\author{H.~T.~Nguyen\altaffilmark{8}}
\author{R.~O'Brient\altaffilmark{7,8}}
\author{R.~W.~Ogburn~IV\altaffilmark{2,3}}
\author{A.~Orlando\altaffilmark{7}}
\author{C.~Pryke\altaffilmark{9,18}}
\author{S.~Richter\altaffilmark{5}}
\author{R.~Schwarz\altaffilmark{18}}
\author{C.~D.~Sheehy\altaffilmark{16,18}}
\author{Z.~K.~Staniszewski\altaffilmark{7,8}}
\author{B.~Steinbach\altaffilmark{7}}
\author{R.~V.~Sudiwala\altaffilmark{1}}
\author{G.~P.~Teply\altaffilmark{7,15}}
\author{K.~L.~Thompson\altaffilmark{2,3}}
\author{J.~E.~Tolan\altaffilmark{2,21}}
\author{C.~Tucker\altaffilmark{1}}
\author{A.~D.~Turner\altaffilmark{8}}
\author{A.~G.~Vieregg\altaffilmark{5,16,19}}
\author{A.~C.~Weber\altaffilmark{8}}
\author{D.~V.~Wiebe\altaffilmark{13}}
\author{J.~Willmert\altaffilmark{18}}
\author{C.~L.~Wong\altaffilmark{5,10}}
\author{W.~L.~K.~Wu\altaffilmark{2,20}}
\author{K.~W.~Yoon\altaffilmark{2}}

\altaffiltext{1}{School of Physics and Astronomy, Cardiff University, Cardiff, CF24 3AA, United Kingdom}
\altaffiltext{2}{Department of Physics, Stanford University, Stanford, CA 94305, USA}
\altaffiltext{3}{Kavli Institute for Particle Astrophysics and Cosmology, SLAC National Accelerator Laboratory, 2575 Sand Hill Rd, Menlo Park, CA 94025, USA}
\altaffiltext{4}{Department of Physics, California Institute of Technology, Pasadena, California 91125, USA}
\altaffiltext{5}{Harvard-Smithsonian Center for Astrophysics, 60 Garden Street MS 42, Cambridge, Massachusetts 02138, USA}
\altaffiltext{6}{Department of Physics, University of Toronto, Toronto, Ontario, M5S 1A7, Canada}
\altaffiltext{7}{Department of Physics, California Institute of Technology, Pasadena, California 91125, USA}
\altaffiltext{8}{Jet Propulsion Laboratory, Pasadena, California 91109, USA}
\altaffiltext{9}{Minnesota Institute for Astrophysics, University of Minnesota, Minneapolis, Minnesota 55455, USA}
\altaffiltext{10}{Department of Physics, Harvard University, Cambridge, MA 02138, USA}
\altaffiltext{11}{Service des Basses Temp\'{e}ratures, Commissariat \`{a} l'Energie Atomique, 38054 Grenoble, France}
\altaffiltext{12}{Department of Physics, University of Illinois at Urbana-Champaign, Urbana, Illinois 61801, USA}
\altaffiltext{13}{Department of Physics and Astronomy, University of British Columbia, Vancouver, British Columbia, V6T 1Z1, Canada}
\altaffiltext{14}{National Institute of Standards and Technology, Boulder, Colorado 80305, USA}
\altaffiltext{15}{Department of Physics, University of California at San Diego, La Jolla, California 92093, USA}
\altaffiltext{16}{Kavli Institute for Cosmological Physics, University of Chicago, Chicago, IL 60637, USA} 
\altaffiltext{17}{Canadian Institute for Advanced Research, Toronto, Ontario, M5G 1Z8, Canada}
\altaffiltext{18}{School of Physics and Astronomy, University of Minnesota, Minneapolis, Minnesota 55455, USA}
\altaffiltext{19}{Department of Physics, Enrico Fermi Institute, University of Chicago, Chicago, IL 60637, USA}
\altaffiltext{20}{Department of Physics, University of California, Berkeley, CA
94720, USA}
\altaffiltext{21}{Corresponding author: \mbox{jetolan@stanford.edu}}

\begin{abstract}
A linear polarization field on the sphere can be uniquely decomposed into 
an \emode\ and a \bmode\ component. These two components are analytically 
defined in terms of spin-2 spherical harmonics. Maps that contain filtered 
modes on a partial sky can also be decomposed into \emode\ and \bmode\ 
components. However, the lack of full sky information prevents 
orthogonally separating these components using spherical harmonics. 
In this paper, we present a technique for decomposing an incomplete map into 
$E$ and \bmode\ components using $E$ and $B$ eigenmodes of the pixel 
covariance in the observed map. This method
is found to orthogonally define $E$ and $B$ in the presence of both partial 
sky coverage and spatial filtering. This method has been applied to the
\biceptwo\ and the \keckarray\ maps and results in reducing $E$ to $B$ leakage 
from \lcdm\ \emodes\ to a level corresponding to a tensor-to-scalar ratio of 
$r<1\times10^{-4}$.
\end{abstract}


\section{Introduction}

\setcounter{footnote}{1}

Current experiments are producing low noise maps of the polarization of the 
cosmic microwave background (CMB) radiation able to constrain models of 
inflation and measure \bmodes\ from gravitational lensing. These experiments 
include \biceptwo, the \keckarray, \polarbear, \sptpol, \actpol, and \planck\ 
\citep{bicep2_res, keck14, sptlensing13, polarbear14, actpol2014, planckI}. 
These experiments do not measure the CMB over the 
entire sky for a variety of reasons. Galactic foregrounds prevent any 
experiment from producing a map of the CMB over 
the entire sky. Any ground or balloon based experiment has a limited view of 
the full sky. Some experiments, including \biceptwo\ and the \keckarray, choose 
to observe a limited field of view to increase map depth over a small 
region of sky or choose to filter their data so that the maps 
incompletely measure the modes within the field.

The ability to uniquely separate a linear polarization field 
into $E$ and \bmodes\ is critical for measuring gravitational waves 
using the \bmode\ polarization. This separation allows the distinction 
to be made between \emodes\ created by scalar perturbations and \bmodes\ 
coming from tensor perturbations \citep{Kamionkowski_97c,Zaldarriaga98b}.

Unfortunately, the unique decomposition into $E$ 
and $B$ is only possible for maps of the full sky. 
Maps containing a limited view of the sky, or an incomplete measurement of 
the true sky modes, are said to suffer from $E/B$ leakage. 
$E$ to $B$ leakage is defined as measured power for a particular \bmode\ 
estimator whose source is true sky \emode\ power. $B$ to $E$ leakage is 
leakage of power in the opposite direction, but in practice it is less of a 
concern for CMB measurements due to the much fainter \bmode\ signal. 
$E/B$ leakage refers to both types of leakage.

There are several ways to mitigate the effect of $E/B$ leakage in analysis. 
Full pixel-space likelihood methods in principle can
optimally separate $E$ and $B$ contributions for any given map.
These have been applied
mainly to maps of relatively modest pixel count,
including many early detections of CMB polarization 
\citep[for example,][]{kovac2002,readhead2004,bischoff2008}.
Current analyses more commonly apply fixed estimators of 
$E$ and $B$ power spectra to observed CMB polarization maps.
The simplest way to correct such estimators for leakage
is to run an ensemble of simulations through the analysis and 
subtract the mean level of leakage in the angular power spectrum. However, 
the sample variance from the leaked power remains and contributes to the 
final uncertainty of measured power in each angular power spectrum bin, limiting 
an experiment's ability to measure \bmodes\ regardless of its instrumental 
sensitivity. For many experiments, including \biceptwo\ and the \keckarray, 
the sample variance of the leaked \emodes\ is comparable to the instrumental 
noise and is a significant contribution to the uncertainty in the \bmode\ 
power spectrum. 

Solutions to this problem rely on the fact that for most 
\bmode\ science it is not necessary to classify all the modes in the measured 
polarization field. Instead, it is sufficient to find subspaces 
that are caused by either $E$ or $B$ and ignore the modes whose source 
cannot be determined. There are a number of published methods that attempt 
this goal.

\citet{smith06} presents an estimator that does not suffer from $E/B$ leakage 
arising from partial sky coverage. This method has been incorporated into the 
\href{http://www.apc.univ-paris7.fr/APC_CS/Recherche/Adamis/MIDAS09/software/pures2hat/pureS2HAT.html}{\xpure\ and S$^2$HAT packages} \citep{grain09}, 
and the \biceptwo\ and \keckarray\ analysis pipeline contains an option in 
which this algorithm is implemented. 

However, many experiments, including \biceptwo\ and the \keckarray, produce 
maps in which some modes have also been removed by filtering. The estimator 
presented in \citet{smith06} does not prevent filtered modes from creating 
$E/B$ leakage. Another method, presented in \citet{Smith_Zald06}, accounts 
for incomplete mode measurement in partial sky maps. However, we have found 
this method to be computationally infeasible for the \biceptwo\ and \keckarray\ 
observing and filtering strategy.

For \biceptwo\ and the \keckarray, we developed a new method for 
distinguishing true sky \bmode\ polarization from the leaked \emodes\ in 
the observed maps. The method extends the work of 
\citet{Bunn02} and applies it to a real data set. It is a standard 
component of the \biceptwo\ and \keckarray\ 
analysis pipeline and effectively eliminates the uncertainty created by 
$E/B$ leakage. The method reduces the final uncertainty in the measured 
$BB$ power spectrum of the \biceptwo\ results \citep{bicep2_res} by more 
than a factor of two, compared to analysis done with the \citet{smith06} 
method. The method results in a larger improvement for the analysis of the 
combined \biceptwo\ and \keckarray\ maps \citep{keck14}, where the noise 
levels are lower. 

The organization of this paper is as follows: Section \ref{sec:eb_pol} provides 
an abbreviated background of a polarization field on a sphere, 
decomposition into spin-2 spherical harmonics, and analytically defines $E$ and 
\bmodes. Section \ref{sec:matrixEB} outlines the eigenvalue problem used in the 
matrix based $E/B$ separation. Section \ref{sec:reob_matrix} describes how an 
observation matrix is created in the \biceptwo\ and \keckarray\ analysis 
pipeline. Section \ref{section:ct} describes constructing the signal 
covariance matrix and Section \ref{sec:ebseparation} uses the covariance 
matrix to solve the eigenvalue problem and find purification matrices. Section 
\ref{sec:matrix_app} prensents results of matrix based $E/B$ separation 
in the \biceptwo\ data set. Concluding remarks are offered in Section 
\ref{sec:conclusion}. 

Unless otherwise stated, we adopt the 
\healpix\ polarization convention\footnote{\url{http://healpix.jpl.nasa.gov/html/intronode12.htm}} and work in J2000.0 equatorial coordinates 
throughout this paper. Bold font letters and symbols represent vectors or 
matrices, even when containing subscripts, in which case the subscript is 
meant to designate a new matrix or vector. Normal font letters and 
symbols represent scalar quantities.

\section{$E$ and $B$-modes from a polarization field}
\label{sec:eb_pol}
This section demonstrates the decomposition of a polarization field on the 
full sky into $E$ and \bmodes. Much of the discussion follows 
\citet{Zaldarriaga96} and \citet{Bunn02}. 

\subsection{Full sky}

 The values of the Stokes parameters $Q$ and $U$ for a particular location 
on the sky are dependent on the choice of coordinate system. By rotating the 
local coordinate system, $Q$ is rotated into $U$ and vice versa. Under rotation 
by an angle $\phi$, the combinations $Q+iU$ and $Q-iU$ transform as:
\begin{align}
(Q + iU )' &= e^{-2i\phi}(Q+iU)
\nonumber\\
(Q - iU )' &= e^{2i\phi}(Q-iU).
\label{eqn:Q_Uspin2}
\end{align}
The $T$, $Q$, and $U$ fields can be expressed as sums of spin weighted 
spherical harmonics. While the temperature anisotropies can be broken 
down into spin-0 harmonics, the polarization field of $Q$ and $U$ must be 
expressed in terms of spin-2 spherical harmonics \citep{Goldberg67}:

\begin{align}
T(\mathbf{r}) &= \sum_{lm}a^T_{lm}\left(\,_0Y_{lm}(\mathbf{r})\right) 
\nonumber\\
(Q + iU )(\mathbf{r}) &= \sum_{lm}a_{+2,lm}\left(\,_{+2}Y_{lm}(\mathbf{r})\right) 
\nonumber\\
(Q - iU )(\mathbf{r}) &= \sum_{lm}a_{-2,lm}\left(\,_{-2}Y_{lm}(\mathbf{r})\right),
\label{eqn:Q_U}
\end{align}
where $\,_{\pm{2}}Y_{lm}$ are the spin-2 case of spin weighted spherical 
harmonics, and the spin-0 case are the normal spherical 
harmonics, $\,_0Y_{lm}$. Since $Q+iU$ and $Q-iU$ are affected by rotations 
of the coordinate system, it is convenient to express the coefficients of 
the spin-2 spherical harmonics using a set of coordinate independent scalar 
$a^E_{lm}$ coefficients and pseudo-scalar $a^B_{lm}$ coefficients: 
\begin{align}
a^{E}_{lm}&\equiv-(a_{+2,lm}+a_{-2,lm})/2
\nonumber\\
a^{B}_{lm}&\equiv-i(a_{+2,lm}-a_{-2,lm})/2. 
\label{eqn:aeabdef}
\end{align}
We also define two combinations of spin-2 spherical harmonics:
\begin{align}
X_{1,lm}&\equiv(\,_{+2}Y_{lm}+\,_{-2}Y_{lm})/2
\nonumber\\
X_{2,lm}&\equiv(\,_{+2}Y_{lm}-\,_{-2}Y_{lm})/2.
\label{eqn:Xdef}
\end{align}

We can use the coefficients in Equation \ref{eqn:aeabdef} and the 
combinations in Equation \ref{eqn:Xdef} to construct real space forms 
of $T$, $Q$, and $U$ fields, according to Equation \ref{eqn:Q_U}:
\begin{align}
T(\mathbf{r}) &= \sum_{lm}a^{T}_{lm}(\,_0Y_{lm}(\mathbf{r})) 
\nonumber\\
Q(\mathbf{r}) &= - \sum_{lm}\left(a^{E}_{lm}X_{1,lm}(\mathbf{r})+ ia^{B}_{lm}X_{2,lm}(\mathbf{r})\right)
\nonumber\\
U(\mathbf{r}) &= - \sum_{lm}\left(a^{B}_{lm}X_{1,lm}(\mathbf{r})- ia^{E}_{lm}X_{2,lm}(\mathbf{r})\right).
\label{eqn:TQU_asX}
\end{align}
Using these relations, we can write the polarization field as a vector:
\begin{align}
P(\mathbf{r}) &\equiv \left(\begin{matrix}Q(\mathbf{r})\\U(\mathbf{r})\end{matrix}\right)
\nonumber\\
&=-\sum_{lm}\left[
\begin{array}{c}
  a^{E}_{lm}X_{1,lm}(\mathbf{r})+ia^{B}_{lm}X_{2,lm}(\mathbf{r})\\
  a^{B}_{lm}X_{1,lm}(\mathbf{r})-ia^{E}_{lm}X_{2,lm}(\mathbf{r})\\ 
\end{array}
\right]
\nonumber\\
&=-\sum_{lm}\left[a^{E}_{lm}\left(\begin{matrix}X_{1,lm}(\mathbf{r})\\-iX_{2,lm}(\mathbf{r})\end{matrix}\right)
+a^{B}_{lm}\left(\begin{matrix}iX_{2,lm}(\mathbf{r})\\X_{1,lm}(\mathbf{r})\end{matrix}\right)\right]
\nonumber\\
&=-\sum_{lm}\left[a^{E}_{lm}Y^{E}_{lm}(\mathbf{r})+a^{B}_{lm}Y^{B}_{lm}(\mathbf{r})\right],
\label{eqn:pfield}
\end{align}
where $Y^{E}_{lm}$ and $Y^{B}_{lm}$ have been introduced and defined in the 
last step. On the full sphere, $Y^{E}_{lm}$ and $Y^{B}_{lm}$ are orthogonal:
\begin{equation}
\int_{S^2}Y^E_{lm}(\mathbf{r})\cdot{Y}^B_{l'm'}(\mathbf{r})dS=0,
\label{eqn:orthoEB}
\end{equation}
for all $l,l'$ and $m,m'$.

\subsection{Orthogonality of pure $E$ and pure $B$}
\label{sec:orthoEB}
The inner product of two polarization fields is defined as:
\begin{equation}
\mathbf{P}\cdot\mathbf{P}'\equiv\int_\Omega{\mathbf{P}\cdot\mathbf{P}'}d\Omega,
\end{equation}
where $\Omega$ is the manifold on which the polarization field is defined: 
for the full sky it is the celestial sphere. In pixelized maps, the vector 
space of a polarization field has a finite dimension: twice the number of 
pixels in the map.

As demonstrated in Equation \ref{eqn:orthoEB}, $E$ and \bmode\ polarization 
fields on the full sky are orthogonal. However, experiments produce $Q/U$ maps 
of portions of the sky, and often filter spatial modes out of these maps. We 
define the term `observed' maps or modes to refer to these incomplete 
measurements of the true sky. 

The spaces of observed \emodes\ and \bmodes\ are non-orthogonal. The 
overlapping subspace between the two is called the ambiguous space. We cannot 
tell whether signal in the ambiguous subspace came from full sky \emodes\ or 
full sky \bmodes.  

The solution is to decompose vector fields on an observed manifold into three 
subspaces: `pure' \emodes, `pure' \bmodes, and ambiguous 
modes. Pure $E$ and \bmodes\ are subspaces of the polarization 
vector space of a particular manifold, defined as:
\begin{itemize}
\item A pure \bmode\ is orthogonal to observed \emodes.
\item A pure \emode\ is orthogonal to observed \bmodes.
\end{itemize}
Therefore, a pure \bmode\ is one that has no $E$ to $B$ leakage: neither pure 
\emodes\ nor ambiguous modes contribute to it. 

\section{How matrix based $E/B$ separation finds pure $E$ and pure $B$}
\label{sec:matrixEB}

A pure \bmode\ on an observed manifold is defined in Section \ref{sec:orthoEB} 
as being orthogonal to observed \emodes: 
\begin{align}
\mathbf{P}^E\cdot\mathbf{b}=\mathbf{0}.
\end{align}
The vector $\mathbf{b}$ is any linear combination of modes 
in the subspace of the pure \bmodes. For pixelized maps, $\mathbf{b}$ contains 
$Q$ and $U$ values for each of the pixels in the map, and $\mathbf{P}^E$ is the 
pixelized version of the \emode\ spherical harmonics. It is useful to multiply 
the above equation by its conjugate transpose, and sum over $l$ and $m$, 
so that we have a scalar representing the degree of orthogonality:
\begin{equation}
\mathbf{b}^\top\left(\sum_{lm}{a^{E*}_{lm}a^{E}_{lm}\mathbf{Y}^{E}_{lm}}{\mathbf{Y}^{E\dagger}_{lm}}\right)\mathbf{b}=0.
\end{equation}
We have freedom to choose the power spectrum,
 ${C}^{EE}_l=\left<a^{E*}_{lm}a^{E}_{lm}\right>$, which is included in the 
covariance matrix, $\mathbf{C}_E$:
\begin{equation}
\mathbf{C}_E\equiv\sum_{lm}{C}^{EE}_l\mathbf{Y}^{E}_{lm}{\mathbf{Y}^{E\dagger}_{lm}}.
\label{eqn:covariancedef}
\end{equation}
We note that this product is the $2\times2$ [$Q,U$] covariance
 block in the signal covariance matrix: 
\begin{align}
\mathbf{C}_E&=\left<\mathbf{P}^E\left(\mathbf{P}^E\right)^\top\right>
\nonumber\\
&=
\left(\begin{matrix}
\left<{Q^E_i}{Q^E_j}\right>&\left<{Q^E_i}{U^E_j}\right>\\
\left<{U^E_i}{Q^E_j}\right>&\left<{U^E_i}{U^E_j}\right>\\
\end{matrix}\right),
\end{align}
where the superscript denotes the \emode\ component of the full sky 
polarization field and $i,j$ designate pixels in the map. We can evaluate the covariance matrix for a particular 
set of pixels and a chosen spectrum. 

By solving a generalized eigenvalue equation of the form:
\begin{align}
\mathbf{C}_B\mathbf{x_i}={\lambda_i}\mathbf{C}_E\mathbf{x_i},
\label{eqn:firsteig}
\end{align}
and selecting eigenmodes corresponding to the largest eigenvalues, we can find 
eigenmodes $\mathbf{b}$ that are nearly orthogonal to \emodes\ and therefore 
approximate pure $B$. Eigenmodes corresponding to the smallest eigenvalues 
approximate pure $E$. This method is a natural extension to the 
signal to noise truncation discussed in \citet{Bond98} and \citet{bunn97} 
and applied in \citet{Kuo02}. The specific application to $E$ and \bmodes\ 
was first discussed in \citet{Bunn02}.

We say the modes approximate pure $E$ and pure \bmodes\ 
because the degree of orthogonality is proportional to the magnitude of the 
eigenvalues. The level of orthogonality is discussed further in Section 
\ref{sec:ebseparation}. However, for the remainder of the paper, we will use 
the terms pure $B$ and pure $E$ to refer to the largest and smallest eigenmodes 
of Equation \ref{eqn:firsteig}, despite the fact that their inner product is 
not identically zero.

Now suppose that the true sky polarization field, $\mathbf{P}$, is  
transformed into an observed polarization field, $\mathbf{\tilde{P}}$, by a
real space linear operation, $\mathbf{R}$:
\begin{align}
\mathbf{\tilde{P}^E}&=\mathbf{RP^E}
\nonumber\\
&=-\mathbf{R}\sum_{lm}a^E_{lm}{\mathbf{Y}^{E}_{lm}}.
\end{align}
Throughout this paper, transformations into observed quantities are indicated 
by the inclusion of a tilde over the variable, in the above equation, 
$\mathbf{P}\rightarrow\mathbf{\tilde{P}}$.
The operator $R$ will typically represent filtering operations
necessary to suppress noise and/or systematics plus an apodization
of the resulting observed maps.

The condition for pure $E$ and pure $B$ must be the same after 
multiplying by $\mathbf{R}$. We still demand that the vectors of pure 
$B$ be orthogonal to all those in the $E$ space, which includes both the 
pure \emodes\ and the ambiguous modes:
\begin{align}
\left(\mathbf{R}\sum_{lm}a^E_{lm}\mathbf{Y}^E_{lm}\right)\cdot\mathbf{b}=\mathbf{0}.
\label{eqn:rortho}
\end{align}

We create a basis of pure $E$ and pure \bmodes\ by solving the eigenvalue 
problem with the covariances of the form:
\begin{align}
\mathbf{\tilde{C}_E}=\mathbf{R}^{\top}\left(\sum_{lm}{C}^{EE}_l{\mathbf{Y}^{E}_{lm}}{\mathbf{Y}^{E\dagger}_{lm}}\right)\mathbf{R}
\nonumber\\
\mathbf{\tilde{C}_B}=\mathbf{R}^{\top}\left(\sum_{lm}{C}^{BB}_l{\mathbf{Y}^{B}_{lm}}{\mathbf{Y}^{B\dagger}_{lm}}\right)\mathbf{R},
\end{align}
so that Equation \ref{eqn:firsteig} becomes:
\begin{equation}
\mathbf{\tilde{C}}_B\mathbf{x_i}={\lambda}_{i}\mathbf{\tilde{C}}_E\mathbf{x_i}.
\label{eqn:geneigenfirst}
\end{equation}

In the simplest case, the matrix $\mathbf{R}$ is an apodization window and 
filled only on its diagonal. However, Equation \ref{eqn:rortho} does not 
necessitate that the real space operator be a diagonal matrix. Any analysis 
steps that can be expressed as linear operations can be included.

In \biceptwo\ and \keckarray\ analysis a number of filtering operations
are typically performed during the map making process.
In the next section the matrix $\mathbf{R}$ corresponding to
these operations is derived.
The practical implementation of a solution to the eigenvalue equation is 
discussed in Section \ref{sec:ebseparation}.

\section{Observation Matrix}
\label{sec:reob_matrix}

The matrix $\mathbf{R}$ transforms an `input map,' $\mathbf{m}$,  
a vector of the true sky polarization field, into a vector of the observed 
map, $\mathbf{\tilde{m}}$. If the matrix $\mathbf{R}$ represents the apodization and 
linear filtering of an analysis pipeline, it is defined to be the `observation' 
matrix for a particular experiment. This choice of $\mathbf{R}$ ensures the 
eigenspaces of Equation \ref{eqn:geneigenfirst} are pure $E$ and $B$ for the 
observed map. This section describes how the observation matrix is computed 
for \biceptwo\ and the \keckarray.

The steps in constructing the observation matrix mirror functions 
in the data reduction pipeline that was originally developed for \QUAD\ 
\citep{pryke09} and later used in the \bicepone\ \citep{barkats14},
\biceptwo\ \citep{bicep2_res}, and \keckarray\ \citep{keck13} analyses. 
This pipeline consists of a \matlab\ library of procedures which
constructs maps, including several filtering steps, from real data or
simulated timestream data for a given input sky map.

The filtering operations performed sequentially in the standard
pipeline include data selection, polynomial filtering, scan-synchronous signal 
subtraction, weighting, binning into map pixels, and deprojection of leaked 
temperature signal.
To construct the observation matrix, matrices representing each of these steps
are multiplied together to form a final matrix that performs all of the
operations at once.
Since each of the operations is linear, the observation matrix is independent 
of the input map. Therefore, the same matrix can be used on any input map and 
will perform the same operations as the standard pipeline.

If the combined matrix 
of timestream operations is $\boldsymbol{\mathcal{V}}$, then transforming a timestream, 
$\mathbf{d}$, into an observed map, $\mathbf{\tilde{m}}$, is simply:
\begin{equation}
\mathbf{\tilde{m}} = \boldsymbol{\mathcal{V}}\mathbf{d}.
\end{equation}

The signal component of a timestream can be generated from an input map, 
$\mathbf{m}$, using a matrix that contains information about the pointing 
and orientations of the detectors, according to the equation 
$\mathbf{d} = \boldsymbol{\mathcal{A}}\mathbf{m}$. The observation matrix, $\mathbf{R}$, is given by the 
product of $\boldsymbol{\mathcal{V}}$ and $\boldsymbol{\mathcal{A}}$:
\begin{align}
\mathbf{\tilde{m}} &=\boldsymbol{\mathcal{VA}}\mathbf{m}\\
&=\mathbf{Rm}.
\end{align}
It is not necessary for the input maps and observed maps to share the same 
pixelization scheme, since the observation matrix can easily be made to 
transform between the two. 

\subsection{Input  \healpix\ maps}
\label{subsection:timstream_genmatrix}
We choose a \healpix\ pixelization scheme \citep{Gorski04} for the 
input maps, $\mathbf{m}$, because it has equal area pixels on the sphere 
and is widely used in the cosmology community.

A true sky signal is represented by the map 
$\mathbf{m^o}$ = $[T^o_{xy} , Q^o_{xy} , U^o_{xy} ]$, where ($x,y$) are the (RA,Dec) 
coordinates of the map. Using \synfast\footnote{
\synfast\ is a program in the \healpix\ suite that renders sky maps from
sets of input $a_{lm}$'s.},
the unobserved input map is convolved 
with the array averaged beam function, $\boldsymbol{\bar{\mathcal{B}}}$, constructed from
measurements of the beam function of all detectors in the array:
\begin{equation}
\label{eqn:applybeam}
  \left[
   \begin{matrix}
   T_{xy}\\
   Q_{xy}\\
   U_{xy}\\
   \end{matrix}
\right]
=
\boldsymbol{\bar{\mathcal{B}}} * \mathbf{m^o}
=
\bar{\mathcal{B}}_{xy} * 
   \left[
   \begin{matrix}
   T^o_{xy}\\
   Q^o_{xy}\\
   U^o_{xy}\\
   \end{matrix}
\right].
\end{equation}
The input map vector is found by reforming the beam convolved two 
dimensional map into a one dimensional vector, $\mathbf{m}$, of length 
$3j$, where $j=1...n_{p}$, for $n_{p}$ pixels in the input map: 
\begin{equation}
\mathbf{m}\equiv    
   \left[
   \begin{matrix}
   T_j\\
   Q_j\\
   U_j\\
   \end{matrix}
   \right].
\end{equation}

\subsection{\bicep2\ and \keckarray\ scan strategy}
\label{sec:scan_strat}
The observing strategies for \biceptwo\ and the \keckarray\ are very similar 
and borrow heavily from \bicepone. All three experiments target a region of sky 
centered at a right ascension of 0 degrees and declination of -57.5 degrees. A 
detailed description of the scan strategy is contained in \citet{bicep2_instr}.

\begin{itemize}
\item{\bf{Halfscans:}}
   During normal observations, the telescope scans in azimuth at a constant 
elevation. The scan speed of 2.8 deg $\mathrm{s}^{-1}$ in azimuth 
places the targeted multipoles of $20<l<200$ at temporal frequencies less 
than 1 Hz. Each scan covers 64.2 degrees in azimuth, at the end of 
which the telescope stops and reverses direction in azimuth and scans back 
across the field center. A scan in a single direction is known as a `halfscan.'

\item{\bf{Scansets:}}
    Halfscans are grouped into sets of $\approx$ 100 halfscans, which are 
known as `scansets.' The scan pattern deliberately covers a fixed range in 
azimuth within each scanset, rather than a fixed range in right ascension.
Over the course of the 50 minute scanset, Earth's rotation results in a 
relative drift of azimuthal coordinates and right ascension of about 12.5 
degrees. At the end of each scanset the elevation is offset by 0.25 degrees, 
and a new scanset commences. The telescope steps in 0.25 degree elevation 
increments between each scanset. All observations take place at 20 elevation 
steps, with a boresight pointing ranging in elevation between 55 and 59.75 
degrees. The geographic location of the telescope, near the South Pole, means 
that elevation and declination are approximately interchangeable.

\item{\bf{Phases:}}
Scansets are grouped together into sets known as `phases.' For \biceptwo\ and 
the \keckarray, CMB phases consist of ten scansets, comprising 9 
hours of observations. CMB phases are grouped into seven types and each type 
has a unique combination of elevation offset and azimuthal position.

\item{\bf{Schedules:}}
The third degree of freedom in the \biceptwo\ and \keckarray\ telescope mounts 
is a rotation about the boresight, referred to as `deck rotation.' The 
polarization angles relative to the cryostats are fixed, so rotating in deck 
angle allows detector pairs to observe at multiple polarization angles. 

 A `schedule' typically consists of a set of phases at a particular deck 
angle. The deck angle is rotated between schedules. There is typically 
one schedule per fridge cycle, occurring every $\sim3$ days for \biceptwo\ and 
$\sim2$ days for the \keckarray.

\end{itemize}

\subsection{Relationship between timestreams and [$T,Q,U$]}
The \biceptwo\ and \keckarray\ detectors consist of pairs of nominally co-pointed, 
orthogonal, polarization sensitive phased array antennas coupled to 
TES bolometers \citep{bkdets}. The signal in the timestream 
from detector `A' is:
\begin{equation}
\tau^A_t= T_t + \cos(2\Psi^A_t) Q_t +  \sin(2\Psi^A_t) U_t,
\end{equation}
where $[T_t,Q_t,U_t]$ are the Stokes parameters of the beam convolved sky signal for 
timestream sample $t$. A timestream consists of $n_t$ time 
ordered measurements of the sky, $t=1...n_t$. $\Psi^A$ is the angle 
the `A' antenna makes with the $Q,U$ axis on the sky. For the \healpix\ 
polarization convention, this axis is a vector pointing towards the north 
celestial pole. 

The relative gain normalized `A' timestream is summed and 
differenced with the normalized timestream from its orthogonal partner `B':
\begin{align}
    s_t&=\frac{1}{2}(\tau^A_t+\tau^B_t) =  T_t + \alpha^+_tQ_t+\beta^+_tU_t
    \nonumber\\
    d_t&=\frac{1}{2}(\tau^A_t-\tau^B_t) = \alpha^-_tQ_t+\beta^-_tU_t.
  \label{eqn:sada}
\end{align}
The variables $\alpha$ and $\beta$ are defined by:
\[
\alpha_t^\pm\equiv \frac{1}{2}\left[\cos(2\Psi^A_t) \pm \cos(2\Psi^B_t)\right]
\]
\begin{equation}
\beta_t^\pm\equiv \frac{1}{2}\left[\sin(2\Psi^A_t) \pm \sin(2\Psi^B_t)\right],
\end{equation}
where $\Psi^B$ is the angle the `B' antenna makes with the $Q,U$ axis 
on the sky. Assuming that `A' and `B' are perfectly co-pointed and
orthogonal, the signal portion of the timestream vectors can be 
described with a transformation, $A_{t{j}}$, from the input map pixel
(with index $j$) to the timestream sample (with index $t$):
\begin{align}
    s_t&=A_{t{j}}T_j
    \nonumber\\
    d_t&=\alpha^-_{t}A_{t{j}}Q_j+\beta^-_{t}A_{t{j}}U_j,
  \label{eqn:sumdiff}
\end{align}
where the terms $\alpha^+$ and $\beta^+$ in the pair sum timestream cancel due to the orthogonal orientation of the `A' and `B' detectors. The signal-plus-noise timestreams, in vector notation, are:

\begin{align}
  \mathbf{s} &= \frac{1}{2}(\mathbf{n_A}+\mathbf{n_B}) + \mathbf{A[T]}
  \\
  \mathbf{d} &= \frac{1}{2}(\mathbf{n_A}-\mathbf{n_B}) +
    \left[\begin{matrix}\boldsymbol{\alpha}^-&\boldsymbol{\beta}^-\\\end{matrix}\right]
    \left[\begin{matrix}\mathbf{A}&\mathbf{0}\\\mathbf{0}&\mathbf{A}\\\end{matrix}\right]
    \left[\begin{matrix}\mathbf{Q}\\\mathbf{U}\\\end{matrix}\right].
    \label{eqn:difftimestream}
\end{align}
where $\mathbf{n_A}$ and $\mathbf{n_B}$ are the time ordered noise components of 
detector `A' and detector `B,' assuming the noise is uncorrelated with the 
pointing of the detector pair. For signal only simulations, $\mathbf{n_A}$ and 
$\mathbf{n_B}$ can be ignored. 

\begin{figure}[h!] 
\begin{center}
\resizebox{0.7\columnwidth}{!}
{\includegraphics{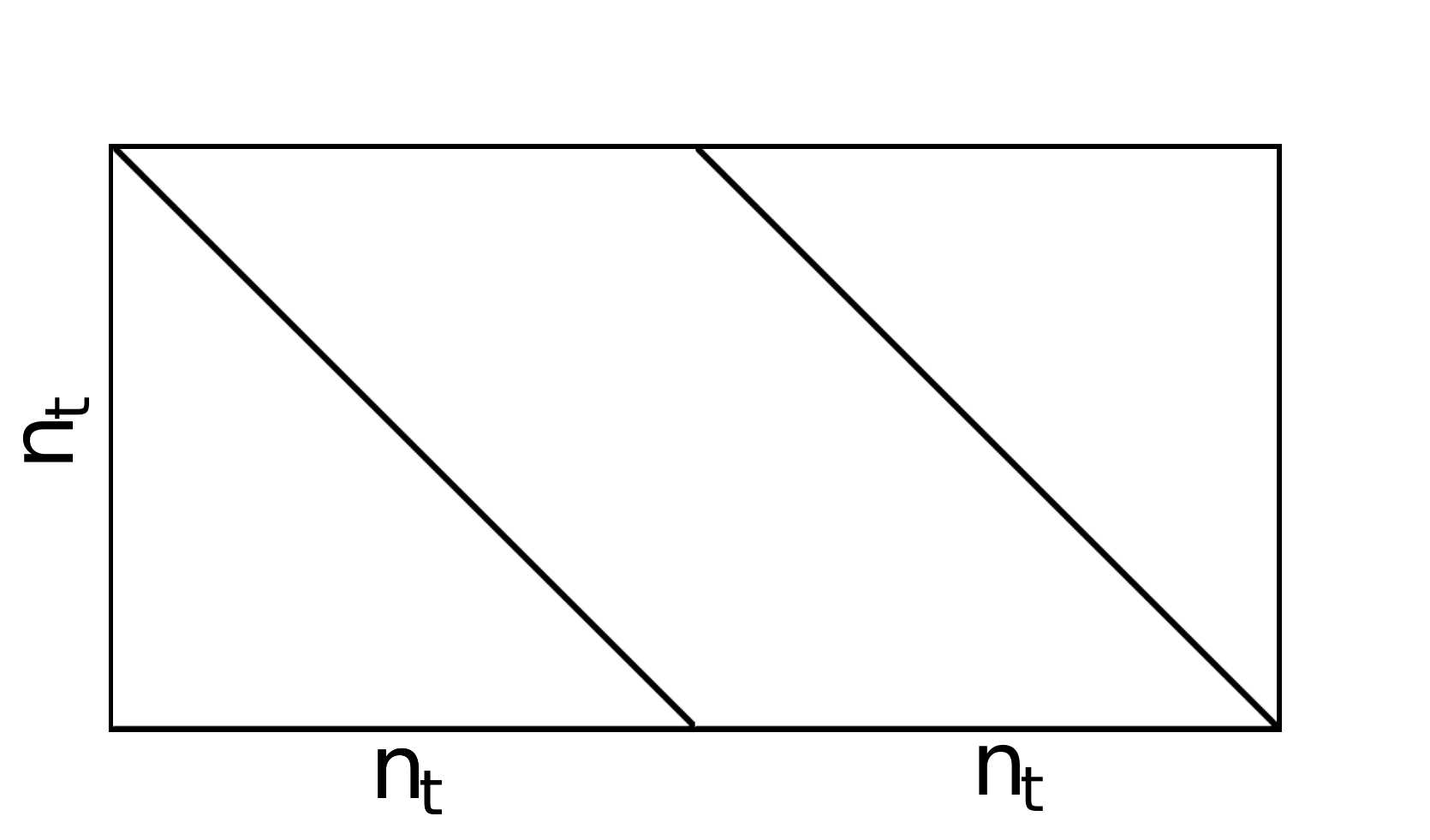}}
\end{center}
  \caption[Detector orientation matrix]
{Detector orientation matrix,
$\left[\begin{matrix}\boldsymbol{\alpha}^-&\boldsymbol{\beta}^-\\\end{matrix}\right]$.
The matrix is only filled on the diagonals of the two 
sub-blocks, $\boldsymbol{\alpha}$ and $\boldsymbol{\beta}$.}
   \label{fig:detect_matrix}
\end{figure}

The matrix $\left[\begin{matrix}\boldsymbol{\alpha}^-&\boldsymbol{\beta}^-\\\end{matrix}\right]$ contains the information about the orientation of a 
pair's antennas relative to $Q$ and $U$ defined on the sky. We call it the 
detector orientation matrix. The combination: 
\begin{equation}
\left[\begin{matrix}\boldsymbol{\alpha}^-&\boldsymbol{\beta}^-\\\end{matrix}\right]
\left[\begin{matrix}\mathbf{A} & \mathbf{0}\\\mathbf{0}&\mathbf{A}\\\end{matrix}\right]
\end{equation}
transforms input $Q,U$ maps into a pair difference timestream. 
$\left[\begin{matrix}\boldsymbol{\alpha}^-&\boldsymbol{\beta}^-\\\end{matrix}\right]$ 
is constructed from two diagonal matrices, $\boldsymbol{\alpha}^-$ and $\boldsymbol{\beta}^-$,
which are filled with the sine and 
cosine of the detector orientations at each time sample.
A graphical representation of the detector orientation matrix is shown in
Figure \ref{fig:detect_matrix}.
(Additional steps accounting for polarization efficiency and
pair non-orthogonality are absorbed into a normalization correction
to the pair difference timestream.)

\subsection{Timestream forming matrix, $\mathbf{A}$}
\begin{figure}[h!] 
\begin{center}
\resizebox{0.9\columnwidth}{!}
{\includegraphics{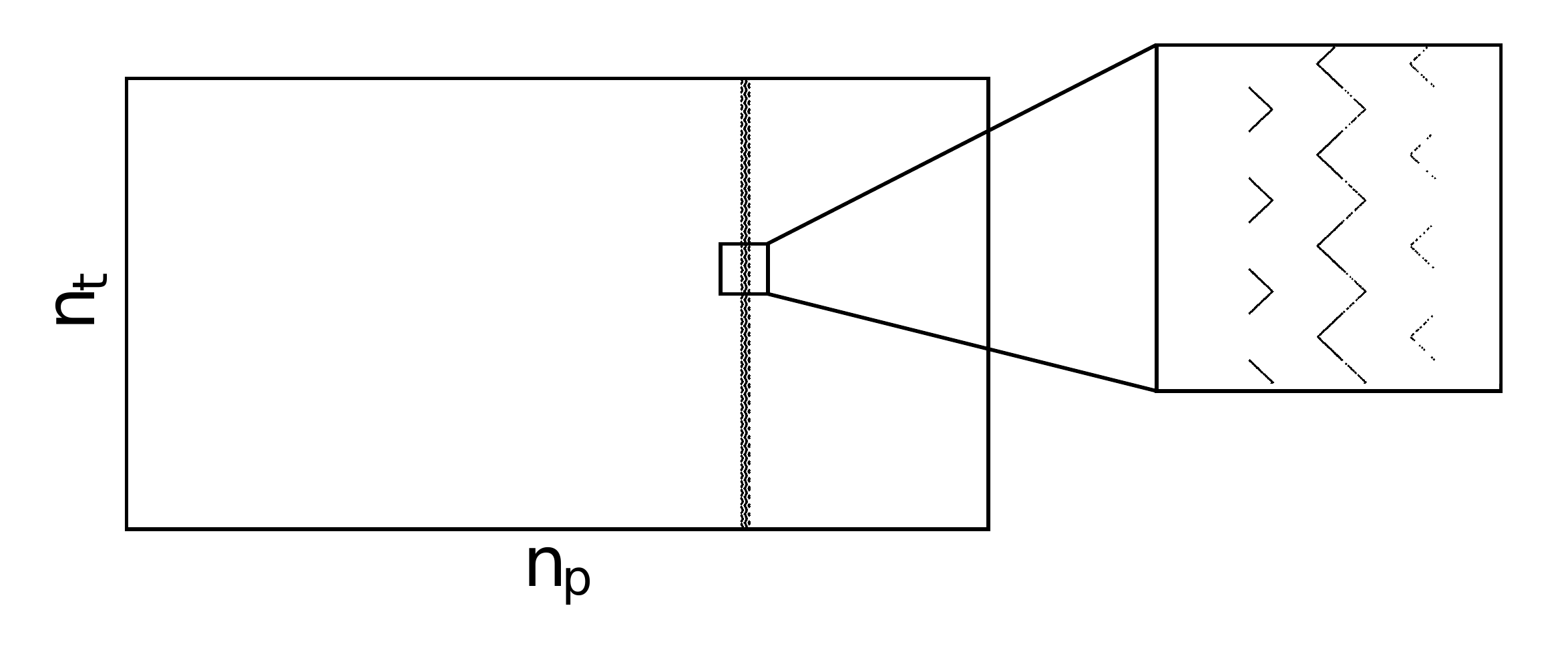}}
\end{center}
  \caption[Timestream forming matrix, $\mathbf{A}$]
{Timestream forming matrix, $\mathbf{A}$: 
filled elements of the matrix that takes \healpix\ 
maps to timestreams. This matrix contains the pointing  of a single 
detector pair over one scanset within a $N_{side}$=512 \healpix\ map. 
The pattern of the filled elements is determined by the particular \healpix\
pixel indexing scheme. There are $n_t$ filled entries, consisting of a 1 for 
each timestream sample. Note that although the above image appears to have 
multiple pointing locations for a single timestream sample, $n_t$, this 
is merely a result of limited resolution in the image. The timestream 
forming matrix contains only one \healpix\ pixel location for each time 
sample.}
   \label{fig:pointhm_matrix}
\end{figure}

The matrix $\mathbf{A}$ = $A_{t{j}}$  represents the timestream forming 
matrix for a detector pair. It transforms the input temperature map, 
$T_{j}$, into the signal component of the pair sum timestream, $s_t$. 
A graphical representation of the timestream forming matrix is shown in 
Figure \ref{fig:pointhm_matrix}.

To create timestreams with smooth transitions at pixel boundary 
crossings, the input maps should have a resolution higher than the spatial 
band limit imposed by the beam function. For this reason, $N_{side}$=512 
\healpix\ maps are used, whose pixels have a Nyquist frequency $\gtrsim2\times$ 
the band limit of the \biceptwo\ and \keckarray\ 150~GHz beam function.

The current \biceptwo\ and \keckarray\ CMB observations fall within the region of sky 
bounded in right ascension by 
$-3^{\mathrm{h}}40^{\mathrm{m}}<\alpha<3^{\mathrm{h}}40^{\mathrm{m}}$ and in 
declination by $-70^\circ<\delta<-45^\circ$.
This region contains $n_p=111,593$ pixels in an $N_{side}$=512 \healpix\ map.
The number of samples in a scanset is typically $n_t\approx{}$43,000.

The simplest form of $\mathbf{A}$ performs nearest neighbor 
interpolation of the \healpix\  maps, in which case $\mathbf{A}$ is 
($n_t \times n_{p}$) and is filled with ones where the detector pair is 
pointed and zeros otherwise.

A more sophisticated form of $\mathbf{A}$ 
performs Taylor interpolation on the \healpix\ map, in which case $\mathbf{A}$ 
is ($n_t \times \frac{\lambda(\lambda+1)}{2} n_{p}$), where $\lambda$ is the 
order of the Taylor polynomial used in interpolation. In this case, 
$\mathbf{A}$ is a matrix that performs Taylor interpolation, allowing 
sub-pixel accuracy to be recovered from the input map, and 
$\mathbf{m}$ must also contain derivatives of the true sky temperature 
and polarization field. This matrix is used to build the deprojection 
templates in Section \ref{subsec:depro_matrix} but is not used for forming 
timestreams because it increases the dimensions of the observation matrix, 
making the computation of the observation matrix more difficult.

\subsection{Polynomial filtering matrix, $\mathbf{F}$}
\label{subsec:filter_matrix} 
To remove low frequency atmospheric noise from the data, a third order 
polynomial is fit and subtracted from each halfscan in the timestreams. 
Since each halfscan traces an approximately constant elevation trajectory across the target field, the 
polynomial filter removes power only in the right ascension direction 
of the maps. In multipole, $l$, the third order polynomial filter typically 
rolls off power below $l<40$. This can be represented by a `filtering matrix,' 
$\mathbf{F}$, which is block diagonal with the block size being the temporal 
length of a halfscan. Each block is composed of a matrix:
\begin{equation}
\mathbf{F=I-V(V^{\top}V)^{-1}V^{\top}},
\end{equation}
where $\mathbf{I}$ is the identity matrix and $\mathbf{V}$ is the same third order 
Vandermonde matrix for each halfscan of equal length. The Vandermonde matrix 
is defined as:
\begin{equation}
V_{t{j}}=x_t^{j-1},
\label{eqn:vandermonde}
\end{equation}
where $j=4$ for a third order filter and $x_{t}$ are the 
coordinate locations.

\begin{figure}[h!] 
\begin{center}
\resizebox{0.8\columnwidth}{!}
{\includegraphics{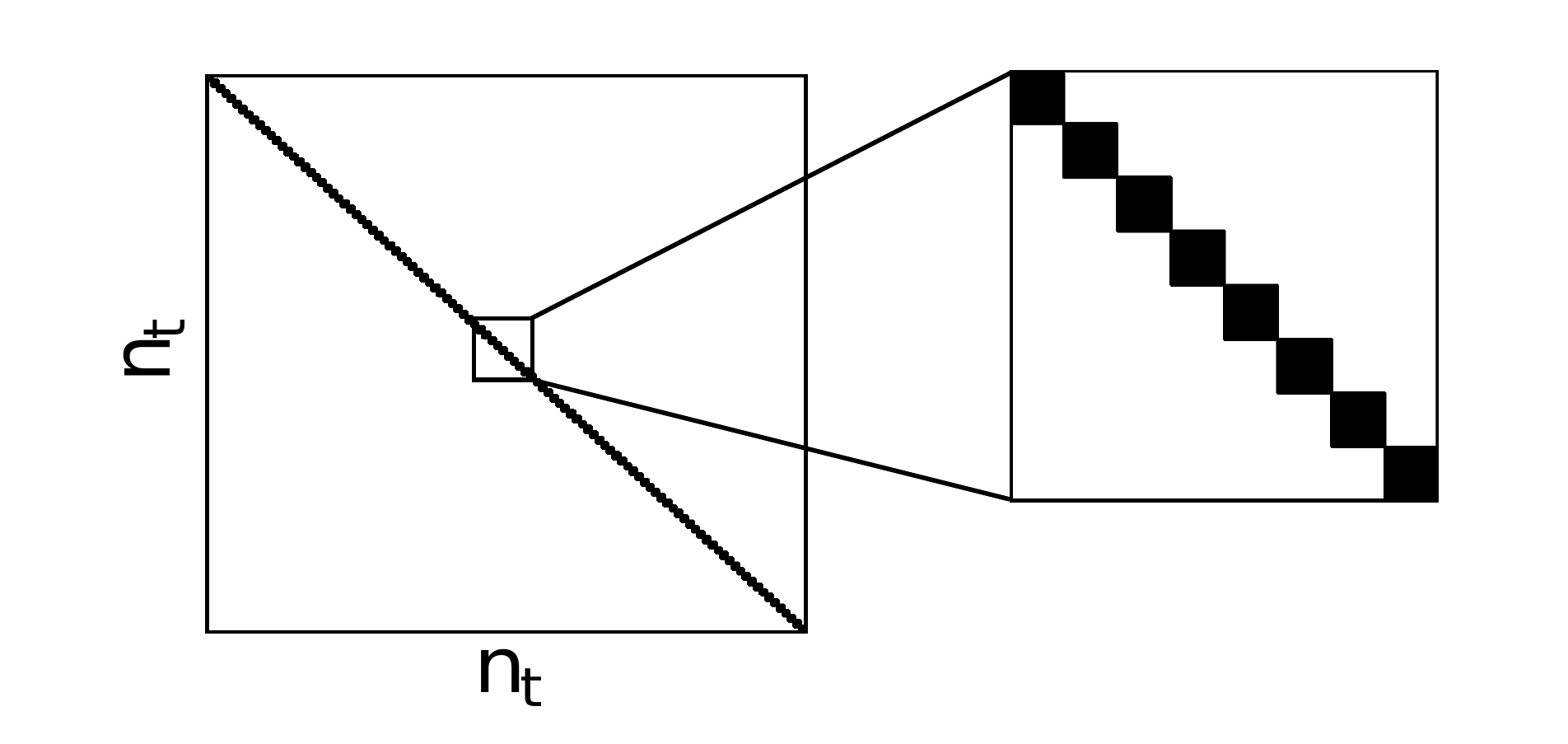}}
\end{center}
  \caption[Polynomial filtering matrix, $\mathbf{F}$]
{Polynomial filtering matrix, $\mathbf{F}$, showing the 
filled elements of the matrix. The matrix is 
very sparse and is block diagonal with blocks the size of a halfscan 
($\approx{}$404 samples).}
   \label{fig:poly_filter}
\end{figure}

For \biceptwo\ and the \keckarray, $x_{t}$ is a vector of the relative 
azimuthal location of each sample in the halfscan. A representation of 
the polynomial filtering matrix is shown in Figure \ref{fig:poly_filter}.

\subsection{Scan-synchronous signal removal matrix, $\mathbf{G}$}
\label{subsec:gsmatrix}
Scan-synchronous subtraction removes signal in the timestreams that is fixed 
relative to the ground rather than moving with the sky. These azimuthally 
fixed signals are decoupled from signals rotating with the sky by the scan 
strategy, which observes over a fixed range in azimuth as the sky slides by
(as described in Section~\ref{sec:scan_strat}). 
A template of the mean azimuthal signal is subtracted from the timestreams for 
each scan direction. This procedure can be represented as a matrix operator, 
referred to as a `scan-synchronous signal matrix.'

\begin{figure}[h!] 
\begin{center}
\resizebox{0.8\columnwidth}{!}
{\includegraphics{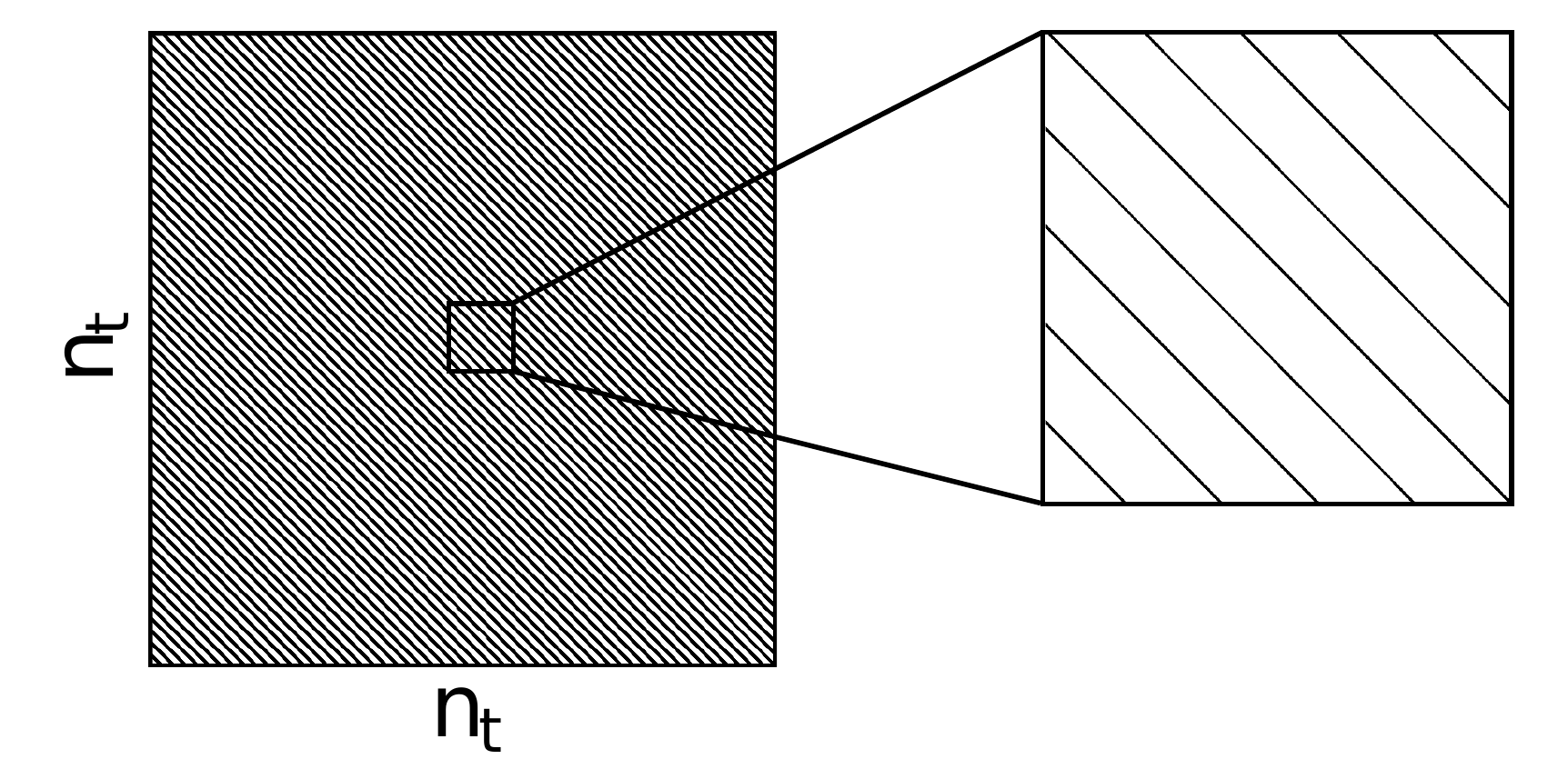}}
\end{center} 
  \caption[Scan-synchronous signal removal matrix, $\mathbf{G}$]
{Scan-synchronous signal removal matrix, $\mathbf{G}$, showing the 
filled elements. The 
scan-synchronous signal matrix is sparse Toeplitz, with off diagonal 
components that subtract the average scan-synchronous signal for one of the 
two scan directions in a scanset.}
   \label{fig:ground_subtract}
\end{figure}

The mean azimuthal signal can be found using a matrix $\mathbf{X}$=$X_{tt'}$, 
for which each row is only filled for entries containing the same 
azimuthal pointing as the diagonal entry. The scan-synchronous signal 
matrix subtracts off the mean azimuthal signal: 
\begin{equation}
\mathbf{G=I-X},
\end{equation}  
where $\mathbf{I}$ is the identity matrix. A graphical representation of the 
scan-synchronous signal removal matrix is shown in Figure \ref{fig:ground_subtract}. 
Note that while the $\mathbf{F}$ matrix is block diagonal and sparse, and the 
$\mathbf{G}$ matrix is sparse, once the two are combined, the resulting filter 
matrix is neither sparse nor block diagonal, making matrix operations more 
computationally demanding.

\subsection{Inverse variance weighting matrices, $\mathbf{w}^\pm$}
\label{subsec:weightmatrix}
\begin{figure}[h!] 
\begin{center}
\resizebox{0.3\columnwidth}{!}
{\includegraphics{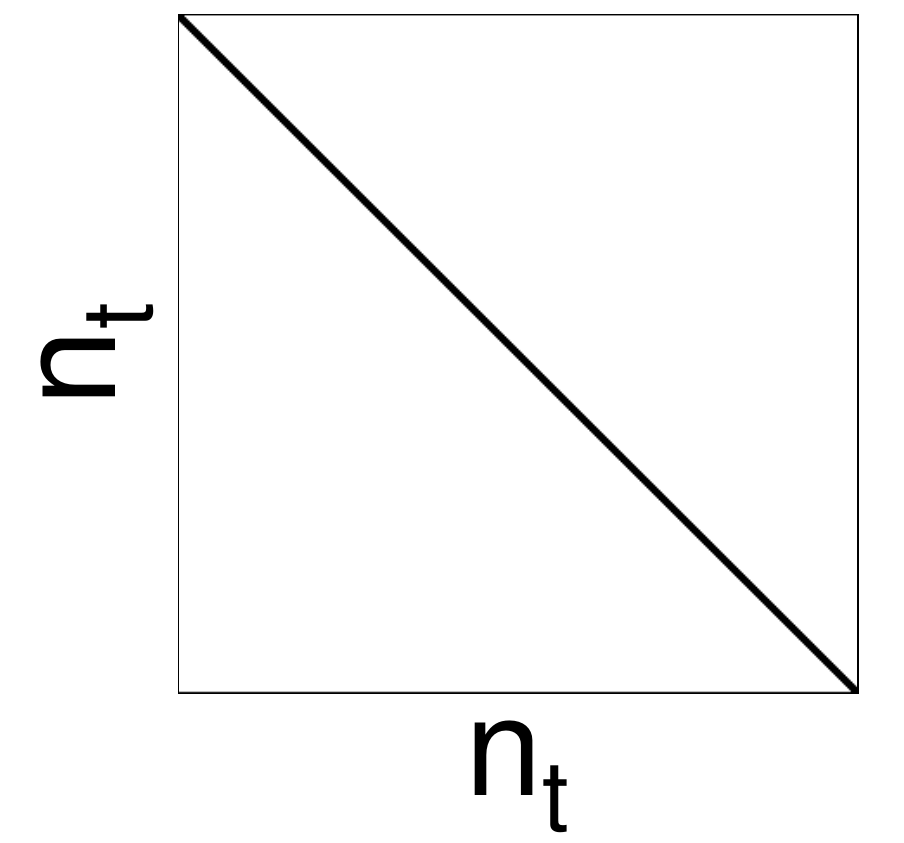} }
\end{center}
  \caption[Weighting matrix]{Weighting matrices $\mathbf{w^{\pm}}$, showing the 
filled elements. The weighting matrices are zero except 
on the diagonal, where they contain the weights based on the inverse variance of the timestream.}
   \label{fig:weight_vect}
\end{figure}

The timestreams are weighted based on the measured inverse variance of each 
scanset. Pair sum and pair difference are weighted separately from weights 
calculated from the two timestreams, $\mathbf{w^+}$ and $\mathbf{w^-}$. The scheme assigns 
lower weight to particularly noisy channels and periods of bad weather. 
This choice of weighting is not a fully ``optimal'' map maker 
\citep{tegmark97}, but instead represents a practical solution that 
avoids calculating and inverting a large noise covariance matrix. The 
weighting is represented by a matrix whose diagonal is filled with the 
vector $\mathbf{w}^+$=$w^+_{tt}$ for pair sum and $\mathbf{w}^-$=$w^-_{tt}$ for 
pair difference, shown in Figure \ref{fig:weight_vect}.

\subsection{Filtered signal timestream generation} 
\begin{figure}[h!] 
\begin{center}
\resizebox{\columnwidth}{!}
{ \includegraphics{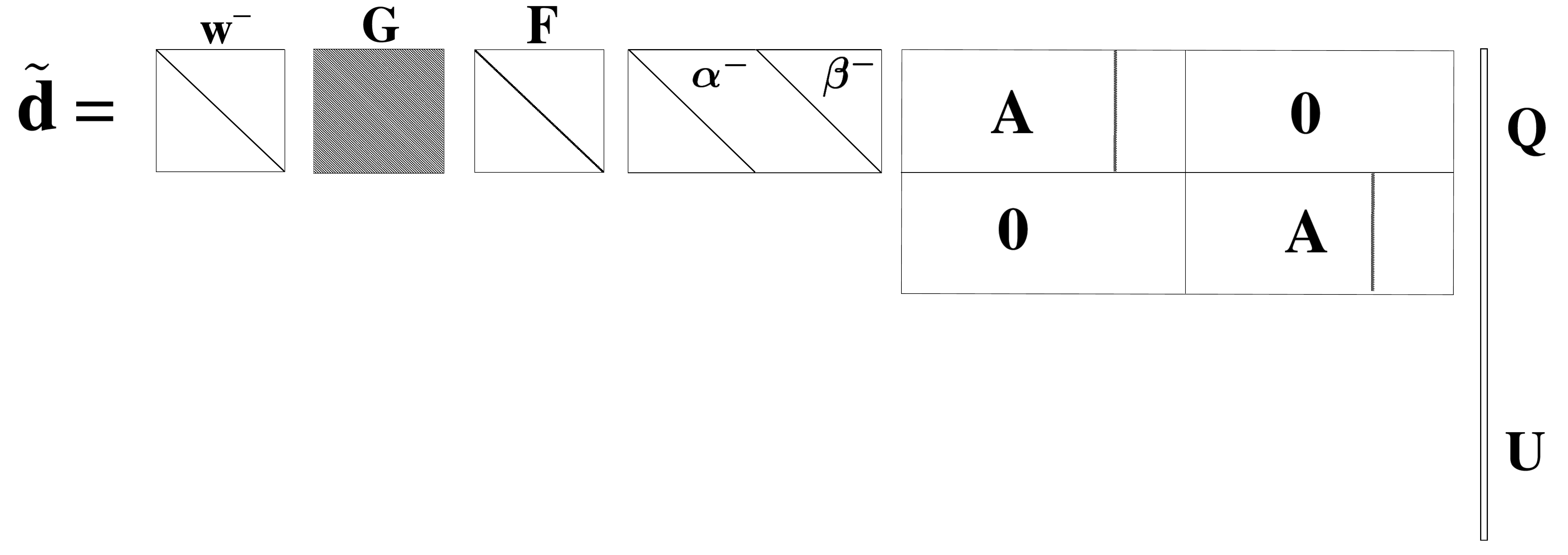} }
\end{center}
  \caption[Matrix generation of simulated timestreams]{Matrix generation of simulated timestreams corresponding to Equation \ref{eqn:simtimestream}.}
   \label{fig:summary_timest}
\end{figure}

Ignoring noise and combining all the operators of Sections 
\ref{subsec:filter_matrix}, \ref{subsec:gsmatrix} and 
\ref{subsec:weightmatrix}, the sum and difference timestreams in Equation \ref{eqn:sumdiff} are tranformed to the filtered timestreams:
\begin{align}
  \begin{aligned}
    \tilde{\mathbf{s}} &= \mathbf{w}^+\mathbf{GFA[T]} \\
    \tilde{\mathbf{d}} &= \mathbf{w}^-\mathbf{GF}
      \left[\begin{matrix}\boldsymbol{\alpha}^-&\boldsymbol{\beta}^-\\\end{matrix}\right]
      \left[\begin{matrix}\mathbf{A}&\mathbf{0}\\\mathbf{0}&\mathbf{A}\\\end{matrix}\right]
      \left[\begin{matrix}\mathbf{Q}\\\mathbf{U}\\\end{matrix}\right],
  \end{aligned}
  \label{eqn:simtimestream}
\end{align}
the second of which is graphically represented in Figure \ref{fig:summary_timest}.

\subsection{Pointing matrix, $\boldsymbol{\Lambda}$}
\label{subsec:pointingmatrix}

\begin{figure}[h!] 
\begin{center}
\resizebox{0.9\columnwidth}{!}
{ \includegraphics{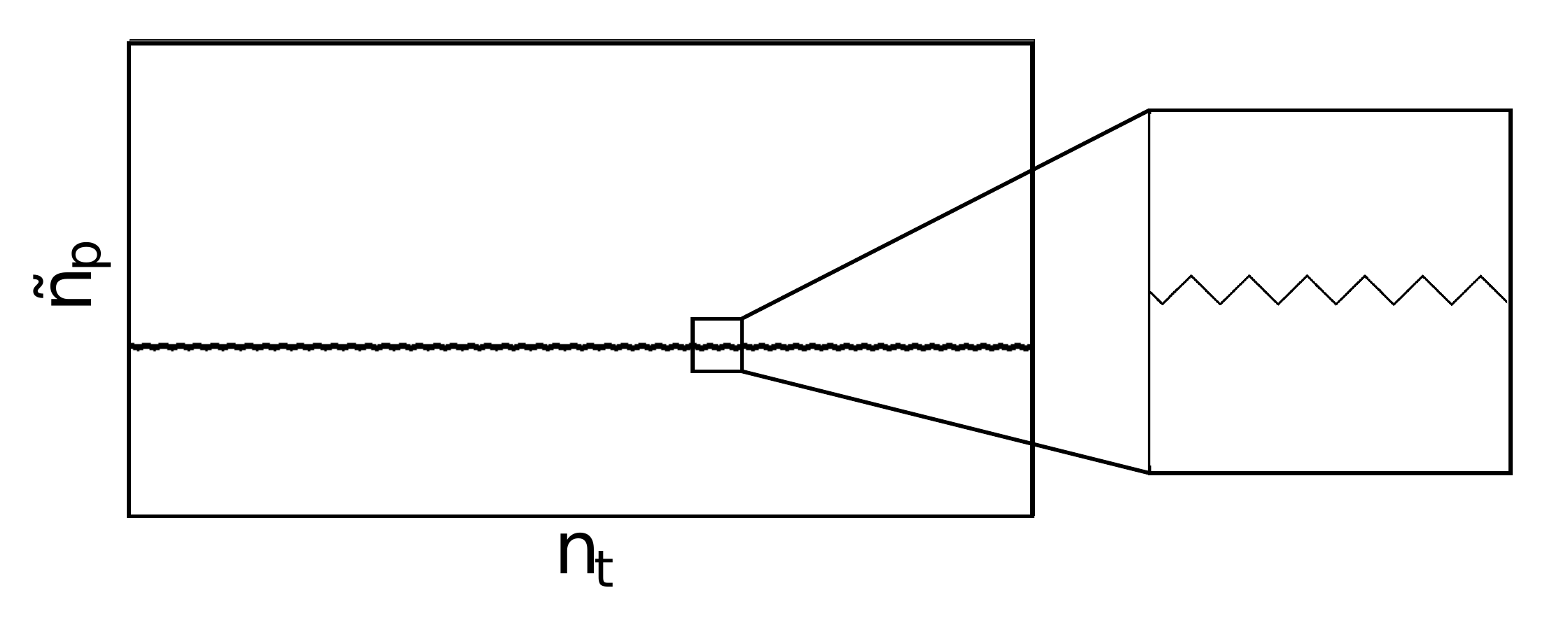} }
\end{center}
  \caption[Pointing matrix]{Pointing matrix,  $\boldsymbol{\Lambda}$:
filled elements of the 
pointing matrix that transforms timestreams to an observed map
in the \bicep\ pixelization. This matrix 
contains the mapping between the pointing of a single detector pair over one 
scanset and the output map pixels. There are 23,600 pixels in a \bicep\ map, 
denoted as $\tilde{n}_p$. There are $n_t$ filled entries, consisting of a 1 
for each timestream sample. Each leg of the zigzag pattern corresponds to a 
halfsan within the scanset, where the telescope is scanning back and forth at 
a fixed elevation.}
   \label{fig:pointm_matrix}
\end{figure}

The timestream quantities $\mathbf{s}$ and $\mathbf{d}$ are converted to maps 
by the pointing matrix, $\boldsymbol{\Lambda}$ = $\Lambda_{it}$. If the pixelization 
of the input maps were identical to the output maps, the pointing matrix would 
be the transpose of the timestream forming matrix:
\[
\boldsymbol{\Lambda}=\mathbf{A^{\top}}.
\]
As discussed in Section \ref{subsection:timstream_genmatrix}, the input
maps are \healpix\ $N_{side}$=512.
However, the \bicep\ maps instead use a simple rectangular grid of pixels
in RA and Dec:
the size of the pixels is 0.25 degrees in Dec, with the pixel size in RA
set to be equivalent to 0.25 degrees of arc at the mid-declination of the
map, resulting in $236\times100=23,600$ pixels.
If the \bicep\ maps are naively used as ``flat maps'' then projection
distortions are inherent.
However, note that the $\mathbf{A}$ and $\boldsymbol{\Lambda}$ matrices
together fully encode the mapping from the underlying curved sky to the observed map pixels,
allowing such distortions to be accounted for.
Figure \ref{fig:pointhm_matrix} 
shows $\mathbf{A}$ for a single detector over a scanset and Figure 
\ref{fig:pointm_matrix} shows $\boldsymbol{\Lambda}$ for a single detector over 
a scanset.

The pointing matrix for a single detector pair can be used to construct a 
pair sum `pairmap:'
\begin{equation}
 \mathbf{m}_{T} = \boldsymbol{\Lambda}\mathbf{s} = \boldsymbol{\Lambda} \mathbf{w}^+\mathbf{GFA[T]}.
\end{equation}
The pair difference timestream is converted into pairmaps using 
two copies of the pointing matrix. The two pair difference 
pairmaps correspond to linear combinations of Stokes $Q$ and $U$:
\begin{equation}
\label{eqn:qu}
\left[\begin{matrix}\mathbf{m}_{\boldsymbol{\alpha}^-}\\\mathbf{m}_{\boldsymbol{\beta}^-}\\\end{matrix}\right]
=
\left[\begin{matrix}\boldsymbol{\Lambda}&\mathbf{0}\\\mathbf{0}&\boldsymbol{\Lambda}\\\end{matrix}\right]
\left[\begin{matrix}\boldsymbol{\alpha}^-\\\boldsymbol{\beta}^-\\\end{matrix}\right]
\mathbf{w}^-\mathbf{GF}
\left[\begin{matrix}\boldsymbol{\alpha}^-&\boldsymbol{\beta}^-\\\end{matrix}\right]
\left[\begin{matrix}\mathbf{A}&\mathbf{0}\\\mathbf{0}&\mathbf{A}\\\end{matrix}\right]
\left[\begin{matrix}\mathbf{Q}\\\mathbf{U}\\\end{matrix}\right].
\end{equation}
For later convenience in abbreviating this equation, we define:
\begin{equation}
\label{eqn:p_def}
\boldsymbol{\mathcal{P}}
\equiv
\left[\begin{matrix}\boldsymbol{\Lambda}&\mathbf{0}\\\mathbf{0}&\boldsymbol{\Lambda}\\\end{matrix}\right]
\left[\begin{matrix}\boldsymbol{\alpha}^-\\\boldsymbol{\beta}^-\\\end{matrix}\right]
\mathbf{w}^-\mathbf{GF}
\left[\begin{matrix}\boldsymbol{\alpha}^-&\boldsymbol{\beta}^-\\\end{matrix}\right]
\left[\begin{matrix}\mathbf{A}&\mathbf{0}\\\mathbf{0}&\mathbf{A}\\\end{matrix}\right].
\end{equation} 

\subsection{Deprojection matrix, $\mathbf{D}$}
\label{subsec:depro_matrix}

A potential systematic concerning polarization measurements is the 
leakage of unpolarized signal into polarized signal. In the case of CMB 
polarization, this takes the form of the relatively bright temperature 
anisotropy leaking into the much fainter polarization anisotropy. The 
leakage is caused by imperfect differencing between the orthogonal 
pairs of detectors. The beam functions can be well approximated by 
elliptical Gaussians, the difference of which correspond to gain, pointing,
width and ellipticity~\citep{hu03, shimon08}.

The \biceptwo\ and \keckarray\ pipeline removes leaked 
temperature signal from the polarization signal using linear regression 
to fit leakage templates to the polarization data.
This method allows the beam mismatch parameters to be 
fitted directly from the CMB data itself, rather than relying on external
calibration data sets, and is robust to temporal variations of the 
beam mismatch.

The templates used in the regression are constructed from \planck\ 143~GHz 
temperature maps\footnote{For the \keckarray\ 95~GHz and 220~GHz bands, we use \planck\ 100~GHz and 217~GHz maps.}. These maps contain both CMB and foreground emission at 
approximately the \biceptwo\ band. The noise in \planck\ 143~GHz is 
significantly subdominant to the CMB temperature anisotropy.
 For a full description and derivation of the deprojection technique, see 
\citet{aikinthesis}, \citet{sheehythesis}, \citet{barkats14}, and 
\citet{bicep2_syst}. In this section, the entire deprojection algorithm 
is re-cast as a matrix operation.   

For the purposes of generating deprojection 
templates, we use a timestream forming matrix that performs Taylor 
expansion around the nearest pixel center to the detector pointing location. 
The Taylor interpolating matrix produces higher fidelity timestreams than a 
nearest neighbor matrix. This is important for the deprojection algorithm since 
small displacements in beam position are responsible for the systematic 
effect that is removed. Without Taylor interpolation, pixel boundary 
discontinuities introduce noise and limit the effectiveness of deprojection. 

A Taylor polynomial of order $\lambda$ has 
$\frac{\lambda(\lambda+1)}{2}$ terms, so the dimensions of the input map 
vector for second order interpolation is $1\times{6n_p}$.
Using Equation \ref{eqn:applybeam}, the input maps are convolved 
with the array averaged beam function. The smoothing is done using \synfast, 
which contains the ability to output derivatives of the temperature (and polarization)
field. 
Because the beam is applied first, the output derivatives are less noisy 
than they would be in the raw maps. 

The maps are of the form:
\begin{equation}
\label{eqn:healptemp}
\boldsymbol{\Theta}=\left[\begin{matrix}\mathbf{T}\\{\grad_{\theta}\mathbf{T}}\\{\grad_{\phi}\mathbf{T}}\\{\grad_{\theta\theta}\mathbf{T}}\\{\grad_{\phi\phi}\mathbf{T}}\\{\grad_{\theta\phi}\mathbf{T}}\end{matrix}\right],
\end{equation}
where $\theta$ and $\phi$ are the \healpix\ map's latitude and longitude.

Using the temperature map and its derivatives, we can 
find the Taylor interpolated temperature timestream by 
replacing $\mathbf{A}$ with a Taylor interpolating matrix, $\mathbf{A'}$:
\begin{align}
\mathbf{A'}=\left[\begin{matrix}
\mathbf{A}&\mathbf{A}\boldsymbol{\Delta\theta}&\mathbf{A}\boldsymbol{\Delta\phi}&\mathbf{A}\frac{\boldsymbol{\Delta\theta}^2}{2}&\mathbf{A}\frac{\boldsymbol{\Delta\phi}^2}{2}&\mathbf{A}\boldsymbol{\Delta\theta{\Delta}\phi}
\end{matrix}\right],
\end{align}
where $\boldsymbol{\Delta\theta}$ and $\boldsymbol{\Delta\phi}$ are diagonal matrices giving the 
difference between the detector pair's pointing and the nearest 
\healpix\ pixel center.

A differential beam generating operator is applied to the 
timestreams to create differential beam timestreams. For example, 
the differential gain timestream is just the beam convolved temperature field:
\begin{equation}
\mathbf{d}_{\delta{g}}={\delta}_g\mathbf{A'}\boldsymbol{\Theta},
\end{equation}
where the fit coefficient for the gain mismatch is $\delta_g$. The 
differential pointing components are found from the first derivatives 
of the temperature field with respect to the focal plane coordinates, 
$x$ and $y$:
\begin{align}
\mathbf{d}_{\delta{x}}&={\delta}_x\grad_x\mathbf{A'}\boldsymbol{\Theta}\\
\mathbf{d}_{\delta{y}}&={\delta}_y\grad_y\mathbf{A'}\boldsymbol{\Theta},
\end{align}
where ${\delta}_x$ and ${\delta}_y$  are the differential beam coefficients 
and $\grad_x$ and $\grad_y$ are partial differential operators with respect 
to the focal plane coordinates. Further details of this calculation and 
derivations for other beam modes are discussed in Appendix~C of 
\citet{bicep2_syst}. 

The differential beam timestreams are transformed into maps 
analogously to Equation \ref{eqn:qu}, creating a pairmap template,
 $\boldsymbol{\tilde{\mathcal{T}}_j}$, for each differential beam mode, ${j}$. 
The template pairmaps for each scanset, $\scriptstyle{\mathbb{S}}$, are then 
coadded over phases. For instance, the template pairmap for differential gain, 
is $\boldsymbol{\tilde{\mathcal{T}_1}}$:
\begin{equation}
\label{eqn:gentemplates}
\boldsymbol{\tilde{\mathcal{T}_1}}=
\sum^{{\scanset}\in{phase}}_{\scanset}
\left(
\left[\begin{matrix}\boldsymbol{\Lambda}&\mathbf{0}\\\mathbf{0}&\boldsymbol{\Lambda}\\\end{matrix}\right]
\left[\begin{matrix}\boldsymbol{\alpha}^-\\\boldsymbol{\beta}^-\\\end{matrix}\right]
\mathbf{w}^-\mathbf{GF}
\left[\begin{matrix}\boldsymbol{\alpha}^-&\boldsymbol{\beta}^-\\\end{matrix}\right]
\left[\begin{matrix}\mathbf{A'}&\mathbf{0}\\\mathbf{0}&\mathbf{A'}\\\end{matrix}\right]
\left[\begin{matrix}\boldsymbol{\Theta}\\\boldsymbol{\Theta}\end{matrix}\right]
\right)_{\scanset}.
\end{equation}
A matrix performing weighted linear least-squares regression against real 
pairmaps produces the fitted coefficients for each of the differential beam 
modes, $\mathbf{c}\equiv[{\delta}_g,{\delta}_x,{\delta}_y,...]$:
\begin{equation}
\mathbf{c}=\left(\boldsymbol{\tilde{\mathcal{T}}}^\top(\boldsymbol{\mathcal{W}}^-)^{-1}\boldsymbol{\tilde{\mathcal{T}}}\right)^{-1}\boldsymbol{\tilde{\mathcal{T}}}^\top(\boldsymbol{\mathcal{W}}^-)^{-1}
\left[\begin{matrix}\mathbf{m}_{\boldsymbol{\alpha}^-}\\\mathbf{m}_{\boldsymbol{\beta}^-}\\\end{matrix}\right],
\label{eqn:deproc}
\end{equation}
where $\left[\begin{matrix}\mathbf{m}_{\boldsymbol{\alpha}^-}\\\mathbf{m}_{\boldsymbol{\beta}^-}\\\end{matrix}\right]$ 
is the real data pairmap coadded over a phase, $\boldsymbol{\mathcal{\tilde{T}}}$ is 
a vector of pairmap templates, and $\boldsymbol{\mathcal{W}}^-$ is the pair 
difference weight map, created from the weight matrix according to:
\begin{equation}
   \boldsymbol{\mathcal{W^-}} = 
\sum^{{\scanset}\in{phase}}_{\scanset}
\left(
      \begin{bmatrix} \boldsymbol{\Lambda} \boldsymbol{\alpha^-}\boldsymbol{\alpha^-}\mathbf{w^-}\boldsymbol{\Lambda}^\top\\ \boldsymbol{\Lambda}\boldsymbol{\beta^-}\boldsymbol{\beta^-}\mathbf{w^-} \boldsymbol{\Lambda}^\top \end{bmatrix}
\right)_{\scanset}.
\label{eqn:weightmap}
\end{equation}
The pairmap templates weighted by $\mathbf{c}$ are then subtracted 
from the real data pairmap:
\begin{equation}
\left[\begin{matrix}\mathbf{\tilde{m}_{\boldsymbol{\alpha}^-}}\\\mathbf{\tilde{m}_{\boldsymbol{\beta}^-}}\\\end{matrix}\right]
=
\left[\begin{matrix}\mathbf{m_{\boldsymbol{\alpha}^-}}\\\mathbf{m_{\boldsymbol{\beta}^-}}\\\end{matrix}\right]
\boldsymbol{-\tilde{\mathcal{T}}}\mathbf{c}.
\end{equation}

This process takes the form of a matrix operator that includes each of the 
beam systematics, giving the deprojection matrix:
\begin{equation}
\mathbf{D} \equiv \mathbf{I}-\boldsymbol{\tilde{\mathcal{T}}}\left(\boldsymbol{\tilde{\mathcal{T}}}^{\top}(\boldsymbol{\mathcal{W}}^-)^{-1}\boldsymbol{\tilde{\mathcal{T}}}\right)^{-1}\boldsymbol{\tilde{\mathcal{T}}}^{\top}\left(\boldsymbol{\mathcal{W}}^-\right)^{-1}. 
\label{eqn:deprosub}
\end{equation}
Deprojected pairmaps are then found according to:
\begin{equation}
\left[\begin{matrix}\mathbf{\tilde{m}_{\boldsymbol{\alpha}^-}}\\\mathbf{\tilde{m}_{\boldsymbol{\beta}^-}}\\\end{matrix}\right]
=
\mathbf{D}
\sum_{\scanset}^{{\scanset}\in phase}
\boldsymbol{\mathcal{P}}_{\scanset}
\left[\begin{matrix}\mathbf{Q}\\\mathbf{U}\\\end{matrix}\right].
\end{equation}

The regression in Equation \ref{eqn:deproc} operates simultaneously over all 
the modes to be deprojected. Because the templates for different modes 
are not in general orthogonal, the coefficient for each mode depends on 
the full set of modes. Therefore, the subtraction in Equation 
\ref{eqn:deprosub} must include the same mode list used in the regression 
in Equation \ref{eqn:deproc}. If the regression included more modes than 
the subtraction step, the regression would have extra degrees of freedom. 
This could result in incomplete removal of leakage signal. We avoid this 
possibility by deferring the regression step until immediately before the 
subtraction step, and explicitly using the same mode selection for both.

As in the standard pipeline, we deproject for each detector pair, after 
coadding scanset to phases. 
To reduce the computational demands, the matrix deprojection pairmaps 
have additionally been coadded over scan direction, whereas the standard
pipeline performs regression separately for left going and right going scans. 
This is the only difference between simulations run with the standard
pipeline and those calculated from the observation matrix and leads 
to a negligible difference, see Figure \ref{fig:reob_mapcomp} and 
Figure \ref{fig:reob_apscomp}. 
   
\begin{figure}[h!] 
\begin{center}
\resizebox{0.5\columnwidth}{!}
{ \includegraphics{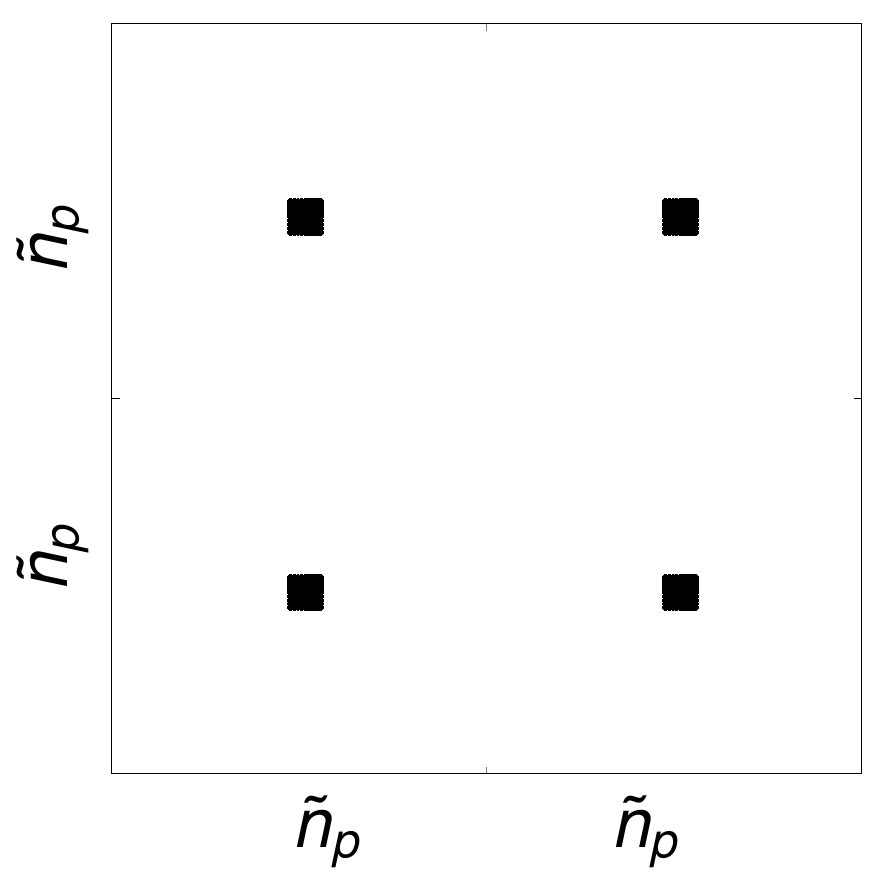}}
\end{center}
  \caption[Deprojection matrix]{Deprojection matrix, $\mathbf{D}$: filled 
elements of the deprojection matrix for one pair, for one phase of data. 
The overall dimensions are $2\tilde{n}_p\times2\tilde{n}_p$, 
twice the number of pixels in a \bicep\ map.}
   \label{fig:deproj_matrix}
\end{figure}

The deprojection matrix made for a phase is less sparse than one made for 
a scanset because over the course of a phase a particular pair will 
observe a larger range of elevation than it would in a scanset. The 
filled elements of the matrix $\mathbf{D}$, for one pair across one phase, is 
shown in Figure \ref{fig:deproj_matrix}.

\subsection{Coadding over scansets and detector pairs to form the observation matrix}
\label{subsec:coadding}

An observed temperature map, $\mathbf{\tilde{T}'}$, can be found by summing 
the pair sum pairmaps temporally over scansets ($\scriptstyle{\mathbb{S}}$) 
and over detector pairs ($\scriptstyle{\mathbb{P}}$):
\begin{equation}
\label{eqn:fullt}
 \mathbf{\tilde{T}'} = \sum_{{\pairs},{\scanset}}\Lambda \mathbf{w}^+\mathbf{GFA[T]}.
\end{equation}
The matrix performing this transformation is defined as $\mathbf{R'_{TT}}$, where the prime indicates the apodization comes from the inverse variance of the pair sum timestream, $\mathbf{w^+}$. The final apodization is applied in Section \ref{sec:modapod}.

The transformation from pair difference pairmaps to $Q,U$ maps 
depends on the detector orientations during the observations. This 
transformation relies on an inversion of a $2\times2$ detector 
orientation matrix. We will now derive the matrix that performs this 
transformation.

Ignoring filtering, the pair difference timestream is 
found using the timestream forming matrix, $A_{t{j}}$:
\begin{equation}
d_t=\frac{1}{2}(\tau^A_t-\tau^B_t) = \alpha^-_{t}A_{t{j}}Q_j+\beta^-_{t}A_{t{j}}U_j.
\end{equation}
Forming linear combinations of the pair difference timestream,
\begin{equation}
\left[\begin{matrix}\alpha^-_t d_t\\\beta^-_t d_t\\\end{matrix}\right]
=
\left[\begin{matrix}
\alpha^-_{t}\alpha^-_{t}A_{t{j}} & \alpha^-_{t}\beta^-_{t}A_{t{j}}\\
\alpha^-_{t}\beta^-_{t}A_{t{j}}  & \beta^-_{t}\beta^-_{t}A_{t{j}}\\
\end{matrix}\right]
\left[\begin{matrix}Q_j\\U_j\\\end{matrix}\right],
\end{equation}
and applying the pointing matrix, $\boldsymbol{\Lambda}$, the vectors 
$\boldsymbol{\alpha}^-\mathbf{d}$ and $\boldsymbol{\beta}^-\mathbf{d}$ are 
binned into map pixels, $i$. At this point we coadd over scansets and detector 
pairs, and apply a weighting, $\mathbf{w}^-$, equal to the inverse of the 
variance of the timestreams during a scanset:

\begin{align}
\label{eqn:beforeinv}
\sum_{\pairs, \scanset}
\left[\begin{matrix}\Lambda_{it}w^-_t\alpha^-_td_t\\\Lambda_{it}w^-_t\beta^-_td_t\\\end{matrix}\right]
=
\left(
\sum_{\pairs, \scanset}
\left[\begin{matrix}
\Lambda_{it}w^-_t\alpha^-_{t}\alpha^-_{t}A_{t{j}} & \Lambda_{it}w^-_t\alpha^-_{t}\beta^-_{t}A_{t{j}}\\
\Lambda_{it}w^-_t\alpha^-_{t}\beta^-_{t}A_{t{j}}  & \Lambda_{it}w^-_t\beta^-_{t}\beta^-_{t}A_{t{j}}\\
\end{matrix}\right]
\right)
\left[\begin{matrix}Q_j\\U_j\\\end{matrix}\right].
\end{align}

We invert the matrix on the right hand side of Equation \ref{eqn:beforeinv} 
to compute a matrix that generates $Q$ and $U$ maps:
\begin{equation}
\label{eqn:inver}
\left[\begin{matrix}e_i&f_i\\f_i&g_i\\\end{matrix}\right]
\equiv
\left(
\sum_{\pairs, \scanset}
\left[\begin{matrix}\Lambda_{it}{w_t}^-\alpha_t^-\alpha_t^-\Lambda_{ti}& \Lambda_{it}{w}_t^-\alpha_t^-\beta_t^-\Lambda_{ti}\\
\Lambda_{it}{w}_t^-\alpha_t^-\beta_t^-\Lambda_{ti}&\Lambda_{it}{w}_t^-\beta_t^-\beta_t^-\Lambda_{ti}\\\end{matrix}\right]
\right)^{-1},
\end{equation} 
where the $t$ index has been summed over to find each of the elements 
in the $2\times2$ matrix on the right hand side and $A_{{tj}}$ has been 
replaced by $\Lambda_{{ti}}$ so the equation 
now determines the $Q,U$ values in the observed map, $Q_i,U_i$. 
There is one $2\times2$ matrix inversion performed for each pixel, $i$, in 
the observed map. In other words, one value of ${e_i}$, ${f_i}$, and ${g_i}$ is 
computed for each pixel in the observed map, and filled into the $i$-th diagonal element of $\mathbf{e}$, $\mathbf{f}$ and $\mathbf{g}$.

The pairmaps $\mathbf{\tilde{m}_{\boldsymbol{\alpha}^-}}$ and 
$\mathbf{\tilde{m}_{\boldsymbol{\beta}^-}}$ are transformed into Stokes 
$Q,U$ by multiplying by 
$\left[\begin{matrix}\mathbf{e}&\mathbf{f}\\\mathbf{f}&\mathbf{g}\\\end{matrix}\right]$.
Observed $Q,U$ maps are found according to:
\begin{equation}
\label{eqn:qu_def}
\left[\begin{matrix}\mathbf{\tilde{Q}'}\\\mathbf{\tilde{U}'}\\\end{matrix}\right]
=
\left[\begin{matrix}\mathbf{e}&\mathbf{f}\\\mathbf{f}&\mathbf{g}\\\end{matrix}\right]
\sum_{{\pairs},{\scanset}}
\boldsymbol{\mathcal{P}}_{{\pairs},{\scanset}}
\left[\begin{matrix}\mathbf{Q}\\\mathbf{U}\\\end{matrix}\right],
\end{equation}
where $\boldsymbol{\mathcal{P}}$ was defined in Equation~\ref{eqn:p_def}.

If the sum of $\boldsymbol{\mathcal{P}}$ matrices could be inverted, it 
would be possible to use this inverse to recover an unbiased estimate of the
original $Q$ and $U$.  However, $\boldsymbol{\mathcal{P}}$ is singular 
because it includes polynomial filtering and scan-synchronous signal 
subtraction, which completely remove some modes that were present in 
the original maps. We therefore use instead the matrix defined by 
the matrix inversion in Equation \ref{eqn:inver}, which does not include 
these filtering operations. Even the inversion in 
Equation \ref{eqn:inver} is singular unless the coadded data contains 
observations at multiple detector angles, $\Psi_t$. Observations at multiple 
detector orientations are made through deck rotations or by coadding over 
receivers in different orientations. As described in Section 
\ref{sec:scan_strat}, deck rotations occur between phases, so coadding over 
phases makes the matrix invertible. 

Including deprojection, Equation \ref{eqn:qu_def} becomes:
\begin{align}
\left[\begin{matrix}\mathbf{\tilde{Q}'}\\\mathbf{\tilde{U}'}\\\end{matrix}\right]
=
\left[\begin{matrix}\mathbf{e}&\mathbf{f}\\\mathbf{f}&\mathbf{g}\\\end{matrix}\right]
\sum_{\pairs}
\left(
\mathbf{D}
\sum_{\scanset}^{{\scanset}\in{phase}}
\boldsymbol{\mathcal{P}}_{\scanset}
\right)_{\pairs}
\left[\begin{matrix}\mathbf{Q}\\\mathbf{U}\\\end{matrix}\right].
\end{align}
This represents the entire $Q,U$ map making process for signal 
simulations: from input maps to observed maps, including filtering 
operations. It can be summarized as:
\begin{equation}
\left[\begin{matrix}\mathbf{\tilde{Q}'}\\\mathbf{\tilde{U}'}\\\end{matrix}\right]
=
\left[\begin{matrix}
 \mathbf{R'_{QQ}} & \mathbf{R'_{QU}} \\ 
 \mathbf{R'_{UQ}} & \mathbf{R'_{UU}} \\ 
\end{matrix}\right]
\left[\begin{matrix}\mathbf{Q}\\\mathbf{U}\\\end{matrix}\right].
\end{equation}

\subsection{Non-apodized observation matrix}
\label{sec:apod}

As constructed, the matrix, $\mathbf{R'}$, contains an 
apodization based on the inverse variance of the timestreams, $\mathbf{w^+}$ 
and $\mathbf{w^-}$. We can, however, choose to remove this apodization, 
producing maps with equal weight across the field in units of $\mu$K. We 
construct the quantities:

\begin{align}
   \mathbf{W^+} &= 
\sum_{\pairs,\scanset}
      \left( \boldsymbol{\Lambda}\mathbf{w^+}\boldsymbol{\Lambda}^\top\right)_{\pairs,\scanset},
\end{align}
 
\begin{align}
   \mathbf{W^-} &= 
\left[\begin{matrix}\mathbf{e}&\mathbf{f}\\\mathbf{f}&\mathbf{g}\\\end{matrix}\right]
\sum_{\pairs,\scanset}
      \left(\begin{bmatrix} \boldsymbol{\Lambda}\boldsymbol{\alpha^-}\boldsymbol{\alpha^-}\mathbf{w^-}\boldsymbol{\Lambda}^\top \\ \boldsymbol{\Lambda}\boldsymbol{\beta^-}\boldsymbol{\beta^-}\mathbf{w^-}\boldsymbol{\Lambda}^\top  \end{bmatrix}\right)_{\pairs,\scanset},
\end{align}
and use these to remove the apodization from the observation matrix, solving 
for the non-apodized observation matrix, $\boldsymbol{\mathcal{R}}$:

\begin{align}
\boldsymbol{\mathcal{R}_{TT}}=\left(\mathbf{W^+}\right)^{-1}\mathbf{{R'}_{TT}}
\end{align}

\begin{equation}
\left[\begin{matrix}
\boldsymbol{\mathcal{R}_{QQ}} & \boldsymbol{\mathcal{R}_{QU}} \\ 
\boldsymbol{\mathcal{R}_{UQ}} & \boldsymbol{\mathcal{R}_{UU}} \\ 
\end{matrix}\right]
=
\left(\mathbf{W^-}\right)^{-1} 
\left[\begin{matrix}
 \mathbf{{R'}_{QQ}} & \mathbf{{R'}_{QU}} \\ 
 \mathbf{{R'}_{UQ}} & \mathbf{{R'}_{UU}} \\ 
\end{matrix}\right]
\end{equation}

\subsection{Selecting the observation matrix's apodization}
\label{sec:modapod}

Using the non-apodized observation matrix of Section \ref{sec:apod}, we can 
create an observation matix with an arbitrary apodization, $\mathbf{Z}$. 
The matrix $\mathbf{R}$ is constructed as follows:

\begin{equation}
 \mathbf{{R}_{TT}}
=
\mathbf{Z}
\boldsymbol{\mathcal{R}_{TT}}
\end{equation}

\begin{equation}
\left[\begin{matrix}
 \mathbf{{R}_{QQ}} & \mathbf{{R}_{QU}} \\ 
 \mathbf{{R}_{UQ}} & \mathbf{{R}_{UU}} \\ 
\end{matrix}\right]
=
\left[\begin{matrix}
\mathbf{Z} & \mathbf{0} \\ 
\mathbf{0} & \mathbf{Z} \\ 
\end{matrix}\right]
 \left[\begin{matrix}
\boldsymbol{\mathcal{R}_{QQ}} & \boldsymbol{\mathcal{R}_{QU}} \\ 
\boldsymbol{\mathcal{R}_{UQ}} & \boldsymbol{\mathcal{R}_{UU}}  
\end{matrix}\right]
\end{equation}

A sensible choice for the apodization, $\mathbf{Z}$, may be the inverse variance 
mask we removed in Section \ref{sec:apod}, or a smoothed version thereof. 
However, there is freedom to choose any apodization at 
this point, and this may prove useful in joint analyses with other experiments, 
where the analysis  combines maps with low noise regions in slightly different 
regions of the sky.

\subsection{Summary}
We have constructed a matrix, $\mathbf{R}$, which performs the linear 
operations of polynomial filtering, scan-synchronous signal subtraction, 
deprojection, weighting and pointing. $\mathbf{R}$ has dimensions 
$(3\tilde{n}_{p},3n_p)$ where $\tilde{n}_p$ is the number of pixels in 
the \bicep\ map and $n_p$ is the number of pixels in the input \healpix\ map.

Using the observation matrix, the entire process of generating a signal 
simulation from an input map is:
\begin{equation}
   \left[
   \begin{matrix}
   \mathbf{\tilde{T}}\\
   \mathbf{\tilde{Q}}\\
   \mathbf{\tilde{U}}\\
   \end{matrix}
   \right]
=
\left[
\begin{array}{c|cc}
  \mathbf{R_{TT}} & \mathbf{0} & \mathbf{0} \\ \hline
  \mathbf{0} & \mathbf{R_{QQ}} & \mathbf{R_{QU}} \\ 
  \mathbf{0} & \mathbf{R_{UQ}} & \mathbf{R_{UU}} \\ 
\end{array}
\right]
   \left[
   \begin{matrix}
   \mathbf{T}\\
   \mathbf{Q}\\
   \mathbf{U}\\
   \end{matrix}
   \right].
\label{eqn:reobmatrix}
\end{equation}
Here the off diagonal terms, $\mathbf{R_{QU}}$ and $\mathbf{R_{UQ}}$, exist 
because the filtering operations are performed on pair difference 
timestreams, which are a combination of $Q$ and $U$. 

The deprojection operator contains regression against a temperature map 
that must be chosen \textit{before} constructing the observation matrix.  
The deprojection operator is a linear filtering operation, 
and it only removes beam systematics arising from one particular temperature
field. One could in principle apply the deprojection operator to $Q,U$ maps
corresponding to a different temperature field. The operator would remove the 
same modes from the polarization field, but these modes would not correspond 
to those which had been mixed between $T$ and $Q,U$ by beam systematics
(or $TE$ correlation).

When constructing an ensemble of \emode\ realizations for use in Monte Carlo 
power spectrum analysis, the $TE$ correlation and the fixed temperature sky 
force us to build constrained realizations. The ensemble of simulations all 
contain identical temperature fields, so we cannot use them for
analysis of temperature, which is acceptable because the focus of our 
analysis is polarization. The ensemble does contain different 
realizations of $Q,U$, constrained for the given temperature field, and 
these can be used in Monte Carlo analysis of polarization. The details 
of constructing these constrained input maps is the subject of Appendix 
\ref{sec:constrained_realizations}.

Because the construction of $\mathbf{R_{QQ}}$, $\mathbf{R_{QU}}$, and $\mathbf{R_{UU}}$ 
depends on a fixed temperature field, the deprojection templates can be 
thought of as numerical constants. $\mathbf{R_{TT}}$ performs a separate 
filtering on the temperature field that is largely decoupled from the 
filtering of $Q$ and $U$.
To include systematics that leak temperature to polarization, the terms 
$\mathbf{R_{TQ}}$ and $\mathbf{R_{TU}}$ would in principle need to be non zero.
However, so long as the leakage corresponded to modes being removed
by the deprojection matrix $\mathbf{D}$ the 
the deprojection elements in the $\mathbf{R_{QQ}}$, 
$\mathbf{R_{QU}}$, and $\mathbf{R_{UU}}$ blocks would
ensure that the output $Q,U$ maps were identical.

\begin{figure}[h!] 
\begin{center}
\resizebox{\columnwidth}{!}
{\includegraphics{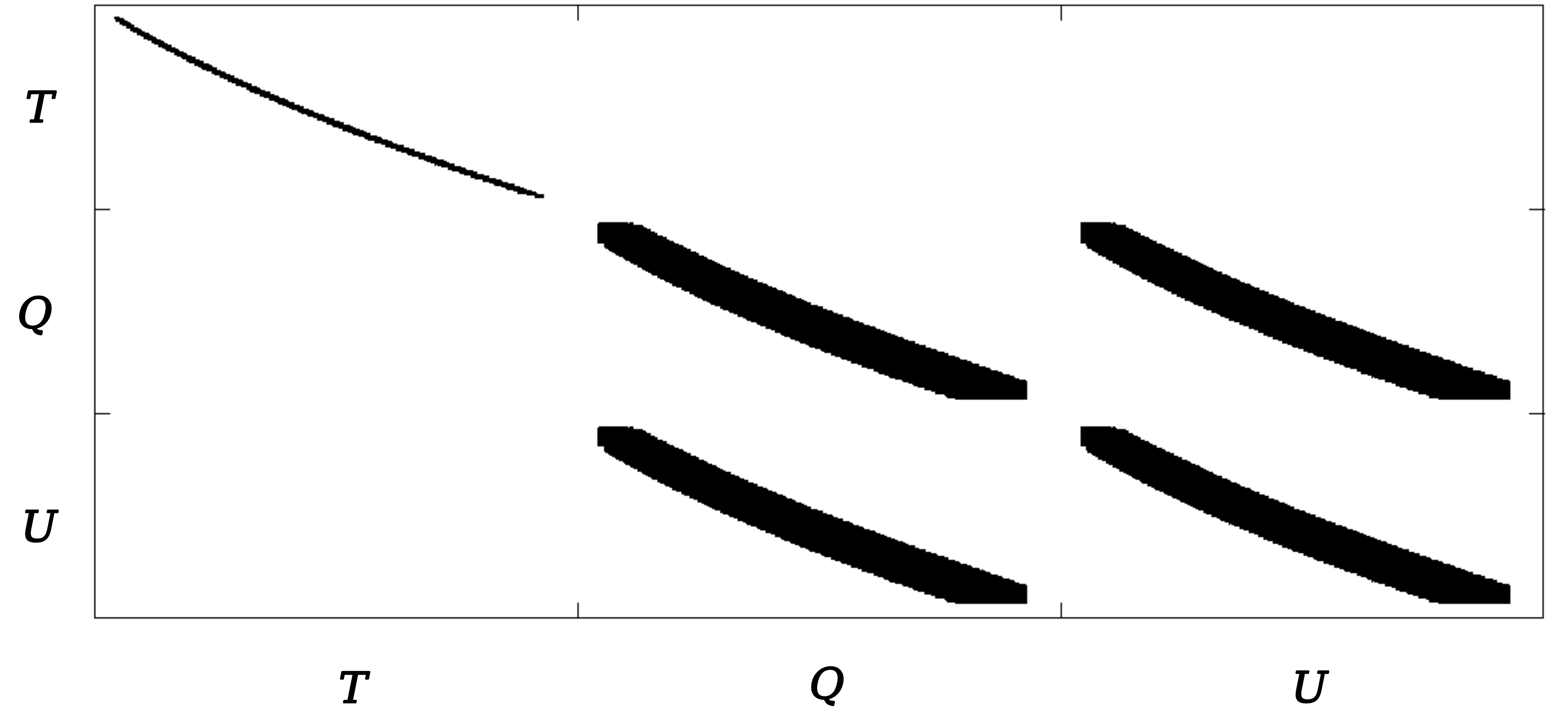} }
\end{center}
   \caption[Observation matrix]{Observation matrix, $\mathbf{R}$: filled elements of 
the observation matrix for the \biceptwo\ 3-year data set. 
$TQ$, $TU$, $UT$, and $QT$ are empty because no T$\rightarrow$P leakage 
is simulated. The horizontal axis corresponds to the \healpix\ 
pixelization and has 3$\times$111,593 elements. The vertical axis corresponds 
to the \bicep\ pixelization, and has 3$\times$23,600 elements. The matrix has 
only $\sim$5\% of its elements filled.}
   \label{fig:reob_matrix}
\end{figure}

Although the nominal dimensions of $\mathbf{R}$ are large, our constant 
elevation scan strategy means that $\mathbf{R}$ is only filled for pixels 
at roughly the same declination. This means that $\mathbf{R}$ is largely 
sparse, as shown in Figure \ref{fig:reob_matrix}.

\begin{figure}[h!] 
\begin{center}
\resizebox{\columnwidth}{!}
{\includegraphics{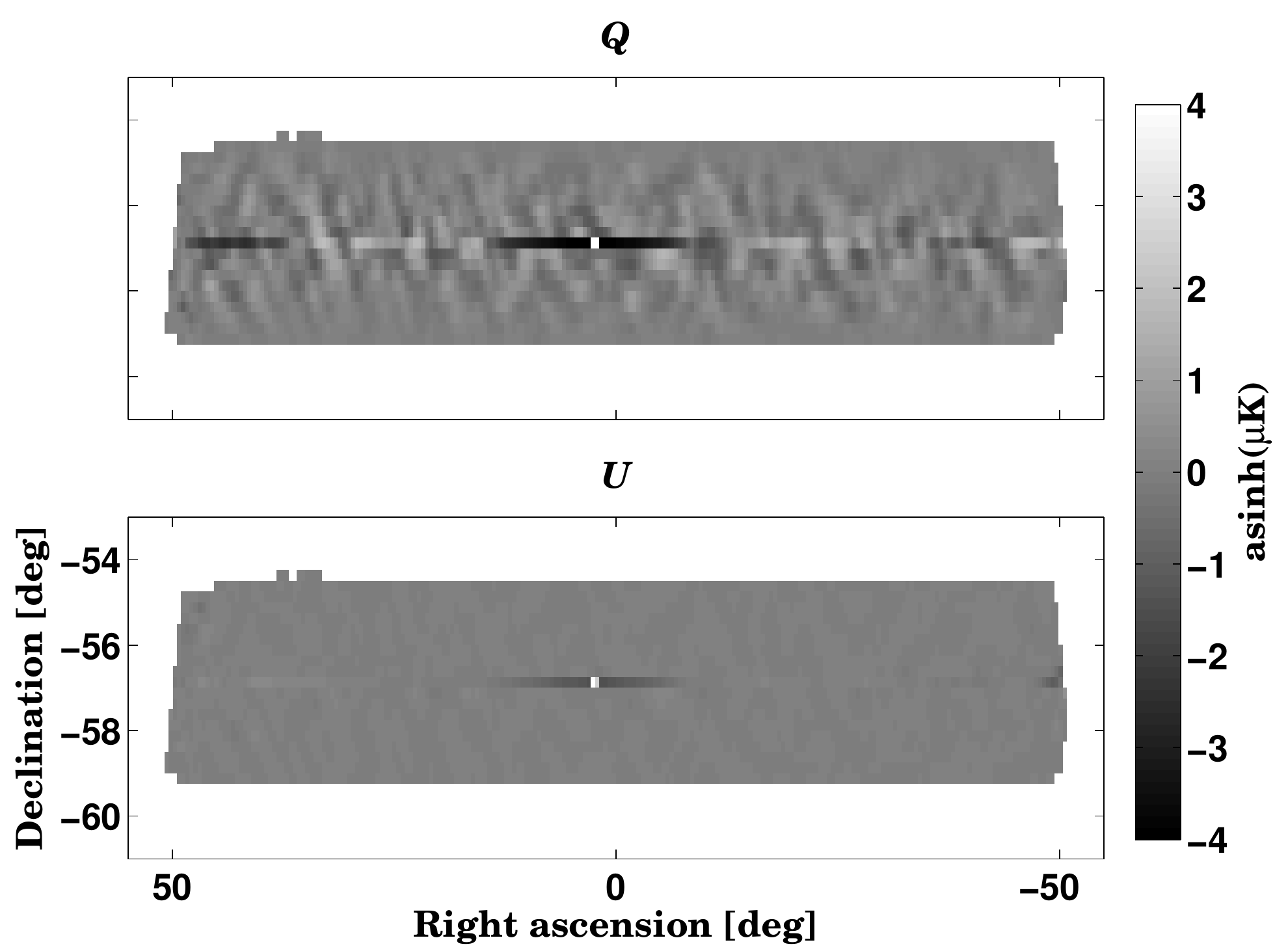}}
\end{center} 
  \caption[A column of the observation matrix]{A single column of the 
observation matrix $\mathbf{R}$, for a \healpix\ $Q$ pixel near the center of 
our field. The value of a single input $Q$ pixel affects both $Q$ and $U$ 
values in the observed map over the range of declinations covered in a phase.} 
   \label{fig:row_of_reob}
\end{figure}

Some intuition about the operations the observation matrix performs can
be gained by plotting a column of the matrix reshaped as maps---see Figure 
\ref{fig:row_of_reob}. The column chosen in this case corresponds 
to a central pixel in the observed field. It shows how $Q$ and $U$ values 
in the observed map are sourced from a $Q$ pixel in the \healpix\ map. 
The bright pixel in the $Q$ observed map corresponds to the location of the 
input $Q$. The effects of polynomial and scan-synchronous signal subtraction 
are visible to the left and right of the bright $Q$ pixel. These two types of 
filtering are performed on scansets and are therefore confined to a row of 
pixels. Deprojection operates on phases, creating the effects seen at other 
declinations. Because all of these filtering operations are 
performed on pair difference data, which contains linear combinations of $Q$ 
and $U$, signal in the observed $U$ map can be created by signal in the 
input \healpix\ $Q$ map. This is why the $U$ map in Figure 
\ref{fig:row_of_reob} is non-zero.

\subsection{Forming maps from real timestreams}
We can form observed maps from the real timestreams using the matrices 
constructed above:
\begin{equation}
\mathbf{\tilde{T}}_{real}
=
\mathbf{Z}
\left(\mathbf{W^+}\right)^{-1} 
\sum_{{\pairs},{\scanset}}
\left(\boldsymbol{\Lambda}\mathbf{w}^+\mathbf{GFs}\right)_{{\pairs},{\scanset}}
\end{equation}
\begin{align}
\label{eqn:realmatrix}
\left[\begin{matrix}\mathbf{\tilde{Q}}_{real}\\\mathbf{\tilde{U}}_{real}\\\end{matrix}\right]
&=
\left[\begin{matrix}
\mathbf{Z} & \mathbf{0} \\ 
\mathbf{0} & \mathbf{Z} \\ 
\end{matrix}\right]
\left(\mathbf{W^-}\right)^{-1} 
\left[\begin{matrix}\mathbf{e}&\mathbf{f}\\\mathbf{f}&\mathbf{g}\\\end{matrix}\right]
\times
\nonumber\\
&
\sum_{\pairs}
\left(
\mathbf{D}
\sum_{\scanset}^{{\scanset}\in{phase}}
\left(
\left[\begin{matrix}\boldsymbol{\Lambda}&\mathbf{0}\\\mathbf{0}&\boldsymbol{\Lambda}\\\end{matrix}\right]
\left[\begin{matrix}\boldsymbol{\alpha}^-\\\boldsymbol{\beta}^-\\\end{matrix}\right]
\mathbf{w}^-\mathbf{GFd}
\right)_{\scanset}
\right)_{\pairs}.
\end{align}
It is important to note that the exact same matrices are used to process 
the real data in Equation \ref{eqn:realmatrix} as are used to construct 
the simulated maps in Equation \ref{eqn:qu}. 

\subsection{Equivalence of observation matrix with standard pipeline}
\label{sec:check_matrix}
The matrix formalism described above is self-contained and complete in 
the sense that it contains the tools necessary to create real data maps and 
simulated maps. 

\begin{figure}[h!] 
\centering
\resizebox{\columnwidth}{!}
{\includegraphics{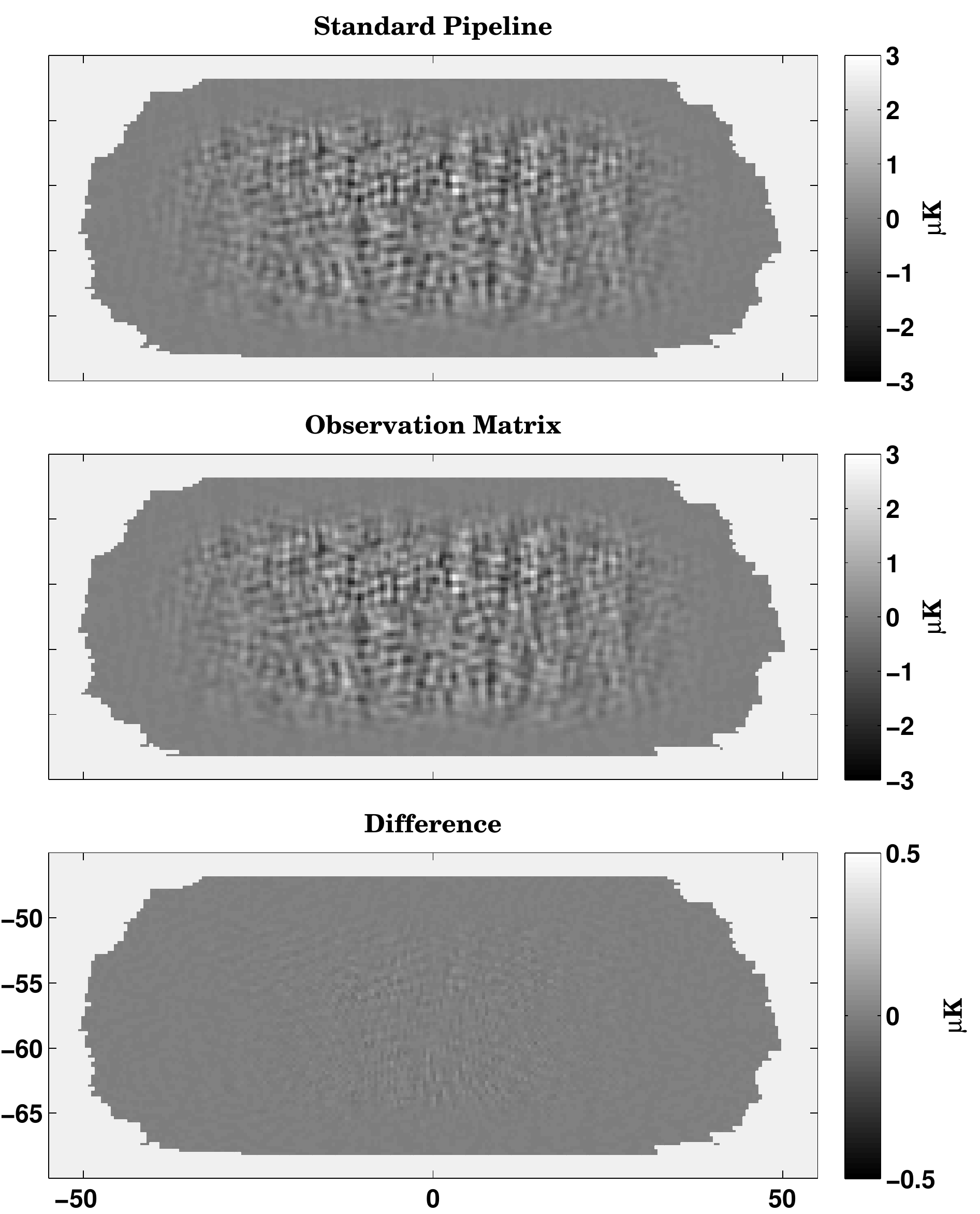}}
  \caption[Comparison of observed $Q$ maps created by the observation 
matrix and the standard pipeline]{Comparison of observed $Q$ maps created by 
the observation matrix and the standard pipeline. The input map for both is 
from the same simulation realization. There are small differences due the 
difference in deprojection timescales, input \healpix\ map resolution, and 
interpolation.}
  \label{fig:reob_mapcomp}
\end{figure}

\begin{figure}[h] 
  \begin{center}
    \resizebox{\columnwidth}{!}
    {\includegraphics{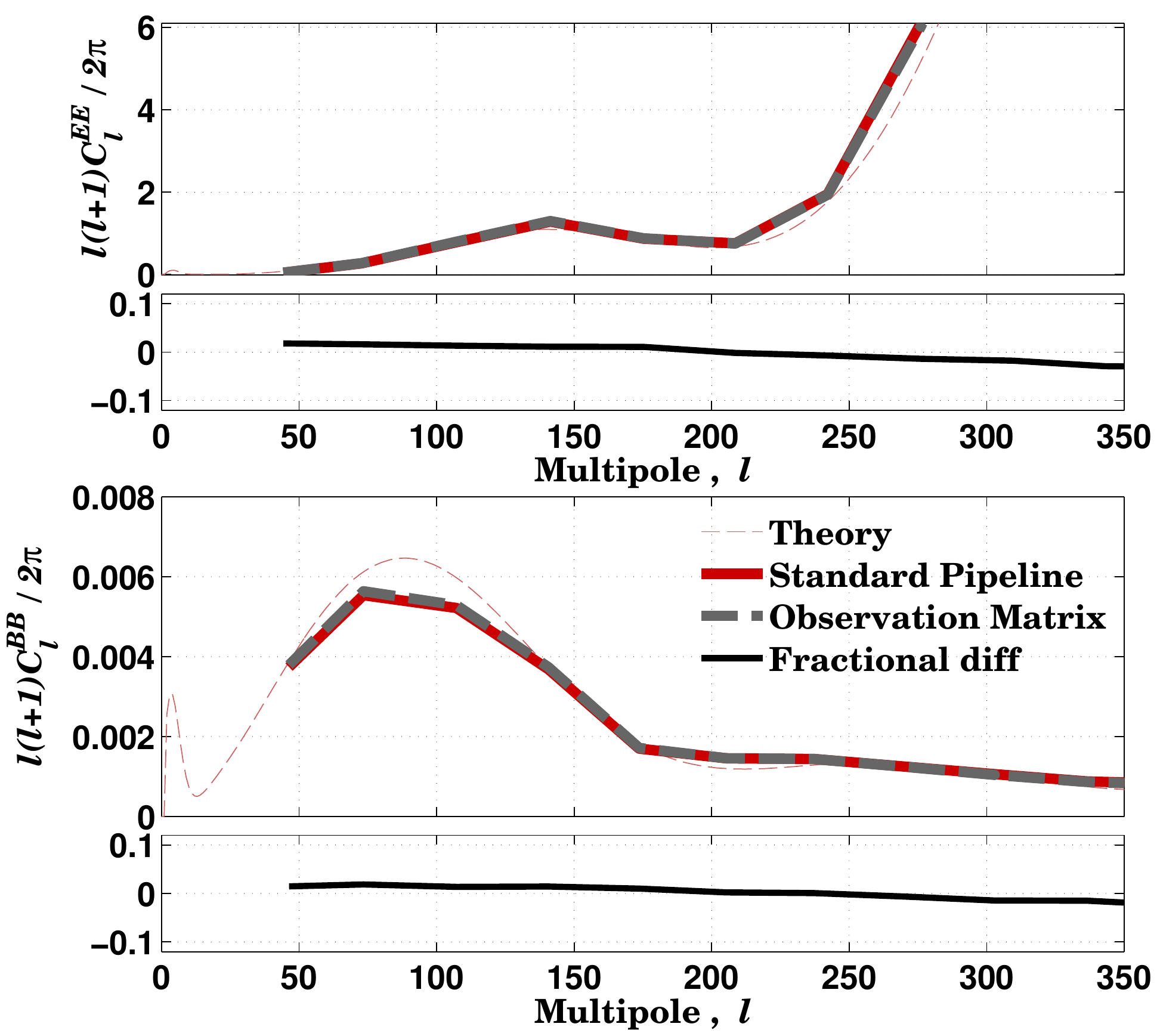}}
  \end{center}
  \caption[Comparison of power spectra of maps created by the 
observation matrix and the standard pipeline] {Comparison of power 
spectra of maps created by the observation matrix and the standard 
pipeline. The input map for both is from the same simulation realization, 
which differs from the theory curve for this particular realization in the 
\bicep\ field. The two methods are fractionally the same to within a few 
percent over the multipoles of interest, $50<l<350$.}
  \label{fig:reob_apscomp}
\end{figure}

We demand that the map making and filtering operations be identical 
between the standard pipeline and the observation matrix. It is 
straightforward to test this equivalence: simulated maps run through 
the standard pipeline must be identical to the maps found with the 
observation matrix. Figure \ref{fig:reob_mapcomp} and Figure 
\ref{fig:reob_apscomp} show that the two match quite well, within a few 
percent over the multipoles of $50<l<350$. The lack of a perfect 
match is due to the difference in deprojection timescale and because the 
standard pipeline uses $N_{side}$=2048 \healpix\ input maps that are Taylor 
interpolated, whereas the observation matrix uses $N_{side}$=512 input 
\healpix\ maps with nearest neighbor interpolation.

\section{Signal covariance matrix, $\mathbf{C}$}
\label{section:ct}
The signal covariance matrix contains the pixel-pixel covariances of a 
map for a given spectrum of Gaussian fluctuations. The diagonal entries 
contain the variance of each pixel, and each row describes the covariance 
of a given pixel with the other pixels in the map. For [$T,Q,U$] maps, the 
covariance matrix contains nine sub-matrices for the correlations 
between $T$,$Q$, and $U$.

\subsection{True sky signal covariance matrix}
\label{subsec:unobs_cov}
A pixel on the sky at location $i$, has values of the Stokes parameters: 
\begin{equation}
{x}_i\equiv\left(
\begin{array}{c}
  T_i\\
  Q_i\\ 
  U_i\\ 
\end{array}
\right).
\end{equation}
The $3\times3$ pixel-pixel covariance between two locations on the sky, $i$ 
and $j$, is given by:
\begin{equation}
\mathbf{C}_{i,j}\equiv\left<{x_i}{x_j}\right> = \boldsymbol{\mathbb{R}}(\alpha)\mathbf{M}({r_i}\cdot{r_j})\boldsymbol{\mathbb{R}}(\alpha)^{\top}.
\end{equation}
The covariance matrix, $\mathbf{M}$, is defined with the $Q,U$ convention 
referenced to the great circle connecting the two points, $i,j$. For a 
particular spectrum $\mathbf{M}$ depends only on the dot product between the pixels, 
${r_i}\cdot{r_j}$. $\mathbf{M}$ contains nine symmetric sub-matrices:
\begin{equation}
\label{eqn:cov_th}
\mathbf{M}(r_i\cdot{r_j})
=
\left(
\begin{array}{ccc}
  \left<T_iT_j\right> & \left<T_iQ_j\right> & \left<T_iU_j\right> \\
  \left<Q_iT_j\right> & \left<Q_iQ_j\right> & \left<Q_iU_j\right> \\ 
  \left<U_iT_j\right> & \left<U_iQ_j\right> & \left<U_iU_j\right> \\ 
\end{array}
   \right)
\end{equation}
The $3\times3$ matrix, $\boldsymbol{\mathbb{R}}$, is applied to rotate 
from this local reference frame to a global frame where $Q,U$ are 
referenced to the North-South meridians.  The angle between the great 
circle connecting any two points and the global frame is given by the 
parameter ${\alpha}$.
\begin{equation}
\boldsymbol{\mathbb{R}}(\alpha) = 
\left(
\begin{array}{ccc}
  {1} & {0} & {0} \\
  {0} & {\cos2\alpha} & {\sin2\alpha} \\ 
  {0} & {-\sin2\alpha}  &  {\cos2\alpha} \\ 
\end{array}
   \right)
\end{equation}
Changing the sign of $\alpha$ allows us to change the 
polarization convention from IAU to \healpix\ ($U$ to $-U$), see 
\citet{Hamaker96}. We have chosen to use the IAU convention for \biceptwo\ 
and \keckarray\ covariance matrices.

The true sky pixel-pixel signal covariance matrix for the Stokes $Q,U$ 
parameters is derived in \citet{Kamionkowski_97c,Zaldarriaga98b}. To 
calculate the covariances, we follow some of the suggestions in Appendix 
A of \citet{tegmark01}.

 We use the \healpix\ ring pixelization, which allows the 
covariance to be calculated simultaneously for all pixels at a particular 
latitude that are separated by the same distance.\footnote{This quality 
of the \healpix\ maps is by design, see \citet{Gorski04}.} This shortcut 
is exploited by simultaneously calculating all equidistant pixels for 
rows in the map that have the same latitude to within $1\times10^{-4}$ 
degrees. This approximation is much smaller than the $\sim$7 arcminute pixels in 
$N_{side}$=512 maps, and the rounding error has been found to be insignificant.

\subsection{Observed signal covariance matrix, $\mathbf{\tilde{C}}$}
\label{sec:reob_theory_cov}

The observed signal covariance matrix contains the pixel-pixel covariance 
in the observed \bicep\ pixelized maps. Theoretically, modifying the true 
sky signal covariance matrix is simple: take the unobserved signal 
covariance of $\mathbf{C}$ of Section \ref{subsec:unobs_cov} and the observation matrix
$\mathbf{R}$ from Section \ref{sec:reob_matrix} and form the product:
\begin{equation}
\mathbf{\tilde{C}}=\mathbf{RCR^{\top}}.
\label{eqn:observeC}
\end{equation}
This equation results in a symmetric positive definite matrix, which is 
rank deficient because of the filtering steps in the observing process. 

Unfortunately, performing this multiplication is computationally demanding: 
The input $\mathbf{C}$ is a square matrix, with 3$\times$111,593 
elements on a side, corresponding to the elements of $T$, $Q$, and $U$. 
To reduce the memory 
requirements of the calculation, we divide the covariance matrix, $\mathbf{C}$, 
into row subsets and calculate in parallel. Once a row subset is calculated, 
the observation matrix is immediately applied to transform the \healpix\ 
covariance to the observed map covariance, which reduces the dimensions of 
the covariance to the 23,600 pixels of the observed maps.

The covariance and observed covariance should both 
be symmetric, which provides a good check on our math. Usually the output is 
slightly (fractionally, $\sim1/10^{7}$) non-symmetric due to rounding errors 
in the multiplication, and we force the final matrix to be symmetric to 
numerical precision by averaging across the diagonal before moving to the 
next steps, since symmetric matrices often allow the use of faster algorithms.  

\begin{figure}[h] 
\begin{center}
\resizebox{\columnwidth}{!}
{\includegraphics{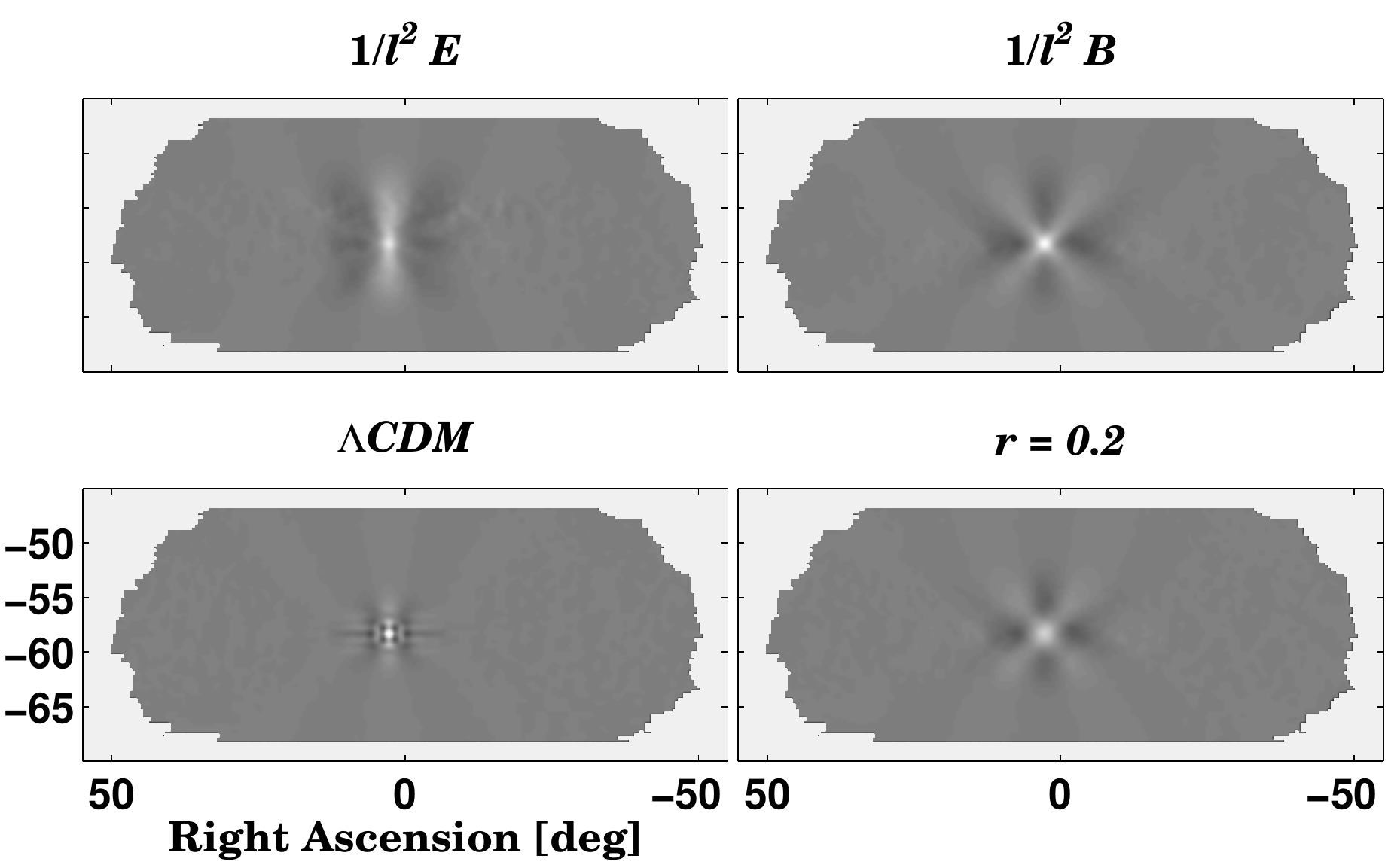}}
\end{center}
  \caption[Pixel-Pixel $Q$ Covariance for one pixel]{Maps showing a row of 
the observed covariance matrix $\mathbf{\tilde{C}}$. The row selected corresponds to the 
covariance of an individual $Q$ pixel at the center of the map. The top 
row shows the covariance used to calculate the pure $E$ and \bmodes\ 
described in Section \ref{sec:ebseparation}. The bottom row shows the 
covariance for an input spectrum corresponding to  $\Lambda$CDM [left], 
and $r = 0.2$ tensors [right].}
   \label{fig:ct_row}
\end{figure}

A row of the observed covariance matrix can be reshaped into a map, which
reveals the structure of the covariance for a particular pixel, see Figure \ref{fig:ct_row}.

\section{$E/B$ separation using a purification matrix}
\label{sec:ebseparation}

The observed covariance matrix contains expected pixel-pixel covariance in 
our observed maps given an initial spectrum. The observation matrix, 
$\mathbf{R}$, has made the $E$ and \bmode\ spaces of the observed covariance 
non-orthogonal. The result of Section \ref{sec:matrixEB} is that 
we can find the orthogonal pure $E$ and pure $B$ spaces by solving the 
eigenvalue problem from Equation \ref{eqn:geneigenfirst}: 
$\mathbf{\tilde{C}_B}\mathbf{x_i}={\lambda_{i}}\mathbf{\tilde{C}_E}\mathbf{x_i}$.

\subsection{Construction of purification matrix}
\begin{figure}[h!] 
\begin{center}
\resizebox{\columnwidth}{!}
{\includegraphics{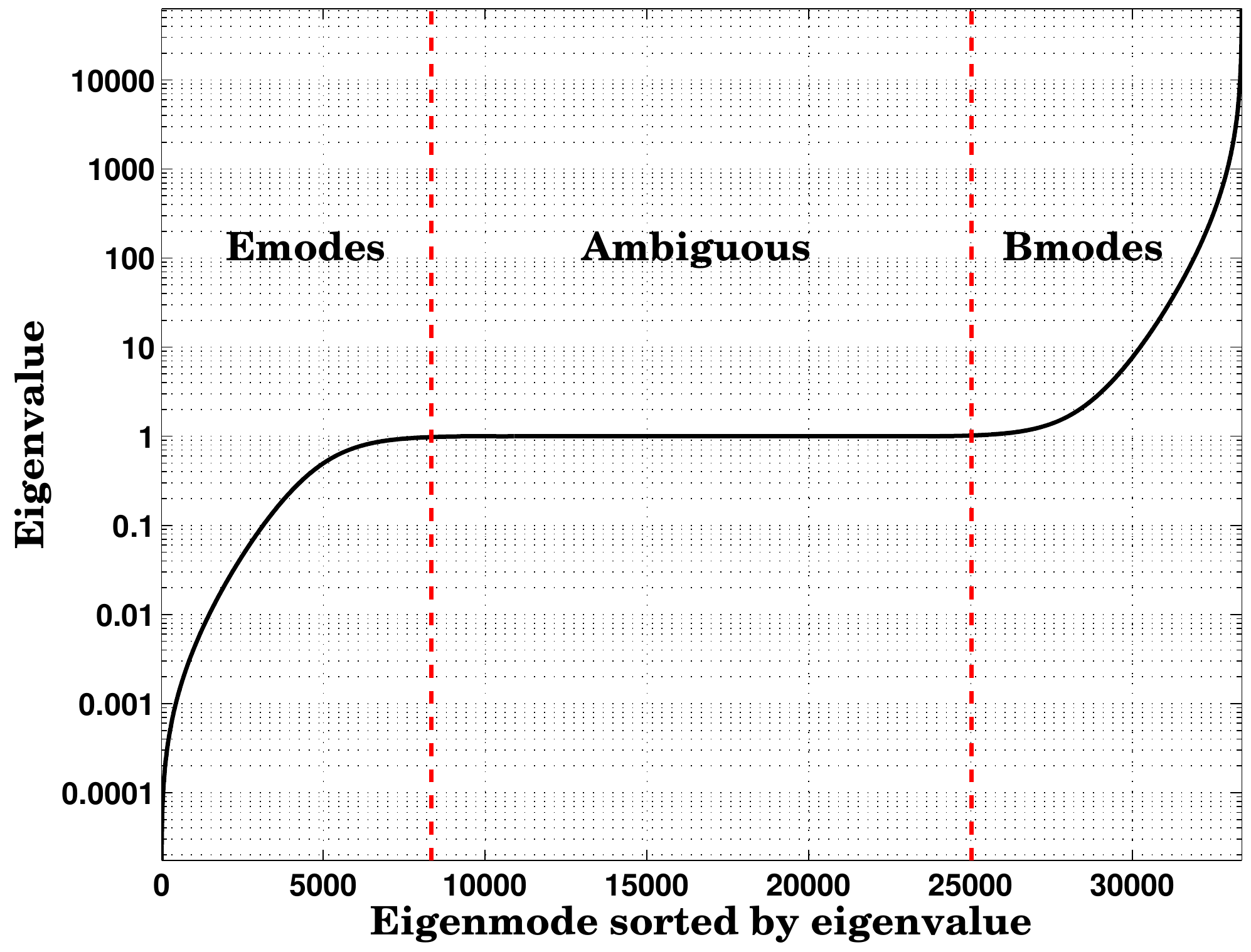}} 
\end{center}
  \caption[Purification eigenvalues]{The generalized eigenvalues for 
the \biceptwo\ observed covariance matrix, sorted by magnitude. Eigenvalues 
near one correspond to ambiguous modes: the modes that are simultaneously 
$E$ and $B$ in the observed space and must be thrown out. 
By selecting eigenmodes with eigenvalues that are the largest and smallest 
$1/4$ of the set of eigenvalues (shown to the left and right of the dashed 
red lines), we can construct subspaces that span the spaces of \bmodes\  
and \emodes\ that can be effectively observed using our scan strategy and 
analysis.}
   \label{fig:eigenvals}
\end{figure}

As written, Equation \ref{eqn:geneigenfirst} is not solvable: 
$\mathbf{\tilde{C}_B}$ has a null space that is the set of pure \emodes. 
Similarly, the space of pure \bmodes\ is the null space of $\mathbf{\tilde{C}_E}$. 
By adding the identity matrix multiplied by a constant, $\sigma^2\mathbf{I}$, to 
the covariance matrices we regularize the problem to find approximate 
solutions and eliminate the null spaces:
\begin{equation}
\label{eqn:gen_eigen}
(\mathbf{\tilde{C}_B}+\sigma^2\mathbf{I})\mathbf{x_i}={\lambda_{i}}(\mathbf{\tilde{C}_E}+\sigma^2{\mathbf{I})\mathbf{x_i}}.
\end{equation}
The amplitude of $\sigma^2$  sets the relative magnitude of the 
ambiguous mode eigenvalues versus the pure $E$ and pure \bmode\ eigenvalues. 
The eigenvalues are shown in Figure \ref{fig:eigenvals}. In our analysis, 
we choose $\sigma^2$ to be 1/100th the mean of the diagonal elements of 
the covariance matrices, $\mathbf{\tilde{C}_E}$ and $\mathbf{\tilde{C}_B}$.

In Equation \ref{eqn:gen_eigen}, $\mathbf{\tilde{C}_E}$ is an observed 
covariance matrix for [$Q,U$] that is constructed according to 
Equations \ref{eqn:covariancedef} and \ref{eqn:observeC}. The input 
spectrum is set to a steeply red \emode\ spectrum, 
$C^{EE}_l=1/l^2$, $C^{BB}_l=0$. $\mathbf{\tilde{C}_B}$ is the same 
except for an input spectrum with only \bmodes, 
$C^{BB}_l=1/l^2$, $C^{EE}_l=0$. The eigenmodes in 
$\mathbf{x_i}$ with the largest eigenvalues comprise a set of 
vectors that span a space of pure \bmodes, $\mathbf{b_i}$. The pure 
$B$ quality of these vectors can be seen in the fact that the product 
$\mathbf{\tilde{C}_Ex}$ is much smaller than the product 
$\mathbf{\tilde{C}_Bx}$.  The eigenmodes in $\mathbf{x_i}$ with the 
smallest eigenvalues comprise a set of vectors that span a 
space of pure \emodes, $\mathbf{e_i}$.  

Using the reddened input spectrum $1/l^2$ causes the magnitude of the 
eigenvalues to be proportional to the band of multipole, $l$, that each mode 
contains. The steepness of the spectrum ensures each mode contains power from 
a limited range of $l$. The particular choice of $1/l^2$ is arbitrary. 

\begin{figure*}[h!]
\begin{center}
\resizebox{\textwidth}{!}
{\includegraphics{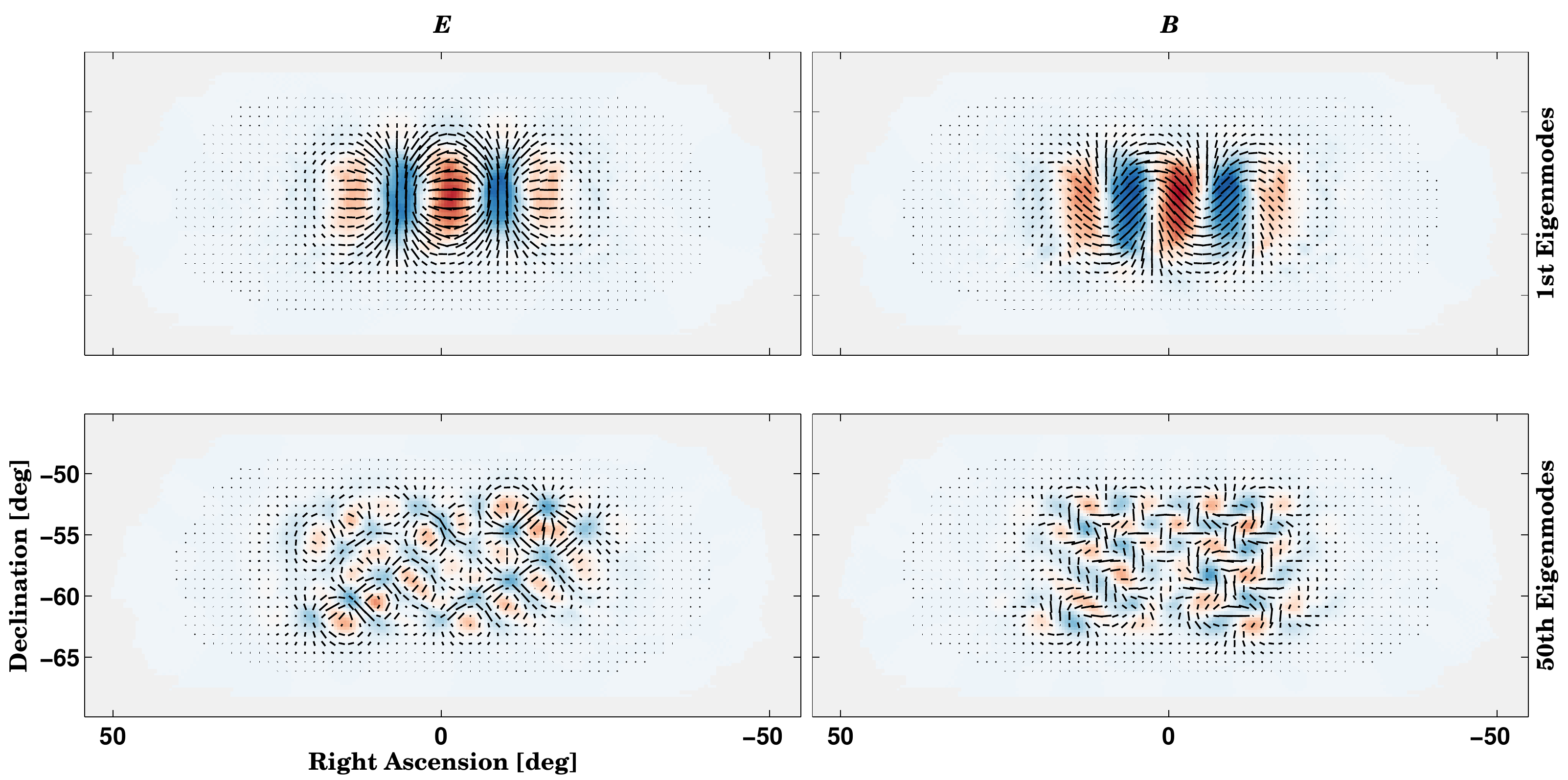}}
\end{center}
  \caption[Purification Eigenmodes]{Eigenmodes of the \biceptwo\ observed 
covariance matrix. Shown are the modes corresponding to the largest and 50th 
largest eigenvalues of Equation \ref{eqn:gen_eigen}. Colormap shows amplitude 
of $E$ and \bmodes. The eigenvalues are shown graphically in Figure 
\ref{fig:eigenvals}.} 
   \label{fig:eigenmodes}
\end{figure*}

A basis constructed from a subset of eigenmodes with large eigenvalues spans a 
subspace of pure \bmodes. We arbitrarily choose the pure $B$ subspace to 
consist of eigenmodes whose corresponding eigenvalues are the largest $1/4$ 
of the set. We find modes contained in this subset adequate for preserving 
power up to $l\sim$700. The pure $E$ subspace is similarly constructed with 
eigenmodes corresponding to the $1/4$ smallest eigenvalues. 
Figure \ref{fig:eigenvals} shows the sorted eigenvalues, and Figure 
\ref{fig:eigenmodes} shows four eigenmodes of the \biceptwo\ observed 
covariance.

Using the set of pure $E$ and pure \bmode\ basis vectors, 
two projection matrices are constructed from the outer products:
\begin{align}
    \boldsymbol{\Pi_E}&={\sum_{i} \mathbf{e_{i} e_{i}}^\top}
    \nonumber\\
    \boldsymbol{\Pi_B}&={\sum_{i} \mathbf{b_{i} b_{i}}^\top},
  \label{eqn:outerprod}
\end{align} 
which we call the purification matrices for pure $E$ and pure $B$. Operating 
the purification matrices on an input map projects onto the space of pure $E$ 
and pure $B$:
\begin{align}
    \mathbf{\tilde{m}_{pureE}}&=\boldsymbol{\Pi_E}\mathbf{\tilde{m}}
    \nonumber\\
    \mathbf{\tilde{m}_{pureB}}&=\boldsymbol{\Pi_B}\mathbf{\tilde{m}}.
  \label{eqn:purification}
\end{align}

From the construction of the pure $B$ basis, $\mathbf{b_i}$, one can see that 
the vector $\mathbf{\tilde{m}_{pureB}}$ vanishes for arbitrary input containing 
only \emodes: $\tilde{\mathbf{m}}=\mathbf{R}\cdot{\sum{a^{E}}_{lm}\mathbf{Y}^E_{lm}}$, as 
desired for a purified $B$ map.

\section{Matrix $E/B$ Separation applied to \biceptwo}
\label{sec:matrix_app}  

This section describes the application of the matrix based $E/B$ separation 
to the \biceptwo\ data set. The technique relies on the existence of the observation 
matrix and purification matrix from the previous sections.

\subsection{Motivation for matrix based $E/B$ separation in \biceptwo}
\label{sec:ebleak}

The \biceptwo\ and \keckarray\ analysis pipeline contains the following 
attributes that can leak \emodes\ to \bmodes : partial sky coverage,
timestream filtering (including deprojection), and choice of map projection+estimator. 
A simulation demonstrating the leaked \bmode\ maps for each of these effects 
is shown in Figure \ref{fig:leaked_bmodes}. These maps are created by applying 
the standard $E$ and $B$ estimator in Fourier space and then an inversion 
back to map space. The observations and data reduction produce three
classes of $E/B$ leakage:

\begin{figure*}
\resizebox{\textwidth}{!}
{\includegraphics{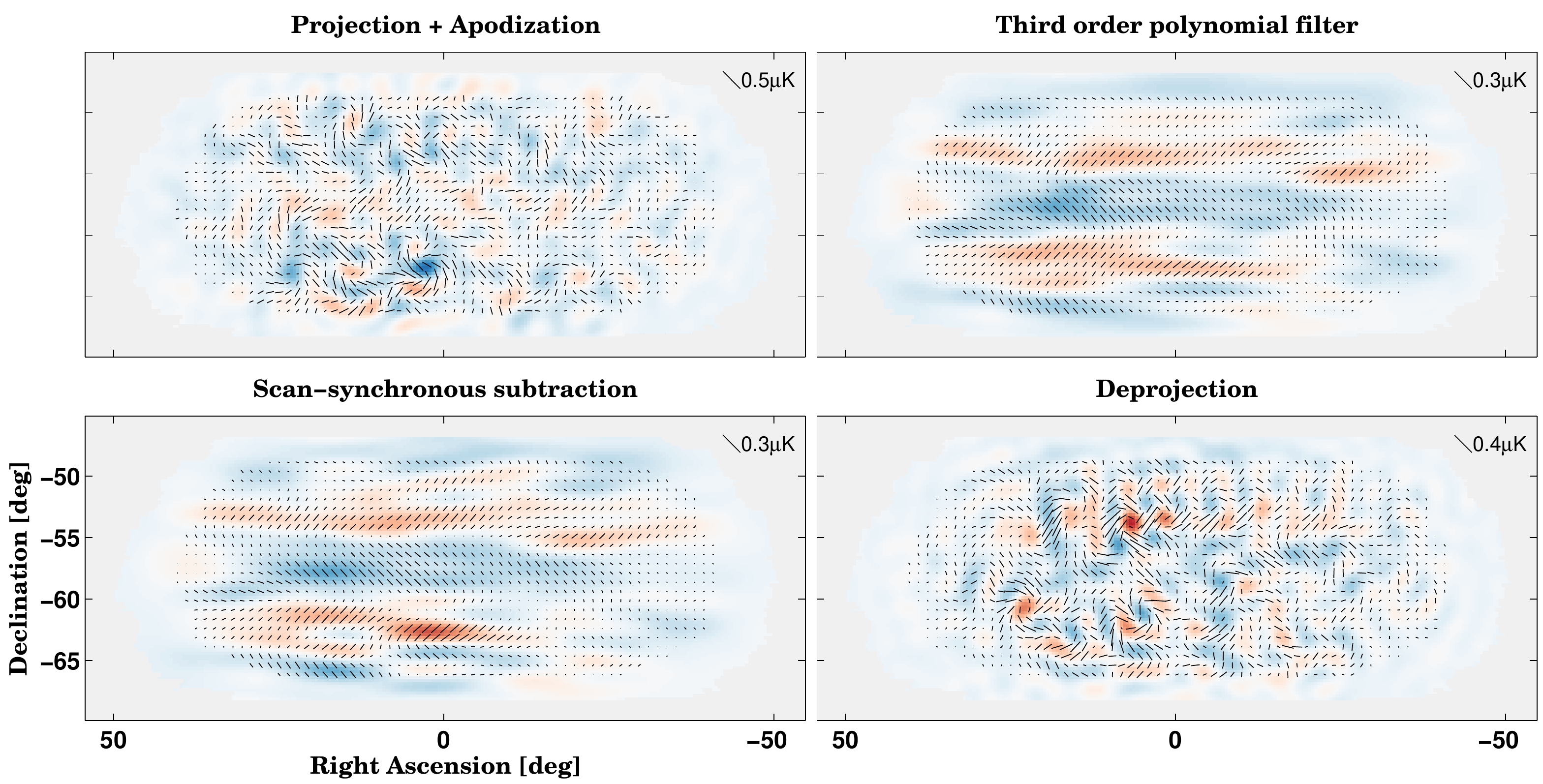} }
  \caption[$E$ to $B$ leakage Maps]{$E$ to $B$ leakage maps: examples of 
leaked \bmodes\ in the \biceptwo\ maps. Top row, left: Leaked \bmodes\ 
due to map projection and apodization. Top row, right: Leaked \bmodes\ due to 
third order polynomial subtraction. Bottom row, left: Leaked \bmodes\ due to
scan-synchronous signal subtraction. Bottom row, right: Leaked \bmodes\ 
due to deprojection of beam systematics.}
   \label{fig:leaked_bmodes}
\end{figure*}

\begin{itemize}
\item{\bf{Apodization}}:
The first obvious deviation from the ideal full sky map is the partial sky 
coverage of \biceptwo\ and \keckarray\ maps. Once a boundary is imposed $E/B$ 
leakage is created. Map boundary effects are reduced by an apodization window 
which tapers the maps smoothly to zero near the edges. Use of apodization 
windows is common practice in any Fourier transform analysis of finite 
regions to prevent `ringing' near boundaries. For small regions of sky, 
effects of the map boundary dominate the leakage even after apodization is 
applied. The apodization window used 
in \biceptwo\ and the \keckarray\ are modified inverse variance maps. We apply 
a smoothing Gaussian with a width of $\sigma=0.5^{\circ}$ to the inverse 
variance map, and, for the combined analyses of \biceptwo\ and the 
\keckarray, we use a geometric mean of the individual experiments' 
apodization maps. 

\item{\bf{Map projection}}:
The total extent of the \biceptwo\ and \keckarray\ maps is
about 50 degrees on the sky in the direction of RA, over a declination
of $-70<\delta<\-45$.
The chosen projection is a simple rectangular grid of pixels
in RA and Dec.
Taking standard discrete Fourier transforms of such maps results
in significant $E/B$ leakage.
While we note that other map projections will have significantly lower 
distortion, such effects will be present for all projections when
subjected to Fourier transform.
However, note that the $\mathbf{A}$ and $\boldsymbol{\Lambda}$ matrices
together fully encode the mapping from the underlying curved sky to the
flat sky of the observed map pixels, and therefore so does the observing matrix
$\mathbf{R}$ derived from these.

\item{\bf{Linear filtering effects}}:
There are three main analytic filters applied in the standard pipeline: 
polynomial filter, scan-synchronous subtraction, and deprojection. With respect 
to $E$ to $B$ leakage, all three filters are similar: by removing modes in the 
$Q,U$ maps, an \emode\ can be turned into an observed \bmode. Polynomial 
filtering and scan-synchronous signal subtraction create comparable leakage, 
both in amplitude and morphology. Deprojection creates more power in the 
leaked $B$ maps, and at smaller angular scales, than either polynomial 
filtering or scan-synchronous signal subtraction.

\end{itemize}

If matrix purification is not used, the
sample variance of $E/B$ leakage in the \biceptwo\ $BB$ power
spectrum is comparable to the uncertainty due to instrumental noise.
This is clearly highly undesirable, and led to the
development of the purification matrix described
in this paper.
The purification matrix `knows about' all of the $E/B$\
mixing effects and how to deal with them.

\subsection{Effectiveness of purification matrix}

\begin{figure*} 
\begin{center}
\resizebox{\textwidth}{!}
{\includegraphics{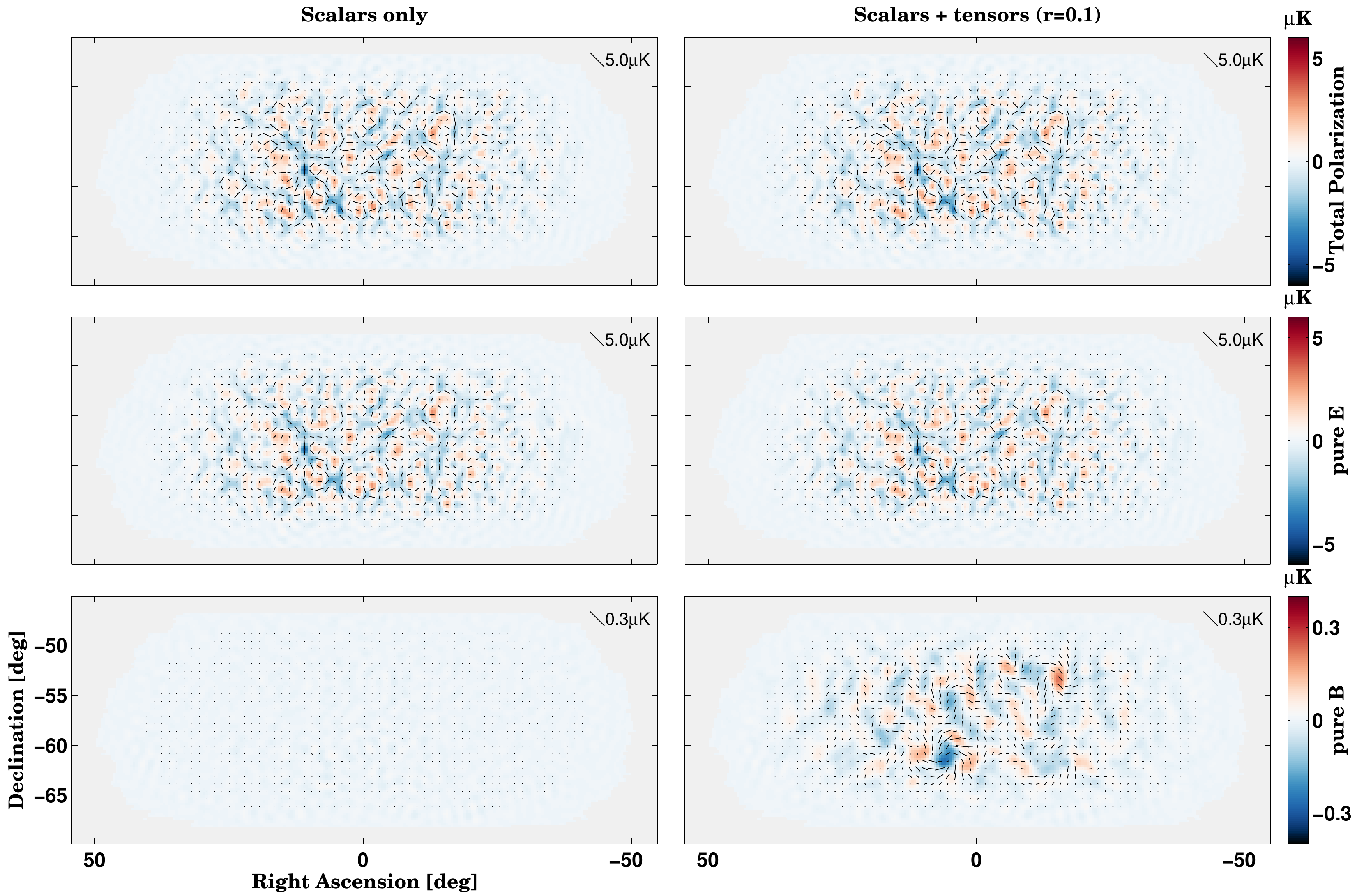}}
\end{center}
  \caption[Effectiveness of $E/B$ separation]{
Polarization maps showing the effectiveness of the \biceptwo\ purification 
matrix at separating noiseless simulations into pure $E$ 
and pure $B$.
Left column: on a scalar only unlensed-\lcdm\ \biceptwo\ 
simulation. Right Column: on the same simulation with the addition of
a small tensor component.
Top Row: the total polarization of the \biceptwo\ observed map, containing 
$E$ and ambiguous modes. 
Center Row: pure \emode\ map, constructed by projecting the total 
polarization map onto the $E$ eigenmodes found in Equation 
\ref{eqn:gen_eigen}. Bottom Row: pure \bmode\ map, constructed by 
projecting the total polarization map onto the $B$ eigenmodes of 
Equation \ref{eqn:gen_eigen}.}
   \label{fig:E_map_proj}
\end{figure*}

The effectiveness of the purification matrix given by Equation 
\ref{eqn:purification} can be immediately tested by applying the 
operator to a vector of [$Q,U$] maps simulated with the standard pipeline.
The upper left map of Figure 
\ref{fig:E_map_proj} shows an observed map whose input is 
unlensed-\lcdm\ \emodes, and the next two rows in that column show the 
resulting $E$ and $B$ maps after projection onto the pure $E$ and $B$ spaces. 
The right column of Figure \ref{fig:E_map_proj} shows an observed input map 
with both unlensed-\lcdm\ and $r=0.1$ projected onto pure $E$ and \bmodes.

Matrix purification is integrated with the existing \bicep\ analysis code 
by applying the purification operator to maps before calculating the 
power spectra. Since the observation matrix is only used for this 
purification step, the purification matrix need only work well enough 
to result in $E/B$ leakage less than the noise level of the experiment. 
Therefore, it is acceptable to use an approximate observation matrix 
constructed from a subset of the full observation list, as long as it 
is representative of the full scan strategy. This shortcut was employed 
in \citet{keck13}, but the results shown in this 
section from \biceptwo\ are from the full set of observations.

\begin{figure} 
\begin{center}
\resizebox{\columnwidth}{!}
{\includegraphics{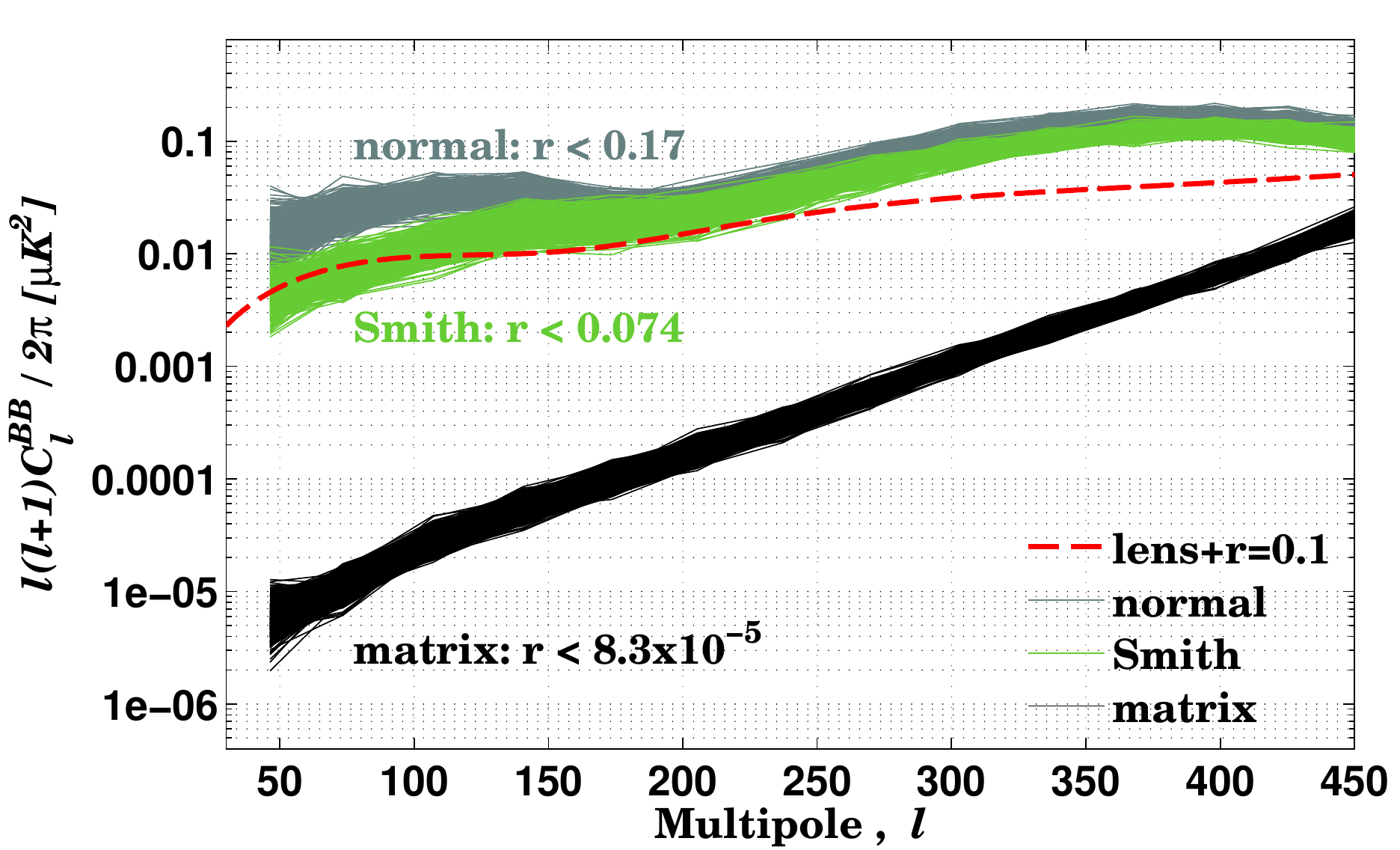}}
\end{center}
  \caption[Purification Effectiveness: \lcdm\ Power Spectra]{$BB$ power 
spectra of noiseless unlensed-\lcdm\ ($r=0$) simulations, estimated using 
various methods, demonstrating the effectiveness of the \biceptwo\ 
purification matrix. The $E/B$ leakage using the matrix estimator is 3 orders 
of magnitude lower than other methods.}
 \label{fig:projection_spectra}
\end{figure}

Figure \ref{fig:projection_spectra} compares the spectra for purified 
maps to the spectra from maps without purification, and to the spectra 
found using the improved estimator suggested in \citet{smith06}. Both the 
purified maps and unpurified maps use the standard $E$ and $B$ estimator in 
Fourier space, and we refer to the unpurified maps processed this way as 
the `normal method.' Figure \ref{fig:projection_spectra} shows the spectra 
for 200 noiseless unlensed-\lcdm\ simulations passed through 
the three estimators. The leaked power is roughly three orders of 
magnitudes smaller when using the matrix purification than when using 
the normal method or Smith estimator.
While the Smith estimator improves over the normal estimator by 
eliminating $E/B$ leakage from apodization, it does not account for 
spatial filtering, which is a significant 
source of $E/B$ leakage in the analysis pipeline.
The mean of the 
leaked power is de-biased from the final power spectra, so what matters is the 
variance of the leaked spectra.
Computing the 95\% confidence limits based 
on the variance in each of the three methods, we find that in the absence 
of \bmode\ signal or instrumental noise, the matrix estimator achieves a 
limit on the tensor-to-scalar ratio of $r<8.3\times10^{-5}$, while the normal 
method and Smith estimator achieve limits of $r<0.17$ and $r<0.074$ 
respectively. 

  \begin{figure}
\begin{center}
\resizebox{\columnwidth}{!}
{ \includegraphics{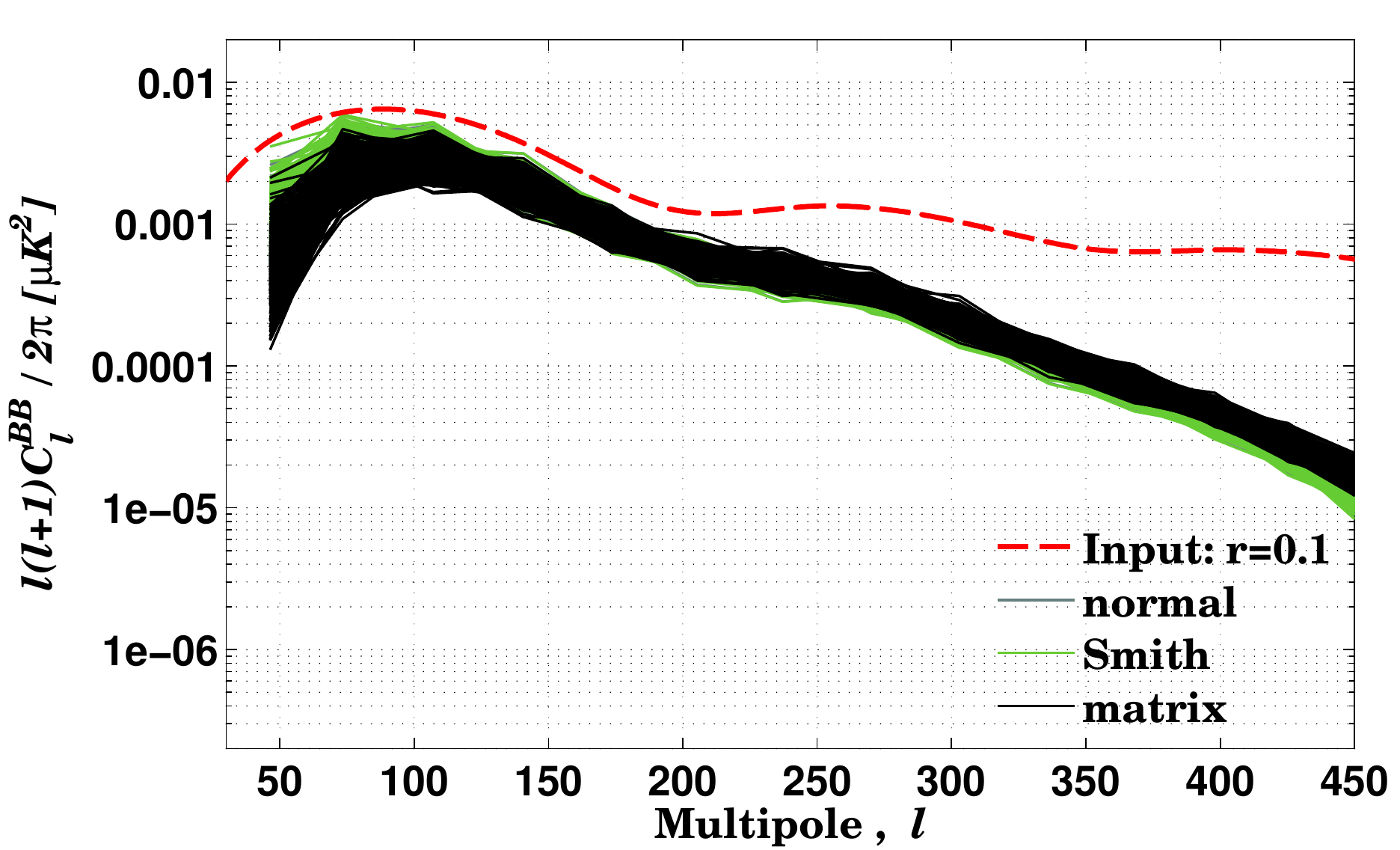} }
\end{center}
 \caption[Purification Effectiveness: $r=0.1$ Power Spectra]{$BB$ power 
spectra of noiseless unlensed ($r=0.1$) tensor only simulations, 
estimated using various methods. All methods suffer from loss of power due 
to filtering and beam effects. The removal of ambiguous modes at low $l$ 
results in a further decrease in power for the matrix method. Note that the 
spectra in this plot have not been corrected for the beam and filter 
suppression factors, but in Figure \ref{fig:projection_spectra} the 
correction is applied.}
   \label{fig:projection_spectra2}
\end{figure}

 Figure \ref{fig:projection_spectra2} shows the spectra from the three 
estimators for input maps containing only input \bmodes\ at the level 
of $r=0.1$. The spectra for all three estimators show beam roll off at
high $l$.
At the lowest $l$, the filtering prevents 
large angular scale modes from being measured. For multipoles around 
$l\sim100$, the matrix estimator recovers slightly less signal than the 
other two methods. However, the extra power measured by the other methods 
near $l\sim100$ largely comes from the ambiguous modes. On the left of 
Figure \ref{fig:projection_spectra}, these ambiguous modes are seen as 
the bump in the normal and Smith method near $l\sim100$.

The total number of degrees of freedom in each band power can be estimated 
according to the formula:
\begin{equation}
N_{l'}=2 \frac{(m_{l'})^2}{\sigma_{l'}^2},
\end{equation}
where $m_{l'}$ is the mean of the simulations in band power $l'$ and 
$\sigma_{l'}^2$ is the variance of the simulations in the band power. 
Table \ref{tab:ndof} shows the number of degrees of freedom for the three 
estimators for a tensor \bmode. The fewer degrees of freedom at 
low $l$ for the matrix estimator are consistent with the decrease in 
recovered power on the left side of Figure \ref{fig:projection_spectra}. 
The highest $l$ bins in Table \ref{tab:ndof} also show fewer degrees of 
freedom in the matrix estimator because the purification 
matrix includes a limited number of pure eigenmodes, as shown in 
Figure \ref{fig:eigenvals}.

\begin{deluxetable}{l c c c}
\tablecolumns{5} 
\tablewidth{0pc} 
\tablecaption{Degrees of freedom in binned $BB$ power spectra for different 
estimators}
\tablehead{
\vspace{3mm}\\
\colhead{} & &\colhead{\textbf{Degrees of Freedom}}  \\
\vspace{2mm}\\
\colhead{Bin center, \textit{l}}& \colhead{Normal} & \colhead{Smith}& \colhead{Matrix}}
\vspace{2mm}
\startdata 
\\[5pt]
37.5&12.9&15.5&8.6\\[2pt]
72.5&40.9&41.4&34.8\\[2pt]
107.5&71.4&69.7&66.8\\[2pt]
142.5&83.7&81.2&81.7\\[2pt]
177.5&120.6&116.4&116.9\\[2pt]
212.5&156.1&153.0&141.9\\[2pt]
247.5&172.0&172.9&145.8\\[2pt]
282.5&202.8&200.8&177.4\\[2pt]
317.5&189.0&185.7&155.0\\[2pt]
\vspace{2mm}\\
\enddata 
\label{tab:ndof}
\end{deluxetable}

\begin{figure}[h!] 
\begin{center}
\resizebox{\columnwidth}{!}
{ \includegraphics{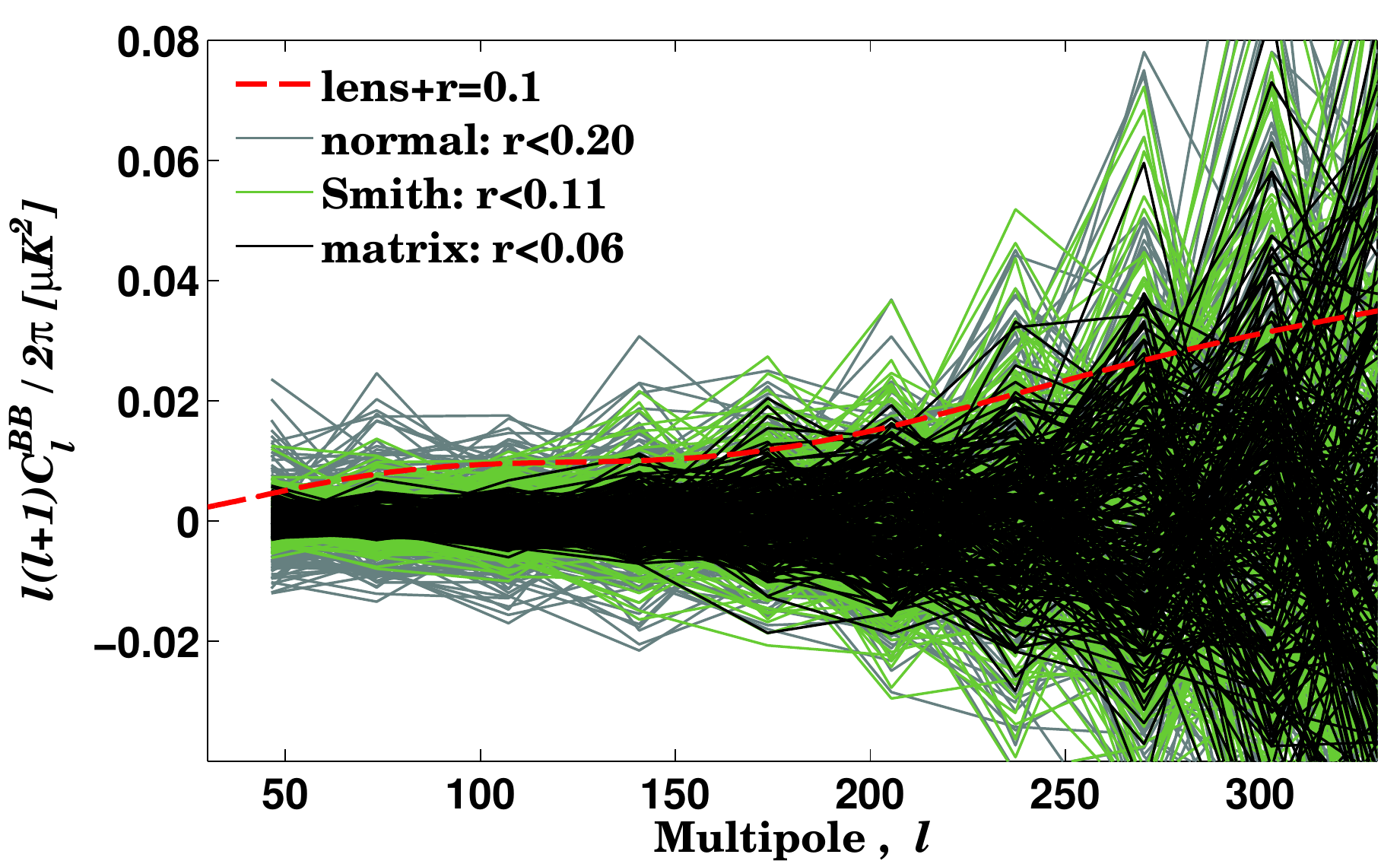} }
\end{center}
  \caption[Purification effectiveness: \lcdm\ + noise Power Spectra]{$BB$ power 
spectra of unlensed-\lcdm\ ($r=0$) + \biceptwo\ noise simulations, estimated 
using various methods, demonstrating the effectiveness of the \biceptwo\ 
purification matrix. For \biceptwo\ noise levels the 
constraint on $r$ (in the absence of signal) is improved by about a factor of 
two. (The mean of the noise and leakage have been de-biased in each case.)}
   \label{fig:projection_spectra_spn}
\end{figure}

Figure \ref{fig:projection_spectra_spn} shows signal plus noise spectra 
for a set of 200 unlensed-\lcdm+noise spectra. The noise 
simulations are the standard \biceptwo\ sign flip realizations discussed in 
\citet{bicep2_res}. In Figure \ref{fig:projection_spectra_spn}, 
the mean noise level and leaked $BB$ power are de-biased. The resulting 
ensemble of simulations is used to construct the errorbars in the final 
spectra. The tighter distribution of the matrix estimator simulations 
is a result of the decrease in $E/B$ leakage. Using the matrix estimator 
results in an improvement in the $r$ limit, for \biceptwo\ noise level and 
filtering in the absence of \bmode\ signal, of about a factor of two 
over the Smith method. The remaining variance in the BB spectrum of the 
matrix estimator is instrumental noise.

\subsection{Transfer functions}
\label{sec:transfer}

   The observation matrix transforms an input \healpix\ map into an observed 
map with a simple matrix multiplication. The speed of the operation 
facilitates the calculation of analysis transfer functions, which are a 
necessary component of the pseudo-$C_l$ \master\ algorithm \citep{hivon02}. 

  We start with input maps, $\mathbf{m_l}$, which are delta functions in a 
particular multipole. These maps are then observed using the matrix 
$\mathbf{R}$. The spectra calculated from these maps represent the response 
in our analysis pipeline to the input delta function, in a manner 
conceptually analogous to Green's functions. 

  Our procedure uses two sets of \healpix\ maps, one set corresponding 
to $TT=TE=EE=1$ and one set with $TT=BB=1$. The observation matrix is 
used to create maps for $l$=$1$ through $700$, with 100 random realizations 
for each $l$. Processing the 140,000 maps would be infeasible without the 
observation matrix, but using the observation matrix it can be 
accomplished in a few hours.

\subsubsection{Band power window functions}
\label{sec:bpwf}
  The power spectra of the output maps for a particular $l$ are 
averaged over the $N=100$ realizations. The averaged spectra are used to 
form a band power window function, $\mathbf{\mathcal{M}}^{XX}_{ll'}$, for a 
particular band power, ${l'}$, which is a function of the input multipole 
of the delta function, ${l}$:
\begin{equation}
\mathbf{\mathcal{M}}^{XX}_{ll'}=\frac{\sum_{r=1}^{N}\boldsymbol{\mathcal{F}}_{ll'}(\mathbf{R}\mathbf{m_l})}{N},
\end{equation}
 where  $\boldsymbol{\mathcal{F}}_{ll'}$ is the analysis pipeline's transformation from 
map to power spectra\footnote{Including the steps: apply matrix 
purification, two dimensional Fourier transform, 
construction of $E$ and $B$, and binning to one dimensional spectra.} 
and $XX=\{TT\rightarrow{TT},TE\rightarrow{TE},EE\rightarrow{EE},EE\rightarrow{BB},BB\rightarrow{BB},BB\rightarrow{EE}\}$. 
When the input maps contain \emodes, and we measure the $BB$ spectra, 
the result is the $EE\rightarrow{BB}$ band power window. The use of the 
purification matrix prevents leakage and makes these band power windows 
have much lower amplitude than the $EE\rightarrow{EE}$ or $BB\rightarrow{BB}$ ones. In \biceptwo, it has been 
standard procedure not to use the \emode\ purification matrix, hence 
the $BB\rightarrow{EE}$ band power window contains (irrelevant)
leakage from $B$ into $E$.

  Calculating the band power window functions in this way accounts for 
all aspects of our instrument and analysis: beam convolution,
sky cut, map projection, polynomial filtering, scan-synchronous signal 
subtraction, deprojection, and power spectrum estimation.

\begin{figure*} 
\begin{center}
\resizebox{\textwidth}{!}
{ \includegraphics{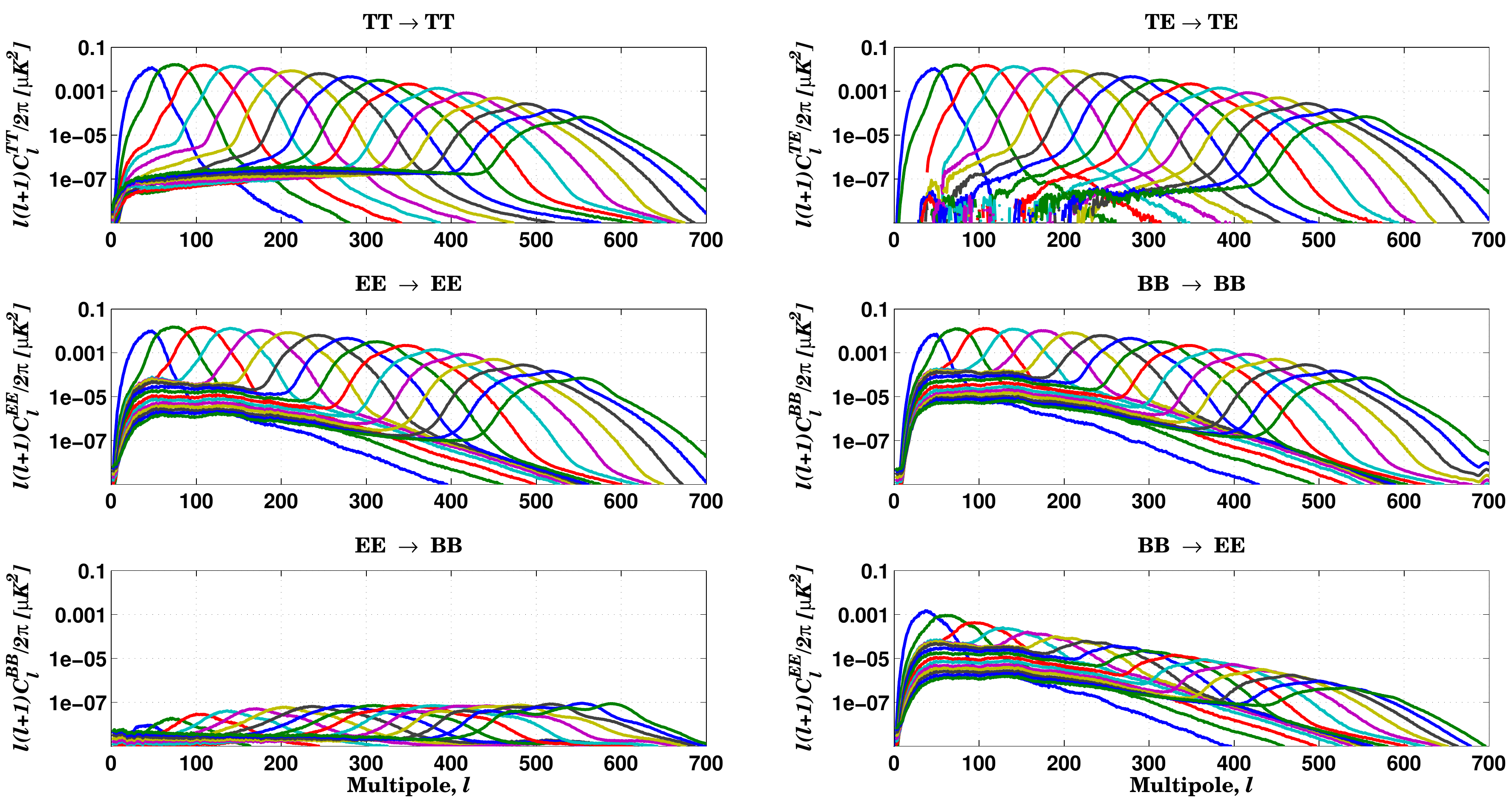}}
\end{center} 
  \caption[Band power window functions]{Band power window functions, 
$\mathbf{\mathcal{M}}^{XX}_{ll'}$. Filtering causes mixing of power from low 
multipoles up to higher multipoles. As noted in Section \ref{sec:bpwf}, the $BB\rightarrow{EE}$ panel in the bottom right shows significantly more power since matrix purification is not applied to \emodes.}
   \label{fig:bpwf_log}
\end{figure*}

Figure \ref{fig:bpwf_log} shows the results of the calculation. 
Although we bin in annular rings of the two dimensional 
power spectra, the filtering operations move power from other $l$ into 
those bins. This means the band powers are sensitive to a broader range in 
$l$ than the nominal range of multipoles. The 
broad shelf in the  band power window functions at lower $l$ is due to 
filtering. 

Table \ref{tab:bpwf} shows the `nominal' and `measured' centers and edges of 
the band power bins, where `nominal' refers to the defined range of annular 
rings in the two dimensional power spectra and `measured' refers to the 
center and $\pm{1}\sigma$ range found in the end to end calculation 
discussed in this section. The center is the mean of the bandpower window 
function and the $\pm{1}\sigma$ interval corresponds to the percentiles 
of the bandpower window function between $16\%$ and $84\%$.

\begin{deluxetable}{lccccccc}
\tablecolumns{7} 
\tablewidth{0pc} 
\tablecaption{$BB$ Band Power Widths}
\tablehead{
\vspace{3mm}\\
\colhead{} & &\colhead{\textbf{Nominal}}& & &\colhead{\textbf{Measured}} & \\
\vspace{2mm}\\
\colhead{Bin Number} & \colhead{low} & \colhead{center}& \colhead{high} & \colhead{low} & \colhead{center}& \colhead{high}}
\vspace{2mm}
\startdata 
\\[5pt]
1&20.0&37.5&55.0&37.0&46.4&54.0\\[2pt]
2&55.0&72.5&90.0&59.0&73.4&86.0\\[2pt]
3&90.0&107.5&125.0&92.0&107.2&121.0\\[2pt]
4&125.0&142.5&160.0&125.0&140.7&157.0\\[2pt]
5&160.0&177.5&195.0&158.0&173.7&192.0\\[2pt]
6&195.0&212.5&230.0&189.0&205.4&227.0\\[2pt]
7&230.0&247.5&265.0&220.0&237.2&262.0\\[2pt]
8&265.0&282.5&300.0&253.0&270.2&298.0\\[2pt]
9&300.0&317.5&335.0&285.0&302.9&333.0\\[2pt]
\vspace{2mm}\\
\enddata 
\tablecomments{Nominal and measured centers and edges of the
band power bins.
The measured values are extracted from the band power window
functions shown in Figure \ref{fig:bpwf_log}.
The low/high values for the latter are the $\pm1\sigma$ points.}
\label{tab:bpwf}
\end{deluxetable}

\subsubsection{Suppression factor}

The integrated area under the curve of each band power window function 
represents the response of that band power measurement to an input spectrum. 
We call this set of values the suppression factor, $S_{l'}$, since they 
approximate how our analysis pipeline suppresses power. 

\begin{figure}[h!] 
\begin{center}
\resizebox{\columnwidth}{!}
 {\includegraphics{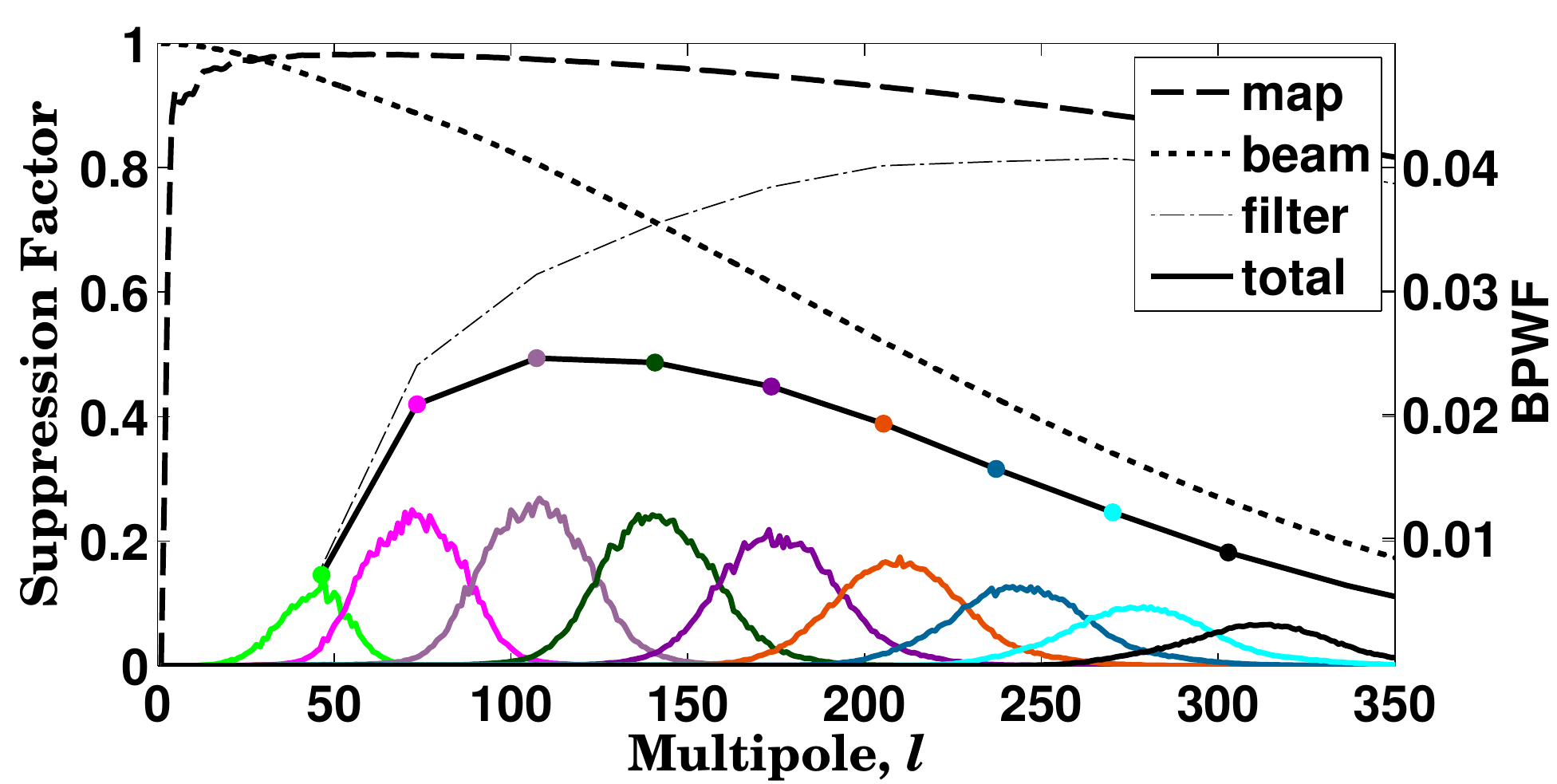} }
\end{center}
  \caption[$BB$ suppression factor]{The $BB$ suppression factor. At 
high $l$ the beam function dominates. At low $l$ the effects of filtering 
dominates. The map window function for the \bicep\ 0.25 degree square pixels
and finite map size is shown as a dashed line. The band power window functions 
are plotted in colors corresponding to individual band powers on a different 
scale.}
   \label{fig:supfac}
\end{figure}

The suppression factor is plotted in Figure \ref{fig:supfac}. At small 
angular scales, the suppression factor is dominated by the roll-off 
of \bicep2's 31 arcminute beam.  The measured array average 150~GHz beam 
function is shown as the dotted line in Figure \ref{fig:supfac}. The 
`map window function,' which includes the finite size of the map and the  
pixel window function for the $\sim0.25$ degree pixels, is shown as the 
dashed line\footnote{Calculated according to \url{http://healpix.jpl.nasa.gov/html/intronode14.htm}}.
At high $l$, the pixel window function is sub-dominant 
to the beam window function. At low $l$ the suppression factor is dominated 
by the timestream filtering effects.

\subsection{Computing challenges}

Building the observation matrix 
requires constructing,  multiplying, and finally averaging a large number 
of sparse matrices. Applying the observation matrix to the true sky 
signal covariance matrix requires a large matrix multiplication to 
find the product, $\mathbf{R}\mathbf{C}\mathbf{R}^{\top}$.

Matrix multiplication is the dominant contributor to computation time. 
We use matrix multiplication routines built in to \matlab. These routines 
incorporate a number of optimized algorithms for computing matrix products, 
including BLAS \citep{Lawson1979}. At this time we have not compared 
the run time on GPUs with that on CPUs, but are aware of this as a 
possible avenue for reducing the compute time.

Solving the eigenvalue problem using the \matlab\ function 
\href{http://www.mathworks.com/help/matlab/ref/eig.html}{\tt{eig()}} 
takes about 48 hrs and 80 GB of RAM for the full \bicep2 observed 
covariance. For \bicep2, this is small fraction of the total computation 
time. However, for experiments whose maps contain more pixels, the 
difficulty of the eigenvalue problem increases. In these cases, the
use of distributed memory parallel code may be necessary. 

For the \biceptwo\ results, we used computing resources provided by the 
Odyssey cluster at Harvard\footnote{\url{https://rc.fas.harvard.edu/odyssey/}}. 
The Odyssey cluster contains 54K CPUs with 190 terabytes of RAM and 
10 petabytes of storage. High memory nodes have access to 256GB of RAM, 
which is useful for large matrix multiplications. Odyssey uses 
SLURM\footnote{\url{http://slurm.schedmd.com/}} as its queue manager, allowing 
our analysis to utilize the large number of cores available.

Although the raw \biceptwo\ data set comprises roughly 3 TB of data, 
the data products from the steps in the matrix analysis chain use 17 TB 
of storage. Processing during the \biceptwo\ matrix analysis steps 
required roughly 1 million CPU hours. This represents a significant 
portion of the total computing demand of the entire \bicep2 analysis 
effort, which comprised roughly 6 million CPU hours.

\section{Conclusions}
\label{sec:conclusion}

We have described a method for decomposing an observed polarization
field into orthogonal components coming from celestial $E$ and \bmodes. 
The method relies on numerically calculating an observation matrix.
In our case the 
observation matrix encodes the mathematical steps translating the true sky to 
an observed map including polynomial filtering, scan-synchronous subtraction, 
pointing of individual detectors, and linear regression of beam systematics. 

Applying the observation matrix to pixel-pixel covariance matrices for $E$ and 
\bmodes\ transforms the true sky covariance into the observed space. We then
solve for the $E$ and $B$ eigenmodes and select those modes that are 
orthogonal. In this way, the orthogonality relationship of the true sky is 
translated to the observed maps. The method accounts for all types of $E/B$ 
leakage: boundary effects, polynomial filtering, linear regression,
etc.---as long as these properties of the observing strategy and analysis 
have been encoded in the observation matrix, making the method more general than the method presented in \citet{smith06}, which only accounts for boundary effects.

The observation matrix has many other possibilities and in principle
allows construction of fully optimal analyses through to power spectra or cosmological parameters
for a single experiment, or a combination of experiments with
partially overlapping sky coverage.
One simple application which we have explored is to use the observation matrix
to directly produce simulated observed maps from input maps
in a single step.
This is dramatically faster than the previous standard pipeline.
However production of observing matrices is sufficiently costly
that we have not generated them for the many alternate
``jackknife'' data split maps and so for the present standard simulations
are still required.

We find that the matrix based $E/B$ separation performs quite well, 
limiting the leakage to a level corresponding 
to $r<1\times10^{-4}$, well below the noise level for any foreseeable 
CMB experiment. The method should prove useful for future ground based 
or balloon based experiments focused on measuring large angular scale \bmodes. 
Experiments measuring the lensing potential using CMB polarization rely on 
cleanly separating $E$ and $B$ as well and may find the technique 
useful. Additionally, the ongoing search for the imprint of gravitational waves 
in CMB polarization will require mitigation of lensing signal from intervening 
structure and foreground removal, both of which are improved by cleanly 
separating $E$ and $B$.

\acknowledgements

\biceptwo\ was supported by the US National Science Foundation under 
grants ANT-0742818 and ANT-1044978 (Caltech/Harvard) 
and ANT-0742592 and ANT-1110087 (Chicago/Minnesota).    
The computations in this paper were run on the Odyssey cluster
supported by the FAS Science Division Research Computing Group at
Harvard University.
The analysis effort at Stanford/SLAC is partially supported by 
the US Department of Energy Office of Science.
Tireless administrative support was provided by Irene Coyle
and Kathy Deniston.

We thank the staff of the US Antarctic Program and in particular 
the South Pole Station without whose help this research would not have 
been possible.
We thank all those who have contributed efforts to the \bicep /\keckarray\
series of experiments, including the \bicepone\ and \bicep3 teams.

\appendix

\section{Generation of constrained realization \healpix\ maps}
\label{sec:constrained_realizations}
An ensemble of signal only simulations is needed for a \master\ pseudo-$C_l$ 
analysis. We use \camb\ with input \planck\ parameters to construct 
power spectra, $C_l$. The power spectra  are used to generate high resolution 
\healpix\ maps that serve as the starting point for each realization of 
the signal simulations. Lacking any constraints on the realizations, the 
maps generated from the power spectra will vary for both the temperature and 
polarization fields. However, \planck\ has measured the temperature field 
of the CMB to high signal to noise. Since the goal of \biceptwo\ and the 
\keckarray\ is to measure the polarization sky and not the temperature sky, our 
ensemble of simulations does not need to contain variation in the well 
measured temperature sky. We therefore use the temperature field of the 
CMB measured by \planck\ as the template for constrained 
realizations of the polarization field.

We have developed a technique for creating realizations of 
the \emode\ sky consistent with the known $TE$ correlation and the 
measured temperature field. These constrained realizations have been 
shown to contain the same $TE$ correlations and $EE$ spectra over many 
realizations of the temperature sky. However, any particular set of 
constrained realizations based on one temperature sky has a slightly 
different distribution of $TE$ and $EE$ than the full ensemble average.

  The primary motivation for fixing the temperature field is to make the 
deprojection operation discussed in Section \ref{subsec:depro_matrix} into 
a linear operator. Recall that deprojection involves a regression against a 
fixed template of the temperature sky. If the temperature sky varies from 
realization to realization, deprojection becomes non-linear and cannot be 
expressed as a matrix operation. 

\subsection{$TE$ correlation}
We start by fixing the coefficients of the spherical harmonics of 
the temperature sky, $a^{T}_{lm}$. The $a_{lm}$ for the constrained \emodes\ 
is found in \citet{Dvorkin07} and can be derived following the steps of a 
simple Cholesky decomposition. The $2\times2$ covariance between $T$ and 
$E$ for each mode ($l,m$) is:
\begin{equation} 
\left[\begin{matrix}C_l^{TT}&C_l^{TE}\\C_l^{TE}&C_l^{EE}\\\end{matrix}\right].
\end{equation}
The lower triangle Cholesky decomposition of this $2\times2$ 
matrix determines the variance and covariance in each mode:
\begin{equation}
\left[\begin{matrix}
{\sqrt{C_l^{TT}}}&{0}\\{C_l^{TE}}/{\sqrt{C_l^{TT}}}&\sqrt{{C_l^{EE}}-{(C_l^{TE})^2}/{C_l^{TT}}}\\
\end{matrix}\right]
\left[\begin{matrix}
{n^T_{lm}}\\{n^E_{lm}}\end{matrix}\right]
=
\left[\begin{matrix}{a^T_{lm}}\\{a^E_{lm}},
\end{matrix}\right]
\end{equation}
where ${n^T_{lm}}$ and ${n^E_{lm}}$ are unit norm, complex random numbers. 
Since $a^T_{lm}$ are known constraints, we can solve the system of equations:
\[
{\sqrt{C_l^{TT}}n^T_{lm}=a^T_{lm}}
\]
\begin{equation}
{C_l^{TE}}/{\sqrt{C_l^{TT}}}n^T_{lm} + {\sqrt{{C_l^{EE}}-{(C_l^{TE})^2}/{C_l^{TT}}}n^E_{lm}=a^E_{lm}},
\label{eq:constrained}
\end{equation}
substituting the first equation into the second to arrive at an expression 
for $a^E_{lm}$ in terms of $a^T_{lm}$ and the power spectra, $C_l$. To ensure 
the maps are real, the condition  ${a_{lm}^*=(-1^m)a_{l-m}}$ is demanded.

\begin{figure*}[!h] 
\resizebox{\textwidth}{!}
{\includegraphics{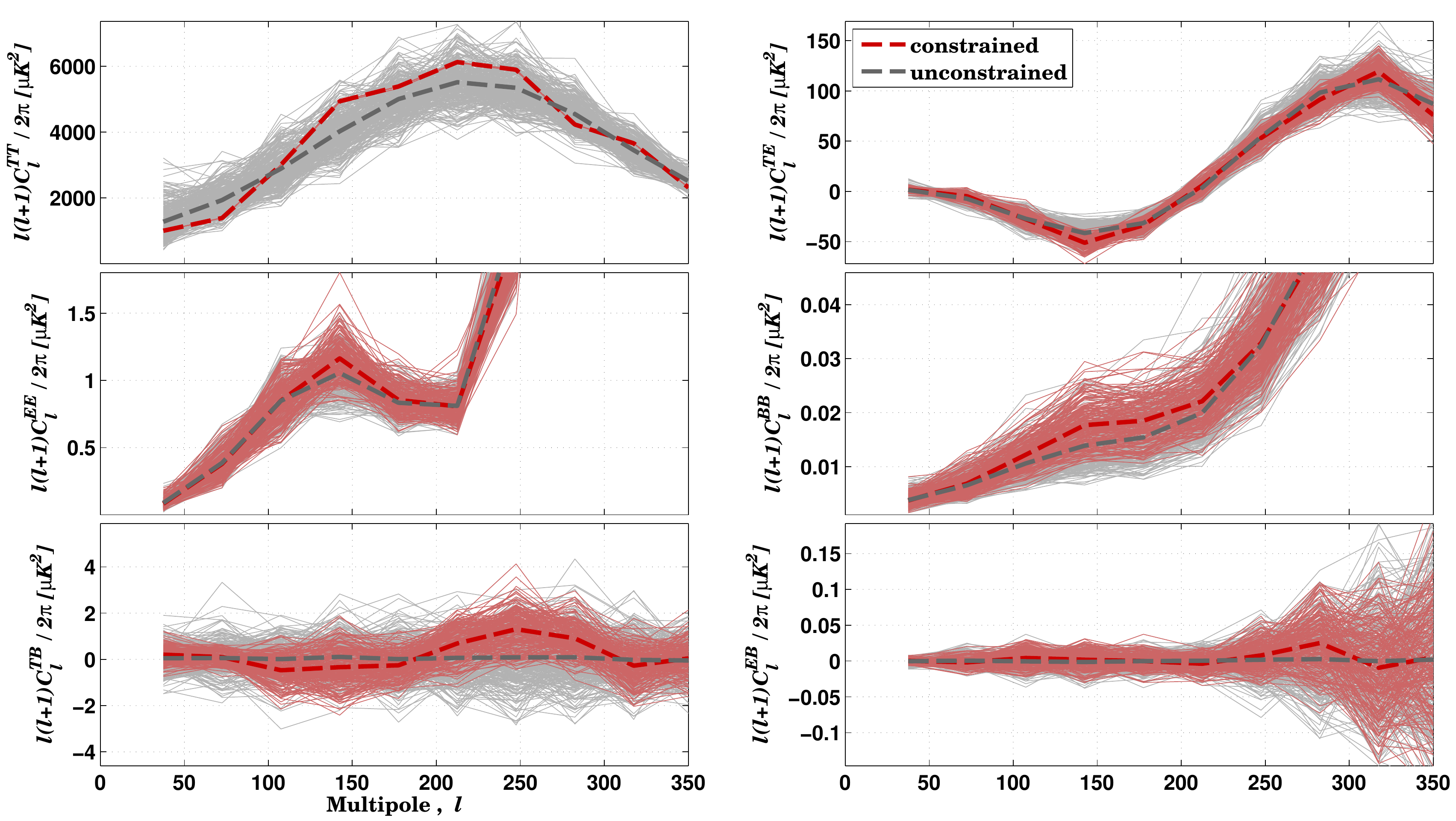} }
  \caption[Power spectra of constrained and unconstrained simulations]{Power 
spectra of unconstrained and constrained simulations. For the constrained 
simulations the $T$ sky is fixed to the Planck NILC map. By chance, 
the $TT$ power in the \bicep\ field is above average near $l = 150$, and 
this leads to increased power and variance in $TE$ and $EE$ for these 
multipoles. The $BB$ spectrum is computed with the Smith estimator for 
both constrained and unconstrained simulations in order to provide an equal comparison between the two as we cannot use the matrix estimator on unconstrained simulations. It therefore contains 
both $E/B$ leakage and lensing signal. The leaked \bmode\ power creates a significant $TB$ signal in the constrained simulations since the leaked \bmodes\ correlate with the temperature template sky.}
   \label{fig:constrained_realizations}
\end{figure*}

Naively, we might expect the constrained $a^E_{lm}$ to have lower 
variance in its power spectrum than the unconstrained $a^E_{lm}$ since 
there is significant $TE$ correlation and the temperature component is 
fixed. However, this is not necessarily the case. When the $a^T_{lm}$ 
fluctuate high, the mean of the constrained ensemble of $a^E_{lm}$'s 
is also high, resulting in increased variance in the ensemble of 
simulations. For particularly high $a^T_{lm}$, this can result in 
larger variance for the constrained simulations than the unconstrained 
simulations. This happens to be the case in the \bicep\ field near $l=150$, as 
seen in Figure \ref{fig:constrained_realizations}. 

In practice, we take the $a^T_{lm}$ from the \planck\ temperature Needlet 
Internal Linear Combination (NILC) map \citep{planckXII}. This map uses the 
multi-frequency coverage of \planck\ to remove the galactic contribution to 
microwave emission, leaving a high signal to noise map of the CMB temperature 
field. The map has some contamination near the galactic plane. However, we have 
found that the impact of this is very local and does not affect the higher 
galactic latitudes where the \bicep\ field is located. The noise level in 
the \planck\ temperature map is fractionally small compared to the 
temperature signal, and we have found that this noise contributes a 
similarly small fraction to the constrained realization of \emodes.

\subsection{Lensing}

The temperature anisotropies in the \planck\ NILC map have been lensed by 
the intervening structure between us and the surface of last scattering. 
This means the $a^T_{lm}$ calculated from the \planck\ NILC map contain the 
effects of lensing and when used in Equation \ref{eq:constrained}, the 
lensing distortion propagates through to the constrained $a^E_{lm}$. Ideally, 
the $a^T_{lm}$ in Equation \ref{eq:constrained} would be from the unlensed 
sky, however, in the absence of an accurate map of the lensing deflection 
field, we have no way of de-lensing the $a^T_{lm}$. 

Because our power spectrum analysis is insensitive to the off-diagonal 
correlations among modes with $l~=l'$ that are produced by lensing, a 
reasonable workaround for this problem is to use the Planck NILC but 
also to use the lensed $C^{TT}_l$ spectrum. The result is an ensemble of 
$a^E_{lm}$ for the lensed $a^T_{lm}$, but which have the correct 
covariance given by $C^{TE}_l$. For the multipole range of interest in 
\bicep2, lensing has a small impact on $a^T_{lm}$. We use the unlensed 
power spectra for $C^{TE}_l$ and $C^{EE}_l$.

Another subtlety incorporating lensing into constrained realizations is 
the question of how to simulate lensing of the polarization sky. This can be 
accomplished using \lenspix\ \citep{lenspix} to numerically lens 
the primordial \emodes, which creates \bmodes\ with the correct statistics. 
For the constrained realizations, this is impossible because 
the NILC map is lensed by the true sky lensing field. The true sky 
lensing field is not known with high signal to noise, and therefore we 
cannot lens the \emodes\ by the same field. 

Our solution to this problem is to lens the \emodes\ with a random 
realization of the lensing field, but use the \planck\ 143~GHz 
map as the temperature field. This procedure ignores lensing correlations 
in $TE$ and $TB$ because the deflection field is different for the 
polarization and temperature. However, the mean of the lensing $TE$ 
and $TB$ is zero, and the additional variance in $TE$ and $TB$ caused 
by lensing is small enough that it can be ignored on degree scales.

\bibliographystyle{apj}
\bibliography{ms}

\end{document}

%% file: defs.tex
\def\bicep{{\sc Bicep}}

\def\bicepone{{\sc Bicep1}}
\def\biceptwo{{\sc Bicep2}}

\def\keckarray{{\it Keck Array}}
\def\planck{{\it Planck}}

\def\QUAD{{\sc QUaD}}

\def\polarbear{{\sc Polarbear}}

\def\actpol{{\sc ACTP}ol}
\def\sptpol{{\sc SPTpol}}

\def\camb{{\tt CAMB}}

\def\synfast{{\tt synfast}}
\def\healpix{{\tt HEALPix}}
\def\lenspix{{\tt LensPix}}
\def\master{{\tt MASTER}}

\def\matlab{{\tt MATLAB}}
\def\lenspix{{\tt LensPix}}
\def\xpure{{\tt Xpure}}

\def\emode{$E$\nobreak-\nobreak{}mode}
\def\bmode{$B$\nobreak-\nobreak{}mode}
\def\emodes{$E$\nobreak-\nobreak{}modes}
\def\bmodes{$B$\nobreak-\nobreak{}modes}

\def\lcdm{$\Lambda$CDM}
\def\grad{{\vec \nabla}}

\def\scanset{\scriptscriptstyle{\mathbb{S}}}
\def\pairs{\scriptscriptstyle{\mathbb{P}}}

\def\aap{Astr. \& Astroph.}

\def\apjl{Astrophys.\ J.\ Lett.}
\def\apjs{Astrophys.\ J.\ Suppl.}